\documentclass[e-print]{aa}  
\usepackage{graphicx}
\usepackage[varg]{txfonts}
\usepackage{xcolor}
\usepackage{subcaption}
\usepackage{mathtools}

\usepackage{natbib,twoopt}
\usepackage[breaklinks=true]{hyperref} 
\makeatletter
  \newcommandtwoopt{\citeads}[3][][]{\href{http://adsabs.harvard.edu/abs/#3}%
    {\def\hyper@linkstart##1##2{}%
     \let\hyper@linkend\@empty\citealp[#1][#2]{#3}}}
  \newcommandtwoopt{\citepads}[3][][]{\href{http://adsabs.harvard.edu/abs/#3}%
    {\def\hyper@linkstart##1##2{}%
     \let\hyper@linkend\@empty\citep[#1][#2]{#3}}}
  \newcommandtwoopt{\citetads}[3][][]{\href{http://adsabs.harvard.edu/abs/#3}%
    {\def\hyper@linkstart##1##2{}%
     \let\hyper@linkend\@empty\citet[#1][#2]{#3}}}
  \newcommandtwoopt{\citeyearads}[3][][]%
    {\href{http://adsabs.harvard.edu/abs/#3}
    {\def\hyper@linkstart##1##2{}%
     \let\hyper@linkend\@empty\citeyear[#1][#2]{#3}}}
\makeatother

\bibpunct{(}{)}{;}{a}{}{,} 

\nolinenumbers

\begin{document} 

   \title{Energy equipartition in Globular Clusters through the eyes of dynamical models}

   \subtitle{}

   \author{M. Teodori\inst{\ref{inst1},\ref{inst2}}
   \and
   O. Straniero\inst{\ref{inst2},\ref{inst3}}
          \and
          M. Merafina\inst{\ref{inst4}}}

   \institute{Department of Mathematics and Physics, University of Campania \textit{Luigi Vanvitelli}, viale Lincoln 5, 81100 Caserta, Italy, \email{matteo.teodori@unicampania.it}\label{inst1}
         \and
             INAF - Osservatorio Astronomico d'Abruzzo, Via M. Maggini, 64100 Teramo, Italy,  \label{inst2}
        \and
             INFN - sezione di Roma, Piazzale Aldo Moro 2, 000185, Rome, Italy,
             \email{oscar.straniero@inaf.it}\label{inst3}
        \and
            Department of Physics, University of Rome La Sapienza, Piazzale Aldo Moro 2, I-000185
Rome, Italy, \email{marco.merafina@roma1.infn.it}\label{inst4}
             }

    \date{\null}
    {

  \abstract
  {Since their birth, globular clusters experience a very peculiar dynamical evolution.
Gravitational encounters drive these systems toward energy equipartition, mass segregation and evaporation altering structural, spatial and kinematic features.}
 {We determine the dynamical state of a few globular clusters by means of a multi-mass King-like dynamical model. Our work focuses on the prediction of the energy equipartition degree and its relation with model parameters.}
{The dynamical model parameters are adjusted in order to reproduce the observed velocity dispersion, as derived from HST proper motion data, as a function of the stellar mass.
In such a way, we estimate $\Phi_0$, a measure of the gravitational potential well. The same fit is repeated by means of the Bianchini relation, a function obtained by interpolating on N-body simulations results. The relationship between $\Phi_0$ and the Bianchini equipartition mass $m_\mathrm{eq}$ has been studied and the structural properties, such as concentration $c$, number of core relaxation timescales $N_\mathrm{core}$ and core radius $r_\mathrm{c}$, are discussed.
To obtain an independent estimate of $\Phi_0$, we also fit observed surface brightness profiles by using the predicted surface density and a mass-luminosity relation from isochrones.}
{The quality of the fits of the velocity dispersion-mass relationship obtained by means of our dynamical model is comparable to those obtained with the Bianchini function.
Nonetheless, when the Bianchini function is used to fit the projected velocity dispersion, the resulting degree of equipartition is underestimated.
On the contrary, our approach provides the equipartition degree at any radial or projected distance by means of $\Phi_0$. 
As a result, a cluster in a more advanced dynamical state shows a larger $\Phi_0$, as well as larger $N_\mathrm{core}$ and $c$, while $r_\mathrm{c}$ decreases.
The estimates of $\Phi_0$ obtained by fitting surface brightness profiles result compatible at $2\sigma$ confidence level with those from internal kinematics, although further investigation of statistical and systematic errors is required.}
{Our work illustrates the predicting power of dynamical models to determine the energy equipartition degree of globular clusters. 
They are a unique tool in determining structural and kinematic properties, also in case of poor observational data, as for the most crowded regions of a cluster, where stars are barely resolved.}

   \maketitle
%
\nolinenumbers

\section{Introduction}
Searching in the surrounding of the Milky Way, one could be stuck in stellar systems spherically shaped, with an increasing density of stars toward their center, named Globular Clusters (GCs). They are commonly known for their very old age, which reaches $\sim 10$ Gyrs in most cases or even more. \\
GCs were thought to consist of a single stellar population with low metallicity. 
However, the detection of chemical and photometrical differences first suggested, and then confirmed, the presence of different generations of stars.
A variety of questions raised, establishing a renewed interest in GCs.
Astronomers are studying the possible formation scenarios of such multiple stellar generations, as well as their initial properties, using the enormous amount of observational data collected recently and developing advanced models \citep[see the reviews by][]{BastianLardo2018, Gratton2019, MiloneMarino2022}. 
Behind their intriguing formation and stellar generations, GCs have a very articulated dynamical evolution. 
The high density environment makes them, among the stellar clusters, the only known collisional system in astronomy.
The relaxation process driven by gravitational encounters between stars plays a fundamental role in determining GCs structure and evolution since their birth. 
The efficiency of relaxation brings these systems toward a degree of kinetic energy equipartition among stars \citep{Spitzer1987}, altering their structural and kinematic features.
Such internal dynamical processes affect the spatial distribution of stars, a phenomenon known as mass segregation. A system close to equipartition has massive stars with a lower velocity dispersion, which consequently sink toward the center. On the contrary, less massive stars are faster and migrate in the outer regions. 
Here, the gravitational pull of our Galaxy subtracts stars from the cluster, leading to a limited phase-space domain, through what is known as evaporation \citep{Ambartsumyan1938,vonHoerner1957,King1958I,King1958II,Spitzer1987}, and producing the tidal streams \citep{Odenkirchen2001,PiattiCarballo-Bello2020}. The joint effect of mass segregation and evaporation flattens the mass function, whose initial shape is largely unknown.
Then, the comprehension of internal dynamics can give hints on the Initial Mass Function.
GCs mass loss also contributes in populating the Galactic environment since their formation. Depending on their orbits, they can suffer shocks and inhomogeneities in the galactic potential \citep{Webb2019}, poorly known in the early GCs life. 
Regarding multiple stellar generations, these dynamical processes coupled with stellar evolution favor a strong and predominant loss of first generation stars in the early evolutionary phases of clusters, suggesting the origin of halo stars. Furthermore, the internal dynamics mixes these populations, mainly erasing their initial spatial and kinematic differences, further complicating the characteristics of the system.
Studying GCs dynamics can thus reveal fundamental insights concerning the early environment of our Galaxy, as well as their contribution to its evolution \citep[see][chap.~8]{Gratton2019}. \\
Although it is well known that massive stars are more centrally concentrated than less massive ones, a theoretical description of energy equipartition, mass segregation and evaporation is still missing and astronomers are trying to quantify these processes. 
A satisfying explanation of such an intricate stellar dynamics is required to extrapolate clues on the primordial environment and the formation scenario of GCs. It can give hints to solve the puzzle of the formation of multiple stellar generations, describing their dynamical evolution.  
However, even single stellar population dynamical phenomenology still needs to be fully understood.\\
Focusing on energy equipartition, several scientists are exploring GCs evolution through N-body simulations, to study the efficiency of this process and its relation with structural and internal properties, as well as initial conditions \citep{TrentiVanDerMarel2013, WebbVesperini2017, PavlikVesperini2021April, PavlikVesperini2021Nov, PavlikVesperini2022}. 
Among such works, the most important in our perspectives is the one from \citet{Bianchini2016}. Here, with a set of N-body simulations, the authors look at the degree of energy equipartition through the velocity dispersion as function of stellar mass, in the inner regions of clusters (namely inside the half-mass radius). They found an analytic formula to fit such a quantity, that is
\begin{equation}
    \sigma(m) = \begin{cases}
       \sigma_\mathrm{0,B} \exp{\left(-\frac{m}{2m_\mathrm{eq}} \right)}  & \mbox{if}\> m\leq m_\mathrm{eq},\\
       \sigma_\mathrm{eq} \left(\frac{m}{m_\mathrm{eq}} \right)^{-1/2} & \text{if}\> m>m_\mathrm{eq} ,
    \end{cases}
    \label{eq:BianchiniFittingFunc}
\end{equation}
where the equipartition mass $m_\mathrm{eq}$ is the parameter that quantifies the degree of energy equipartition, while $\sigma_\mathrm{0,B}$ is a normalization constant and $\sigma_\mathrm{eq}=\sigma_\mathrm{0,B}\exp{(-1/2)}$.
In a subsequent paper, \citet{Bianchini2018} characterize the variation of $m_\mathrm{eq}$ during GCs evolution and its dependence on the radial coordinate.\\
The provided fitting function includes the limit of complete equipartition for masses $m > m_\mathrm{eq}$, that is $\sigma(m) \propto m^{-1/2}$. Then, $m_\mathrm{eq}$ tells which masses reached the complete equipartition, with high-mass stars closer to equipartition than low-mass stars.
Furthermore, since the degree of equipartition is expected to increase with the dynamical state of a cluster, the $m_\mathrm{eq}$ parameter shows a relation with the concentration and the number of relaxation timescales a cluster has seen during its life \citep{Bianchini2016}. Such work gained attention in literature due to the simplicity of the analytical function. It was used in N-body simulations to quantify the degree of energy equipartition in GCs hosting multiple stellar populations \citep{Vesperini2021} and to explore the effect of anisotropy, primordial binaries, tidal field and black holes in equipartition \citep{PavlikVesperini2021April, PavlikVesperini2021Nov,PavlikVesperini2022, ArosVesperini2023}. The exponential fitting function in Eq. \eqref{eq:BianchiniFittingFunc} is often compared with the classical power law $\sigma(m) =\sigma_\mathrm{s} (m/m_\mathrm{s})^{-\eta}$ with $\eta = 0.5$ for complete equipartition, $\sigma_\mathrm{s}$ and $m_\mathrm{s}$ scale values for the velocity dispersion and mass. \\
Thanks to HST proper motion, observers can now both measure internal kinematics and estimate stellar masses, quantifying the equipartition level in a few GCs \citep{Libralato2018, Libralato2019, Libralato2022}.
In \citet{Watkins2022} (hereafter W22), the authors analyzed the degree of energy equipartition in nine GCs, using the \citet{Bianchini2016} fitting function and the classical one, with the respective parameters. The studied clusters have most of their stars, if not all, out of complete equipartition.\\
Although the Bianchini function reproduces well the observable, it does not reveal the physics behind the equipartition process. Our aim is to compare measured data and such a fitting function with theoretical results, which give a higher knowledge level in cluster dynamics. 
Multi-mass King-like dynamical models have the advantages of being self-consistent, that is they are directly obtained from the physics of the system, namely the gravitational interaction between stars, the mass distribution and the limited available phase-space. As a consequence, the velocity dispersion dependence on mass is predicted by the model itself, and it is not related to an interpolation function that fits well simulations outputs and observations. In this paper we carefully discuss the relationship with N-body based results, namely the successful Bianchini fitting function, as compared to model outcomes.
We recall that N-body simulations study the temporal evolution of stars in clusters, numerically integrating the Newtonian equation of motion, after choosing suitable initial conditions. Current state-of-the-art simulations include a variety of phenomena, like the treatment of stellar evolution, binary systems, compact objects formation and tidal forces, that make them a powerful numerical tool to simulate more realistic clusters. 
However, the definition of such recipes and the choice of initial conditions require themselves accurate modeling. Theoretical assumptions and prescriptions can affect significantly the simulations results. This brings most researchers to the exploration of parameters of those models determining initial conditions and describing the considered phenomena. 
As a result, N-body methods are a modeling tool needed to explore the temporal evolution, but they require theoretical references to consistently set up the simulations itself.
Although the computational resources have strongly grown in the last years, realistic simulations require large and advanced computer infrastructures as well as computational time, which grows with the square of the number of simulated stars.\\
On the contrary, exploring dynamical models for GCs is much less computationally expensive. In our approach, a single equilibrium configuration can be obtained in few seconds or minutes, depending on the desired precision. 
Furthermore, they can predict radial and projected profiles for several observables, such as the surface density (both in mass and in number of stars), the surface brightness (if a mass-luminosity relation is known) and the velocity dispersion.
When using models including a mass distribution, the knowledge of the distribution function characterizes not only the phase-space but also the mass function. It results as a description of the seven-dimensional space of masses, positions, and velocities.
Such theoretical framework is clearly limited by the assumptions made, as in our case where isotropic velocities are considered. Furthermore, some theoretical quantities require observational constraints in order to get their numerical values, as usually happens in physical models. 
The temporal evolution is assumed as described by equilibrium configurations with different structural parameters, whose values determine the dynamical state. 
The temporal dependence of such quantities can be addressed with numerical simulations, which offer a unique opportunity in that sense \citep[see][for a relation between the King single-mass model parameter $W_0$ and advanced simulations]{Wang2016}.
Indeed, observing the historical properties of GCs would require to look at the satellites of distant galaxies (i.e., far enough to look back in time into the life of GCs) under the strong assumption that cluster evolution is similar in every galactic environment, which may not be the case. Unfortunately, we are currently limited to GCs of the Milky Way and its satellites and nearby galaxies.\\
Numerical simulations and theoretical modeling can, therefore, provide a broader perspective into the dynamics of GCs.
However, we strongly emphasize that taking advantages of dynamical models is fundamental to understand the physics behind processes. They are an advanced interpretative and predictive tool in the perspective of addressing the major open questions in the field. \\


\section{Methods}
Our main approach consists in comparing the velocity dispersion dependence on stellar mass predicted by our multi-mass dynamical model for GCs, with the Bianchini function in Eq. \eqref{eq:BianchiniFittingFunc}, fitting the observational data by W22.
The fitting procedure leads to an estimate of model parameters, to be related with the Bianchini equipartition mass $m_\mathrm{eq}$.\\
We also look at the surface brightness profiles (SBPs) of the clusters by fitting the measured data by \citet{Trager1995} with our theoretical prediction, which provides an estimate of parameters from a different observable. This allows us to see how confident our model is at reproducing both the observed internal kinematics and luminosity profile.

\subsection{Dynamical model}
Our dynamical model derives from the distribution function (DF) obtained by \citet{King65} as an approximated solution of the Fokker-Planck equation, that comes from the Boltzmann equation for collisional systems \citep{Chandrasekhar1943RevModPhys, Rosenbluth1957, SpitzerHarm1958, Chandra1960}. 
The King model \citep{King66} was a single-mass one, with the DF having a cut-off in the phase-space that takes the Galactic tidal field into account. Such a model describes the equilibrium configuration of the system, once the DF is used to calculate the density profile and solve the Poisson equation for the gravitational potential.
It was used for many years by astronomers to fit the SBPs of GCs and to obtain structural parameters \citep{PetersonKing1975, Trager1995, Harris1996, Miocchi2013}.
A further step on was done first by \citet{DaCostaFreeman1976}, introducing a discrete multi-mass model with ten mass classes, each having its own DF with an energy cut-off and a weight factor to be constrained by observations. They successfully fit M3 surface brightness, where the King model was failing to reproduce both the inner and outer profile.\\
Due to the general good agreement between the King model and observations, little effort was put in developing and exploring multi-mass models, although they are fundamental in the comprehension of several phenomena produced or altered by the mass distribution, like segregation, evaporation, and equipartition 
\citep{Miocchi2006, GielesZocchi2015,Peuten2017, Torniamenti2019, Dickson2023}.
Even if N-body simulations are offering great opportunities in understanding the dynamical evolution of GCs, we underline the importance of developing analytical models to describe physical processes, offering an important tool in the comprehension of such systems, like it was the King DF.\\
The formulation of a continuous multi-mass model was presented by \citet{Merafina2019}, as an extension of the Da Costa \& Freeman one and recovering the King formalism.
The model we present recall the same approach with some improvements concerning the derivation and the relation with the global mass function. 
Our DF is an approximated solution of a generalized expression of the Fokker-Planck equation, valid for collisional systems with a mass distribution (see Eq. \ref{eq:mmFP} in Appendix \ref{sec:App-modelderivation}). The DF keeps the property of being limited in the phase-space. Moreover, not only it brings information on the velocity distribution within the cluster, but it also concerns the mass distribution of the stars, meaning that the mass function is embedded in the DF. Its expression is 
\begin{equation}
    g(r,v,m) = k(m)\,\mathrm{e}^{-m[\varphi(r)-\varphi_\mathrm{0}]/(k_\mathrm{B}\theta)}\left[\mathrm{e}^{-\varepsilon(v,m)/(k_\mathrm{B}\theta)}- \mathrm{e}^{-\varepsilon_\mathrm{c}(r,m)/(k_\mathrm{B}\theta)}\right],
   \label{eq:mmDF}
\end{equation}
where $\varepsilon=mv^2/2$ is the kinetic energy of a star with mass $m$ and $\varepsilon_\mathrm{c}=mv_\mathrm{e}^2/2$ is its cut-off energy with $v_\mathrm{e}=v_\mathrm{e}(r)$ the escape velocity, $\varphi(r)$ is the gravitational potential (with $\varphi_\mathrm{0}$ its value in the cluster center) and $r$ is the radial coordinate. The variable $\theta$ is the thermodynamic temperature, a memory of the Boltzmann DF limit and constant all over the equilibrium configuration \citep{Merafina2017,Merafina2018,Merafina2019} and $k_\mathrm{B}$ is the Boltzmann constant. 
The multiplying factor $k(m)$ weights the DF of each mass $m$, like it was in the \citet{DaCostaFreeman1976} model, although theoretically it gathers some mass-dependent functions resulting from the derivation, related with the mass function and the escape velocity at cluster center (see Appendix \ref{sec:App-modelderivation} for details).\\
The mass density radial profile $\rho(r)$ is computed through an integration over the masses and velocities of the DF in Eq. \eqref{eq:mmDF}, that is
\begin{equation}
    \rho(r) = \int_{m_\mathrm{min}}^{m_\mathrm{max}}\left[\int_{0}^{v_\mathrm{e}(r)} g(r,v,m)\mathrm{d}^3v\right] \> \mathrm{d}m \, ,
    \label{eq:rhor}
\end{equation}
with $m_\mathrm{min}$ and $m_\mathrm{max}$ the extremes of the mass range, where the mass function is valid.
Equation \eqref{eq:rhor} is used to solve the Poisson equation for the gravitational potential. Following the King formalism, one could introduce $W(r) = \varepsilon_\mathrm{c,1}(r)/(k_\mathrm{B}\theta)=m_1 v_\mathrm{e}^2(r)/(2k_\mathrm{B}\theta)=m_1[\varphi_\mathrm{R} -\varphi(r)]/(k_\mathrm{B}\theta)$ and solve the Poisson equation for $W(r)$, obtaining a set of equilibrium configurations each identified by the initial condition $W_{0,1}=m_1[\varphi_R -\varphi_0]/(k_\mathrm{B}\theta)$, like shown in \citet{Merafina2019}.
However, the parameter for determining the configuration should not depend on the scale mass (the models by \citet{DaCostaFreeman1976} and \citet{Merafina2019} were using the greatest mass $m_1$ of the system). Using a generic mass unit $m_\mathrm{u}$ and solving for $W_\mathrm{u}(r) =m_\mathrm{u} [\varphi_\mathrm{R}-\varphi(r)]/(k_\mathrm{B}\theta)$, the expression of the dimensionless Poisson equation keeps the same analytical expression, that is
\begin{equation}
    \frac{1}{x^2}\frac{\mathrm{d}}{\mathrm{d}x}\left( x^2 \frac{\mathrm{d}W_\mathrm{u}}{\mathrm{d}x} \right) = -9\frac{\rho}{\rho_0}, 
    \label{eq:PoissonWu}
\end{equation}
with the dimensionless coordinate $x=r/r_\mathrm{k,u}$ in units of the King radius $r_\mathrm{k,u}=\sqrt{9k_\mathrm{B}\theta/(4\pi G m_\mathrm{u}\rho_0)}$. The initial conditions of Eq. \eqref{eq:PoissonWu} are $W'_\mathrm{u}(x=0)=0$ and $W_\mathrm{u}(x=0) = W_\mathrm{0,u} = m_\mathrm{u} (\varphi_\mathrm{R} -\varphi_0)/(k_\mathrm{B}\theta)$, which determines the solution. 
Here, we use a better parameter to identify the equilibrium configuration, that is $\Phi_0 = (\varphi_\mathrm{R} -\varphi_0)/(k_\mathrm{B}\theta)$, which measures the potential well without depending on a scale mass (as $W_\mathrm{0,u}$ do) and having the dimension of the inverse of a mass.\\
To solve Eq. \eqref{eq:rhor}, the quantities in the DF in Eq. \eqref{eq:mmDF} must be known.
At the present state of development, the factor $k(m)$ has to be constrained from the global mass function $\xi(m)$ with a numerical procedure. Since the integration of $g(r,v,m)$ over the radial coordinate $r$ and the velocity $v$ gives the mass function by definition, we can iteratively constrain $k(m)$. In this work, we consider a single-power law mass function $\xi(m)\propto m^{\alpha}$, with the slope $\alpha$ taken from \citet{Baumgardt2023} or theoretically assumed.
The code that numerically integrates Eq. \eqref{eq:rhor}, once given the slope $\alpha$, the mass range and $\Phi_0$, evaluates $k(m)$ with a convergence procedure and draws radial profiles for each mass composing the system. These include the 3D velocity dispersion profile as function of mass 
\begin{equation}
    \sigma(r,m) = \langle v^2(r,m)\rangle = \frac{\int_0^{v_\mathrm{e}(r)} g(r,v,m)v^2 \mathrm{d}^3v}{\int_0^{v_\mathrm{e}(r)} g(r,v,m) \mathrm{d}^3v},
\end{equation}
 that we project in two dimensions to obtain $\sigma(R,m)$, with $R$ the projected distance from the cluster center. This profile is needed to fit the proper motion observations on $\sigma(m)$.\\
 The model also provides the density profile of each mass, which is used to predict the surface density profile for each mass $\Sigma(R,m)$, by integrating along the line of sight. This allows us to estimate the surface brightness profile $I(R)$, by introducing a mass-luminosity relation. 
 This is a fundamental step when dealing with multi-mass systems in particular. 
 We go through the details of the production of the theoretical SBPs in Sect. \ref{subsubsec:SBPs}, where we discuss the fitting procedure on observational data.

\subsection{Fitting procedure}
\label{subsec:fittingprocedure}

\subsubsection{Velocity dispersion -- mass relationship}
\label{subsubsec:velocitydispersion-mass} 
W22 provide the binned velocity dispersion of stars as function of their estimated mass for a sample of Galactic GCs (see their Table 2).
Their dispersion measures are obtained from proper motion (then in the 2D plane of the sky) and inside a ring in the inner regions of the analyzed clusters.\\
In order to compare the model prediction with the data, we compute $\sigma(m)$ from the projected profile $\sigma(R,m)$ in the same region.
When looking at a projected radial distance from cluster center $R$, the observer intercepts all stars having 3D radius $r \in [R,R_\mathrm{t}]$, where $R_\mathrm{t}=r_\mathrm{t}$ is the tidal radius. Then, we average the 3D profile $\sigma(r,m)$ in that range to obtain the 2D profile $\sigma(R,m)$.
Subsequently, the projected dispersion is averaged in the circular annulus covered by the data sample, drawn around the median radius (see W22, Table 1).  
We use the observed core radius $r_\mathrm{c}$ from the \citet{Harris1996} catalog (2010 edition, hereafter indicated as the Harris catalog) and the theoretical core radius $r_\mathrm{c,th}$ to convert the dimensional coordinates into our dimensionless ones. 
Here, $r_\mathrm{c,th}$ is obtained from the surface density profile, predicted by our model. This coincides with $r_\mathrm{c}$ if the surface brightness profile is proportional to the surface density profile (with a factor only dependent on mass).\\
The overall average procedure returns the velocity dispersion as function of stellar mass, normalized to the dispersion of the lowest mass $\sigma_0$. The shape of $\sigma(m)/\sigma_0$ depends on the model parameters. 
In particular, for an increasing $\Phi_0$ that gives a more advanced dynamical state, the dispersion $\sigma(m)$ is closer to the complete energy equipartition limit $\sigma\propto m^{-1/2}$. Furthermore, more massive stars have a greater degree of equipartition than less massive ones for any value of $\Phi_0$ (see Fig. \ref{fig:sigmam_3D_eqlimit} in Appendix \ref{sec:App-additionalplots}).\\
Concerning the choice of the mass function shape needed to draw theoretical profiles, we set $m_\mathrm{min}=0.1$ and $m_\mathrm{max}=1.0$, while for the slope $\alpha$ we select two approaches:
the first takes it from \citet{Baumgardt2023}, the second one (more general) explores a wide range $\alpha \in [-2.0, 0.0]$, actually adding the slope to the parameter space of the minimization procedure.
As a consequence, the first approach gives the best-fit values of $\sigma_0$ and $\Phi_0$, while the second also returns an estimate for $\alpha$. For the fit, we numerically minimize the $\chi^2$ test value between data and model prediction. 
We apply this procedure to the GCs in our sample, only removing \object{NGC 2808} following W22.
We also fit the velocity dispersion data with the Bianchini function in Eq. \eqref{eq:BianchiniFittingFunc}, to have an estimate of the $\chi^2$, to be compared with our model ones. Although such a fit is less detailed than in W22, we obtain the Bianchini equipartition mass $m_\mathrm{eq}$ and the normalization constant $\sigma_\mathrm{0,B}$.

\subsubsection{Surface brightness profiles}
\label{subsubsec:SBPs}
An extensive catalog of SBPs for Milky Way GCs was released by \citet{Trager1995}. Although more accurate measurements are available from the work conducted by \citet{Noyola2006} with HST photometry, they are restricted to internal regions and to a few subsets of clusters. For these reasons, we choose the \citet{Trager1995} dataset.
To analyze the profiles, we follow a procedure similar to that described by \citet{McLaughlin&vanderMarel2005} and \citet{Zocchi+2012}, that estimated uncertainties on the data and fit SBPs with single-mass dynamical models.
The \citet{Trager1995} SBPs consists of measurements for the logarithm of the projected radial coordinate $R_i$, the surface (apparent) magnitude in the V-band $\mu_\mathrm{V}(R_i)$, as well as its best-fit with Chebyshev polynomials and a weight $w_\mathrm{i}$ for each point according to its by-eye reliability. 
The analysis performed by \citet{McLaughlin&vanderMarel2005} provides an estimate of the uncertainty for each surface brightness measure, namely $\delta \mu_{\mathrm{V}}(R_i) = \sigma_\mu / w_i$, where $\sigma_\mu$ is a constant that varies from cluster-to-cluster.
Following \citet{Zocchi+2012}, we remove data points with $w_i < 0.15$ and we correct each measure for the extinction as derived from the Harris catalog.\\ 
In the context of multi-mass dynamical models, the standard assumption of a constant mass-to-light ratio is too crude. 
To obtain the theoretical profile for the surface brightness, useful to fit the available data, we need a mass-magnitude relation in the V-band. Using theoretical isochrones we can establish such a relationship by assigning to the masses of the dynamical model a value for the corresponding absolute magnitude $M_\mathrm{V}$. 
We use \href{http://basti-iac.oa-abruzzo.inaf.it/index.html}{BaSTI} isochrones \citep{Hidalgo+2018,Pietrinferni+2021,Salaris+2022,Pietrinferni+2024}, considering an age of 13 Gyrs, $[\alpha/\element{Fe}]=+0.4$, the \element{He} mass fraction $Y = 0.247$ and a different metallicity [\element{Fe}/\element{H}] for each cluster as a reference case. The metallicity value is taken from the Harris catalog, where we also get the distance modulus to convert the observed apparent magnitudes into absolute ones.
Due to stellar mass loss, the final mass of each star is naturally different from its initial mass, depending on its evolutionary state. In particular, initially more massive stars are now much brighter than low-mass stars, even if their current mass may be similar. Furthermore, since GCs are very old, and excluding remnants, the most massive stars have around $0.8\,M_\sun$ and their mass loss occurred mainly in the last million years. As a consequence, they did not have the time to adjust their dynamical state, which is better described by their initial mass. 
We then use the initial mass as representative of the dynamical mass, in order to assign the corresponding $M_\mathrm{V}$ value.
We recall that when the stars evolve beyond the main sequence, a large variation in luminosity occurs for a very small variation in their mass. 
These evolved stars are also the more massive ones still alive in the cluster. For these reasons, they play an important role in shaping the surface brightness profile in the more central regions of clusters, because of the mass segregation process.
To account for these effects, we choose a regular mass step, but substantially smaller than that used to determine the velocity dispersion.
Note that the mass range, in particular the maximum mass of stars still alive, is provided by the isochrones and depends on the age of the cluster.
 While this maximum mass has a little effect on the prediction of $\sigma(m)$, it will play an important role in shaping the surface brightness profile.\\
Finally, we follow \citet{Zocchi+2012} regarding the minimization procedure of the $\chi^2$, by converting both the observed and the theoretical SBPs into solar luminosities.\\
Concerning the theoretical prediction, the model gives the surface density profile for each mass $\Sigma(R,m)$. The global SBP $I(R)$ is defined as 
\begin{equation}
    I(R)=\int_{m_\mathrm{min}}^{m_\mathrm{max}} \frac{L_\mathrm{V}}{m}\Sigma(R,m)\> \mathrm{d}m,
\end{equation}
where $L_\mathrm{V}$ is the luminosity obtained by converting the theoretical absolute magnitudes $M_\mathrm{V}$ to solar luminosities, using $M_\mathrm{V,\sun}=4.83$ (see the \href{https://nssdc.gsfc.nasa.gov/planetary/factsheet/sunfact.html}{Sun Fact Sheet} by NASA). 
Although as noted by \citet{Torres2010} $M_\mathrm{V,\sun}$ may suffer from non-negligible uncertainties, its choice is irrelevant in our fits.\\
The computed theoretical profile is normalized to its central value $I_0$. The shape of $I(R)/I_0$ depends on the parameters of the model, and its values are given for the dimensionless radial coordinate. 
However, in our fitting procedure, we must be aware that the observed radial coordinate is in physical units. 
To deal with all these, we define a numerical auxiliary function $h(R/R_\mathrm{c};\Phi_0, \alpha, m_\mathrm{min}, m_\mathrm{max})=I(R)/I_0$ where $R_\mathrm{c}$ is the core radius, $\alpha$ is the mass function slope that we take from \citet{Baumgardt2023}, $m_\mathrm{min}$ and $m_\mathrm{max}$ are the mass extremes given by the isochrone. 
We numerically minimize the $\chi^2$ and obtain an estimate of the best-fit $\Phi_0$, $I_0$ and $R_\mathrm{c}$.
On the theoretical ground, since the 2D and 3D radial grids are the same, $r_\mathrm{c}=R_\mathrm{c}$ and $r_\mathrm{t}=R_\mathrm{t}$.
We apply this method to each cluster for which we have analyzed the velocity dispersion -- mass relation. 
In this way, we can obtain two independent estimations of the $\Phi_0$ parameter.

\subsection{Errors estimation}
\label{subsec:Errors-Estimation}
Our fitting procedure is based on the $\chi^2$ minimization with respect to parameters. In our modeling for the velocity dispersion -- mass relation, such parameters are $\Phi_0$ and $\sigma_0$ for the first approach, while the second one also includes the slope $\alpha$. The minimization of the $\chi^2$ requires that $\partial \chi^2 /\partial a_j = 0$ and $\partial^2 \chi^2 /\partial a_i \partial a_j<0$ where $a_j$ represents the parameters. We obtain such conditions with a numerical procedure. For the uncertainties, we need to compute the matrix $M_{ij} = (1/2)\,\partial^2 \chi^2 /\partial a_i \partial a_j$, whose inverse $M^{-1}$ is the covariance matrix, from which we obtain parameters errors $\delta \Phi_0 = (M^{-1})_{00}^{1/2}$ and $\delta \sigma_0 = (M^{-1})_{11}^{1/2}$ as well as error bands for the velocity dispersion. A similar procedure is done when the slope is considered as a free parameter.\\
Our theoretical prediction for the velocity dispersion as function of stellar mass depends on parameters. We can write that $\sigma(m; \Phi_0, \sigma_0) = \sigma_0 f(m; \Phi_0)$ where $f(m; \Phi_0)$ is the numerical relation obtained from the model (that generally also depends on $\alpha$, fixed in the first approach). To obtain the matrix $M$, we must compute the derivative of the $\chi^2$ with respect to parameters. Specifically, we have an analytical expression for the first and second derivatives with respect to $\sigma_0$.
Concerning the derivatives with respect to $\Phi_0$, we use the Finite Difference Methods for their approximation. For the first derivative, we use
\begin{equation}
    \frac{\partial \sigma}{\partial \Phi_0}= \frac{1}{2}\frac{\sigma(\Phi_0+\Delta \Phi_0) - \sigma(\Phi_0-\Delta \Phi_0)}{\Delta \Phi_0} + \mathcal{O}(\Delta \Phi_0^2),
    \label{eq:dsigmadphi0}
\end{equation}
where $\Delta \Phi_0$ is the numerical step in $\Phi_0$ used to compute the function $f$. Regarding the second derivative, we take
\begin{equation}
     \frac{\partial^2 \sigma}{\partial \Phi_0^2} = \frac{\sigma(\Phi_0+\Delta \Phi_0) -2\sigma(\Phi_0)+\sigma(\Phi_0-\Delta \Phi_0)}{\Delta \Phi_0^2} + \mathcal{O} (\Delta \Phi_0^2).
     \label{eq:d2sigmadphi02}
\end{equation}
Both approximations are considered at second order. 
When the slope is added to the parameter space, we have $\sigma(m; \Phi_0, \alpha, \sigma_0) = \sigma_0 f(m; \Phi_0, \alpha)$. For the derivatives with respect to the slope, we consider the same Finite Difference Eqs. \eqref{eq:dsigmadphi0} and \eqref{eq:d2sigmadphi02} for their approximation, using the numerical step $\Delta \alpha$. The mixed derivatives are obtained similarly, applying the same method.\\
Concerning the fitting procedure on SBPs, we follow the same approach. We determine the covariance matrix by computing (and inverting) $M_{ij} = (1/2)\,\partial^2 \chi^2 /\partial a_i \partial a_j$ where the parameters are $\Phi_0$, $R_\mathrm{c}$ and $I_0$. Except for the latter that has an analytical expression, the first, and second derivative of the $\chi^2$ with respect to $\Phi_0$ and $R_\mathrm{c}$ need the derivatives of $I(R)$ with respect to the same parameters. These are approximated by using the Finite Difference Method, similarly to what already described, as well as for mixed derivatives.
We obtain uncertainties $\delta \Phi_0$, $\delta I_0$ and $\delta R_\mathrm{c}$, from which we can also compute the uncertainty on the tidal radius.

\section{Results} 
We report here the main results concerning our fit of the projected velocity dispersion as function of stellar mass. The estimated model parameters are given and the relation between $\Phi_0$ and the equipartition mass $m_\mathrm{eq}$ is shown and discussed. 
The relations with GCs structural parameters are also presented. \\Finally, we show the estimated $\Phi_0$ parameter by fitting the SBPs and compare them with the values obtained by fitting the $\sigma(m)$ observable.

\subsection{Fitting velocity dispersion to constrain $\Phi_0$}
\label{subsec:VaryingPHI0}
As a first approach, we evaluate $\Phi_0$ by fitting the velocity dispersion data with the model prediction, considering a single power law mass function $\xi(m) \propto m^{\alpha}$ with the slope $\alpha$ taken from \citet{Baumgardt2023}, but with a mass range $0.1<m/M_{\sun}<1.0$, slightly broader than the usual one $[0.2,0.8]\,M_{\sun}$. 
However, it can be shown that the effect of a different range of masses (when reasonable) is negligible with respect to variations in $\Phi_0$ or $\alpha$.\\
In Fig. \ref{fig:NGC6397sigmam_1par} we plot $\sigma(m)$ predicted by the model against dispersion data by W22 and the fit with the Bianchini function for \object{NGC 6397}. 
\begin{figure}[htbp]
    \centering
    \includegraphics[width=\linewidth]{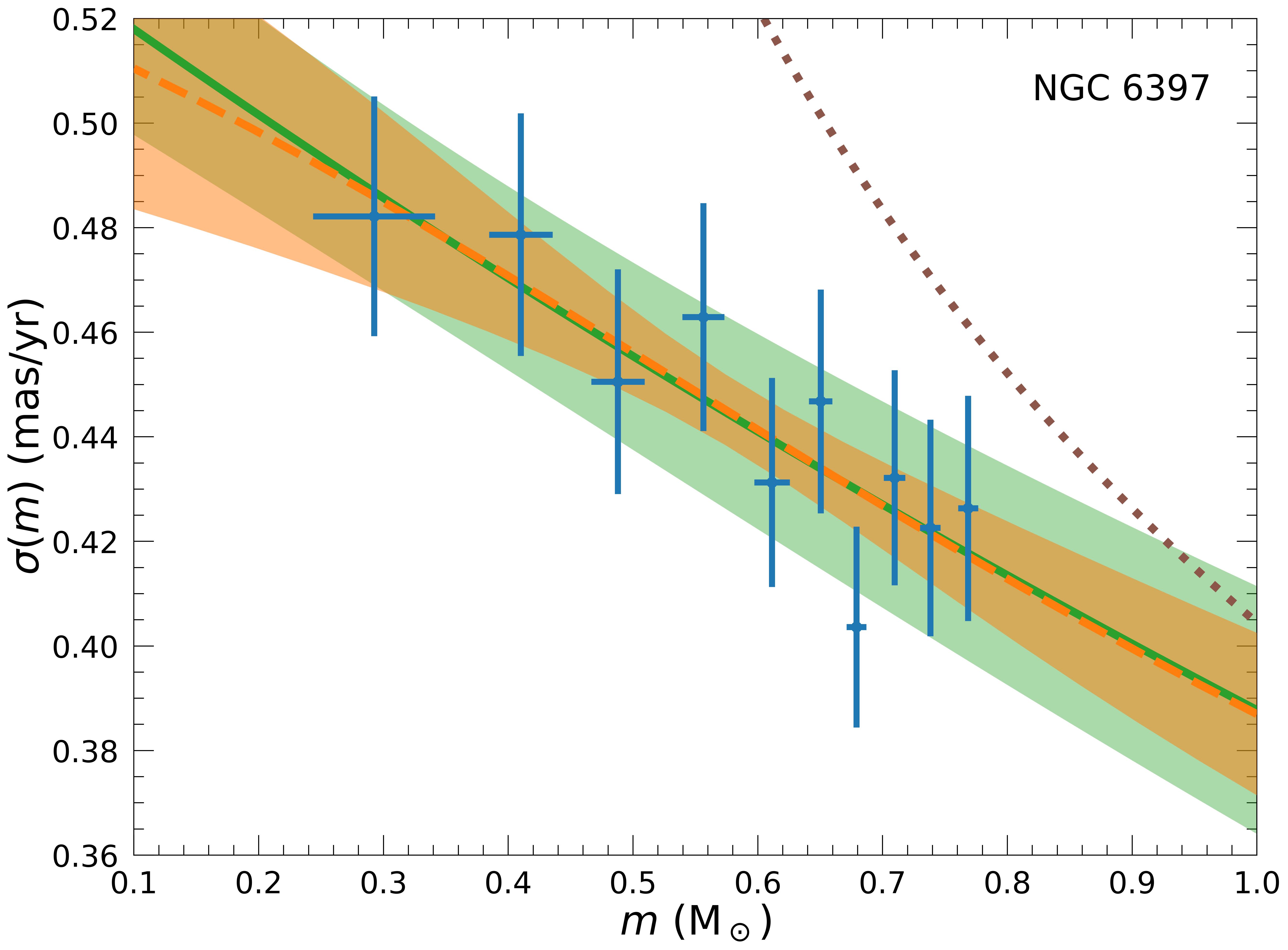}
    \caption{Projected velocity dispersion as function of stellar mass for \object{NGC 6397}. The error bars are the data from \citet{Watkins2022}, the continuous green line is our best-fit with the \citet{Bianchini2016} fitting function in Eq. \eqref{eq:BianchiniFittingFunc} with its error band, the dotted brown line is the complete equipartition limit ($\sigma\propto m^{-1/2}$) and the dashed orange line is our model best-fit with its $68\%$ confidence band.}
    \label{fig:NGC6397sigmam_1par}
\end{figure}

\begin{table*}[htbp]
    \centering
    \caption{Outline of parameters for the analyzed GCs}
    \label{tab:params_alphaBaumgardt2023}
    \begin{tabular}[c]{lccr@{$\pm$}lr@{$\pm$}lr@{$\pm$}lr@{$\pm$}lrr}
    \hline \hline
    \multicolumn{1}{c}{{ID}} 
    & \multicolumn{1}{c}{$\alpha^{(1)}$}
    &\multicolumn{1}{c}{$m_\mathrm{eq}^{(2)}\, (M_{\sun})$}
    &\multicolumn{2}{c}{$m_\mathrm{eq}\, (M_{\sun})$} 
    &\multicolumn{2}{c}{$\Phi_0\, (M_{\sun}^{-1})$} 
    & \multicolumn{2}{c}{$\sigma_\mathrm{0,B}$ (mas yr$^{-1}$)} 
    & \multicolumn{2}{c}{$\sigma_\mathrm{0}$ (mas yr$^{-1}$)} 
    &\multicolumn{1}{c}{$\chi_\mathrm{B}^2$} 
    & \multicolumn{1}{c}{ $\chi^2$}\\
        \hline
\object{NGC 104}  & -0.65 & 1.37$^\mathrm{+0.16}_\mathrm{-0.13}$ & 1.29&0.15 &   15.3&1.1 & 0.727&0.022 & 0.690&0.004 & 8.52 & 8.23\\
\object{NGC 5139} & -0.80 & 2.82$^\mathrm{+0.38}_\mathrm{-0.30}$ & 2.82&0.43 &   5.67&0.58 & 0.772&0.014 & 0.749&0.002  & 13.76 & 14.21\\
\object{NGC 5904} & -0.76 & 1.53$^\mathrm{+1.26}_\mathrm{-0.51}$ & 1.26&0.43 &  14.2&3.7 & 0.275&0.026 & 0.260&0.003  & 7.56 &  7.57\\
\object{NGC 6266} & -1.14 & 1.30$^\mathrm{+0.91}_\mathrm{-0.39}$ & 1.12&0.25 &  17.5&5.3 & 0.584&0.044 & 0.556&0.005  & 4.11 &  4.17\\
\object{NGC 6341} & -0.82 & 1.23$^\mathrm{+0.81}_\mathrm{-0.36}$ & 1.07&0.18 &  18.0&2.4 & 0.254&0.014 & 0.244&0.002  & 2.75 &  2.65\\
\object{NGC 6397} & -0.32 & 1.85$^\mathrm{+1.21}_\mathrm{-0.56}$ & 1.55&0.33 &  17.0&2.4 & 0.535&0.022 & 0.510&0.008  & 3.50 &  3.48\\
\object{NGC 6656} & -0.90 & 1.30$^\mathrm{+0.39}_\mathrm{-0.23}$ & 1.32&0.27 &  10.7&2.0 & 0.758&0.038 & 0.716&0.006  & 6.42 &  6.46\\
\object{NGC 6752} & -0.67 & 2.49$^\mathrm{+1.01}_\mathrm{-0.56}$ & 2.42&0.61 &  12.4&1.0 & 0.432&0.013 & 0.419&0.004  & 6.46 &  6.63\\ \hline
    \end{tabular}
    \tablefoot{Table columns: ID, mass function slope (1), equipartition mass from (2) and our fitting with the Bianchini function, estimated $\Phi_0$, scaling velocity dispersion $\sigma_\mathrm{0,B}$ and $\sigma_{0}$ (in milliarcseconds) and $\chi^2$ test value for both the Bianchini fitting function (with a B subscript) and our model prediction.}
    \tablebib{
(1)~\citet{Baumgardt2023}; (2) \citet{Watkins2022}.
}
\end{table*}

\begin{table*}[htbp]
    \centering
    \caption{Estimated parameters for the classical power law.}
    \label{tab:params_alphaBaumgardt2023_eta}
    \begin{tabular}[c]{lc r@{$\pm$}l r@{$\pm$}l r}
    \hline \hline
    \multicolumn{1}{c}{{ID}} 
    &\multicolumn{1}{c}{\textbf{$\eta^{(1)}$}}
    &\multicolumn{2}{c}{$\eta$} 
    & \multicolumn{2}{c}{$\sigma_\mathrm{s}$ (mas yr$^{-1}$)} 
    &\multicolumn{1}{c}{$\chi_\mathrm{\eta}^2$} \\
        \hline
\object{NGC 104}  & 0.220$^\mathrm{+0.027}_\mathrm{-0.024}$ & 0.239&0.024 &  0.507&0.006  & 6.62 \\
\object{NGC 5139} & 0.107$^\mathrm{+0.013}_\mathrm{-0.012}$ & 0.107&0.016 &  0.665&0.005  &12.73 \\
\object{NGC 5904} & 0.261$^\mathrm{+0.100}_\mathrm{-0.093}$ & 0.271&0.094 &  0.189&0.007  & 7.69 \\
\object{NGC 6266} & 0.316$^\mathrm{+0.102}_\mathrm{-0.102}$ & 0.324&0.075 &  0.380&0.009  & 4.28 \\
\object{NGC 6341} & 0.311$^\mathrm{+0.089}_\mathrm{-0.093}$ & 0.325&0.053 &  0.163&0.003  & 2.49 \\
\object{NGC 6397} & 0.145$^\mathrm{+0.053}_\mathrm{-0.051}$ & 0.159&0.035 &  0.404&0.009  & 3.77 \\
\object{NGC 6656} & 0.256$^\mathrm{+0.051}_\mathrm{-0.053}$ & 0.239&0.047 &  0.533&0.012  & 6.11 \\
\object{NGC 6752} & 0.108$^\mathrm{+0.029}_\mathrm{-0.030}$ & 0.105&0.025 &  0.360&0.007  & 6.13 \\ \hline
    \end{tabular}
    \tablefoot{Table columns: ID, $\eta$ from (1) and our fitting with the classical power law $\sigma(m)=\sigma_s (m/m_\mathrm{s})^{-\eta}$ (using $m_\mathrm{s}=1.0 M_{\sun}$), scaling velocity dispersion $\sigma_\mathrm{s}$ (in milliarcseconds) and $\chi^2$ test value.}
    \tablebib{
(1)~\citet{Watkins2022}.}
\end{table*}

The plot clearly underlines a great compatibility between the model prediction and the Bianchini fitting function on the same data.
We confirm a partial degree of equipartition in all clusters and for all masses (see also Fig. \ref{fig:additional_sigmam} in Appendix \ref{sec:App-additionalplots} for other clusters). We expect that a cluster in an advanced dynamical state is closer to energy equipartition and shows a smaller value of $m_\mathrm{eq}$. In the model framework, dynamically old means a greater $\Phi_0$ value, namely a deeper gravitational potential well.
Then, there should be a relation between these parameters: a cluster showing a higher degree of equipartition should have a greater value of $\Phi_0$ and a lower value of $m_\mathrm{eq}$. This is what we obtain in Fig. \ref{fig:meq-vs-PHI0}, where we plot the relationship between $m_\mathrm{eq}$ by W22 and the estimated $\Phi_0$, colored with the obtained $\chi^2$.
In Table \ref{tab:params_alphaBaumgardt2023} we give the estimated parameters and the $\chi^2$ test value for all analyzed clusters, also for our fit with the Bianchini fitting function. 
Our evaluation of $m_\mathrm{eq}$, conducted on binned data and with a non-linear least squares algorithm (available in the SciPy software package), is not so far from that obtained by W22, that worked on unbinned data, set prior constraints on the parameter values and use a Monte Carlo method to sample the parameter space.
The $\chi^2$ values of the two fits are close to each other, sustaining a similar confidence with the data. Furthermore, the scaling dispersion $\sigma_0$ of our model is systematically lower than $\sigma_\mathrm{0,B}$, a natural behavior due to the different meaning they have. While the first is the dispersion of the lowest model mass, the latter is the dispersion value in the limit $m\rightarrow 0$.\\
The results of the fits made by using the classical power law $\sigma(m) = \sigma_\mathrm{s} (m/m_\mathrm{s})^{-\eta}$, with $m_\mathrm{s}=1.0\,M_\sun$, are shown in Table \ref{tab:params_alphaBaumgardt2023_eta}. 
Our estimates of $\eta$, obtained with the same method as for $m_\mathrm{eq}$, are compatible with those provided by W22, statistically equivalent to the other fitting functions, with small variations in the obtained $\chi^2$. 
Furthermore, $\eta$ follows the expected trend with $m_\mathrm{eq}$ and $\Phi_0$. For a more advanced dynamical state, we have a larger degree of equipartition described by a higher $\eta$ value, as well as a larger $\Phi_0$ and a smaller $m_\mathrm{eq}$, as expected. 
W22 already provided a discussion of the relationship between the classical power law and the Bianchini fitting function, as well as the obtained values for the equipartition mass and $\eta$, showing that they appear equivalent in measuring the equipartition degree.
Here we focus on the link between the Bianchini function and the dynamical model prediction due to the closer statistical compatibility obtained in our fits.\\
\begin{figure}[htbp]
    \centering
    \includegraphics[width=\linewidth]{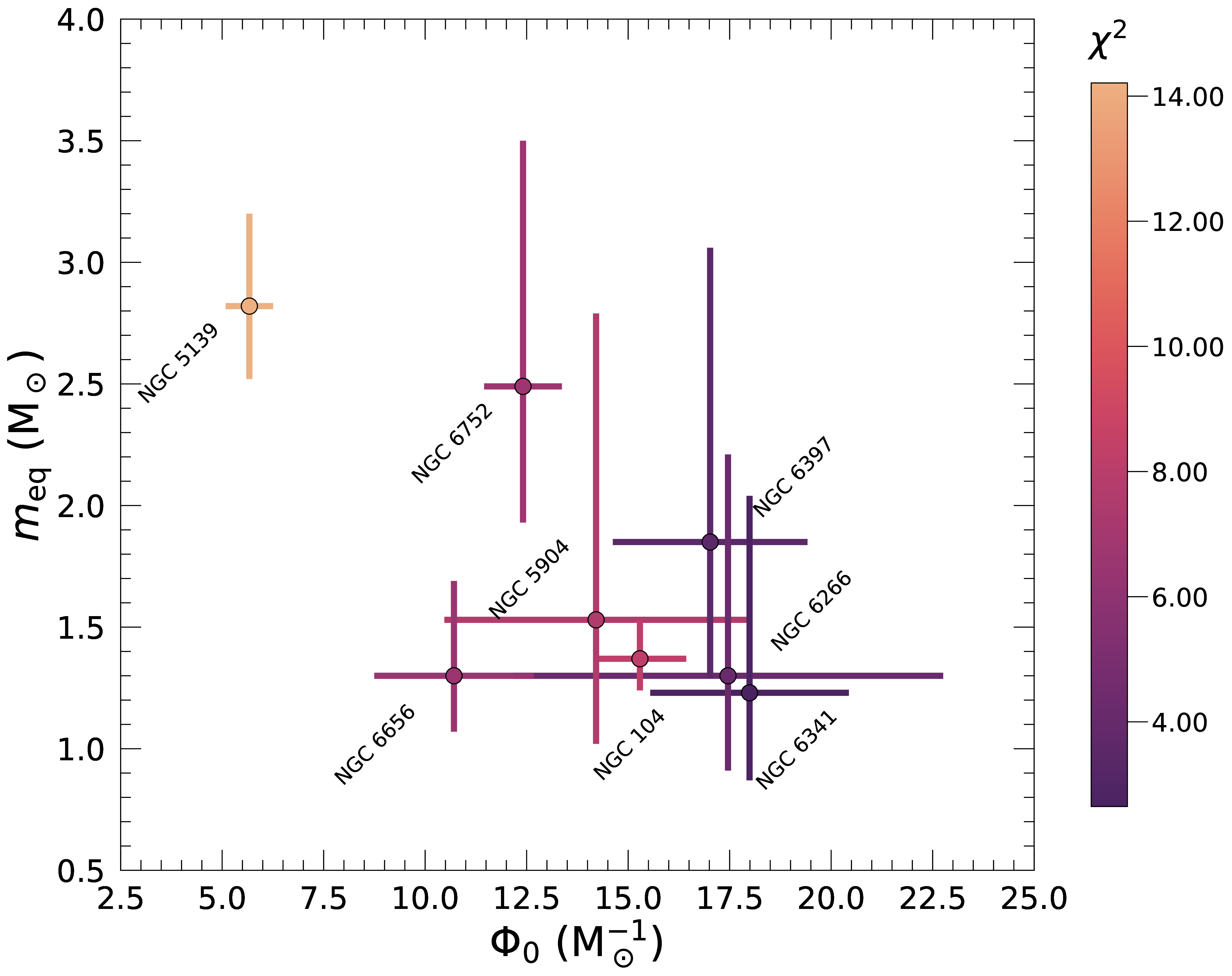}
    \caption{Relation between the estimated $\Phi_0$ and $m_\mathrm{eq}$ from \citet{Watkins2022}. The $\chi^2$ test value is given in colors.}
    \label{fig:meq-vs-PHI0}
\end{figure}
Concerning the results of the fitting procedure and the relation between $m_{\mathrm{eq}}$ and $\Phi_0$, some considerations must be done. First, for \object{NGC 5139} ($\omega$-Cen) we have the lowest value of $\Phi_0$ and the greatest value of $m_\mathrm{eq}$, also coming with a very poor fit quality (largest $\chi^2$ value).  
This cluster must be treated with caution due to its intricate history and structure. It shows evidence of multiple stellar populations that are dynamically interacting and are likely to have different degrees of equipartition. Furthermore, each has its own mass function and chemical pattern. It is also strongly suspected to be the survived core of a disrupted dwarf galaxy. The full picture behind this object requires further study with more sophisticated models, given its complexity. Although the overall trend for $\sigma(m)$ looks promising, we emphasize that the uncertainties may be underestimated due to the fluctuations around the decreasing pattern (we come back to this point in the next lines). 
This suggests an additional source of statistical error that may increase the estimated uncertainty in the corresponding $\Phi_0$ value. However, the long relaxation timescale of such a cluster (see Table 1 in W22 work) may indicate that our estimates are not far from reality, specifically a low energy equipartition level corresponding to a smaller $\Phi_0$ and a larger $m_\mathrm{eq}$.
Furthermore, \object{NGC 6397} and \object{NGC 6752} are post-core collapsed objects according to the Harris catalog, since they have a very high value of the concentration, while \object{NGC 6266} is suspected to be collapsed even if its concentration is not very high. 
Considering that only for three of the analyzed clusters, namely \object{NGC 6266}, \object{NGC 6341} and \object{NGC 6397}, we have a high confidence level coming from the fitting procedure, excluding collapsed objects and poor fit quality results, we are left only with \object{NGC 6341}, which is also the cluster with the lowest $m_\mathrm{eq}$ and the highest $\Phi_0$. 
Moreover, this cluster has a low mass coverage in the velocity dispersion data, with a range of $0.59<m/M_{\sun}<0.78$. 
Looking at data points (see Fig. \ref{fig:additional_sigmam} in Appendix \ref{sec:App-additionalplots}), it looks evident a decreasing trend of $\sigma(m)$ with the mass, but the oscillations around such tendency, which change from cluster to cluster, suggest that uncertainties can be underestimated. Indeed, we obtain a relation between our $\chi^2$ test value and the average relative error on $\sigma(m)$ data (see Fig. \ref{fig:chisq_1par_RelErrsigmam} in Appendix \ref{sec:App-additionalplots}), showing that low relative errors produce a greater $\chi^2$ value in our fitting procedure (with $\omega$-Cen showing the lower relative error and the larger $\chi^2$), while the opposite is vaguely true for greater relative errors.
Furthermore, our $\chi^2$ looks slightly asymmetric with respect to $\Phi_0$ in the minimum, suggesting a more advanced uncertainty estimation method for such parameter, like evaluating the right and left errors. However, this does not affect the results that we present here.\\
The uncertainties given for $\Phi_0$ reflect the observed ones and come from the minimization process. A possible source of error in our procedure is the choice of the core radius. It can affect the theoretical prediction, since it determines the radial shell where the projected velocity dispersion $\sigma(R,m)$ is averaged (see Sect. \ref{subsubsec:velocitydispersion-mass}). 
An underestimation of the core radius would map the observed radial shell into a more outer dimensionless radial shell. As a result, the velocity dispersion would be generally smaller and more dominated by the low-mass stars due to mass segregation. Then, $\sigma(m)$ would be steeper and the corresponding best-fit $\Phi_0$ would be higher. The opposite holds for an overestimate of the core radius. 
This introduces a systematic error into our procedure for determining model parameters, but we do not explore such an effect here. 
Actually, the expected relation between $\Phi_0$ and $r_\mathrm{c}$ suggests that only a large relative uncertainty on the core radius may affect the estimate of $\Phi_0$. 
We quantify the effect of varying $r_\mathrm{c}$ on the $\Phi_0$ determination in Sect. \ref{subsec:RelationStructParams}.\\
All these uncertainties motivate an independent examination of $\Phi_0$, as obtained by fitting the surface brightness profile (see Sect. \ref{subsec:FittingSBPs}).\\
Consequently, our work must be considered as a first guess in the evaluation of multi-mass dynamical model parameters from GCs kinematic observations concerning energy equipartition.

\subsection{Adding the mass function slope to the parameter space}
\label{subsec:VaryingPHI0alpha}
As an extension of the previous analysis, we add to the parameter space the mass function slope $\alpha$. The fitting procedure is naturally more computationally expensive, since we have to iteratively solve equilibrium configurations with a different mass function. As previously done, we minimize the $\chi^2$ for each cluster: we explore $\Phi_0$ around a value close to the one obtained with the first approach and the slope $\alpha$ is taken in the range $[-2.0,0.0]$. 
Such choice includes usual GCs slopes, and it is only a first guess, to be eventually modified to increase the precision.
Unfortunately, that is not the case, because we find a degeneration between $\Phi_0$ and $\alpha$. \\
In Fig. \ref{fig:NGC6397_chisq_2ndguess_contour}
we show the contour plot for \object{NGC 6397} in the $\Phi_0$-$\alpha$ plane, with some values around the $\chi^2$ minimum, together with the mass function assumed in the first approach, with its uncertainty.
\begin{figure}[htbp]
    \centering
    \includegraphics[width=\linewidth]{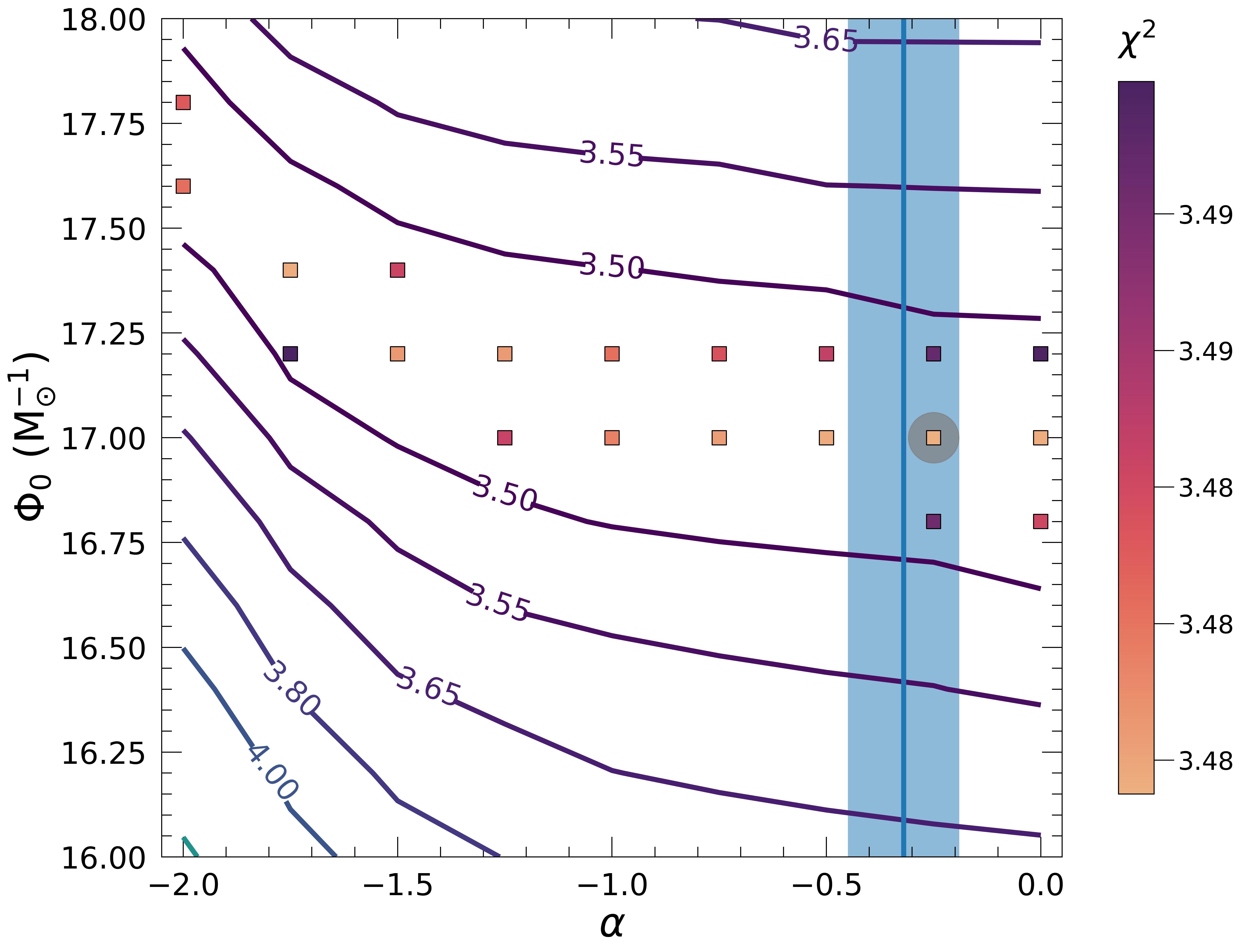}
    \caption{Contour plot for \object{NGC 6397} in the $\Phi_0$-$\alpha$ plane, showing the $\chi^2$ levels. The blue band gives the slope from \citet{Baumgardt2023}. Squares indicate the $\chi^2$ values around the minimum, identified by a big gray circle.}
    \label{fig:NGC6397_chisq_2ndguess_contour}
\end{figure}
Even if our minimum falls in the error band by \citet{Baumgardt2023} for \object{NGC 6397}, the same is not valid for other clusters (see Fig. \ref{fig:additional_contours} in Appendix \ref{sec:App-additionalplots}). Furthermore, the shape of the parameter space is preventing us to constrain the slope of the mass function, whose uncertainties are too big and overcome the considered range.
In Table \ref{tab:params_varyingPhi0Alpha} we give the results for the estimated model parameters, with their errors.
\begin{table}[htbp]
    \centering
     \caption{Results of the second fitting approach}
    \begin{tabular}[c]{lr@{$\pm$}lr@{$\pm$}lr@{$\pm$}l}
    \hline \hline
    \multicolumn{1}{c}{{ID}} &
    \multicolumn{2}{c}{$\alpha$} &
    \multicolumn{2}{c}{$\Phi_0 \, (M_{\sun}^{-1}$)} &
    \multicolumn{2}{c}{$\sigma_0$ (mas yr$^{-1}$)}\\\hline
\object{NGC 104}  &  0.0 & 7.9 & 15.3 & 1.2  & 0.690 & 0.008\\
\object{NGC 5139} &  0 & 31  & 5.64 & 0.57 & 0.749 & 0.003\\
\object{NGC 5904} & -2 & 26  & 14.2 & 7.5  & 0.260 & 0.043\\
\object{NGC 6266} & -2 & 15  & 17.5 & 4.3  & 0.556 & 0.017\\
\object{NGC 6341} &  0 & 18  & 18.0 & 5.9  & 0.244 & 0.033\\
\object{NGC 6397} &  0 & 17  & 17.0 & 1.4  & 0.511 & 0.030\\
\object{NGC 6656} &  0.0 & 5.7 & 11.1 & 3.1  & 0.719 & 0.012\\
\object{NGC 6752} & -1.0 & 8.6 & 14.4 & 3.4  & 0.419 & 0.021\\\hline
    \end{tabular}
    \tablefoot{Table columns: cluster ID and estimated values for $\alpha$, $\Phi_0$ and $\sigma_0$. We report the parameters value according to the two most significant digits of the associated uncertainty.}
    \label{tab:params_varyingPhi0Alpha}
\end{table}

Although the contour plots are not the same for all analyzed clusters, in some cases there is a small tendency of having a greater $\Phi_0$ corresponding to a lower $\alpha$, while in other ones the contour plot lines are almost horizontally oriented.
Indeed, the velocity dispersion dependence on mass is steeper and closer to equipartition for both a larger $\Phi_0$ or a greater slope $\alpha$. Such effect is expected since a larger slope means a flatter mass function, with more massive stars than a steeper one. This results in a greater degree of energy equipartition, because massive stars are more efficient in reaching equipartition due to their higher mass, which cause them to suffer more gravitational encounters than less massive stars. Similarly, a greater $\Phi_0$ implies a deeper potential well, namely a more dynamically evolved cluster, with a higher level of equipartition. As a result, a given degree of equipartition can be reached with more massive stars and a less deep gravitational well or by a stronger gravity and a steeper mass function. 
However, the effect of the gravitational potential through the $\Phi_0$ parameter looks more important, with smaller changes affecting equipartition more than variations in the slope.
This is also remarked by the estimated errors on $\Phi_0$, which are compatible with respect to the first approach, meaning that any reasonable assumption on the mass function slope would work well in setting up the model and constrain $\Phi_0$ from the velocity dispersion $\sigma(m)$.

\subsection{Shell selection and projection effects}
As already outlined, the data we are fitting concerns projected quantities, and it is restricted to a circular ring close to the central regions of the clusters. We explore, by means of our model and the Bianchini fitting function, the effects of projection and shell selection in quantifying the degree of equipartition.\\
Our dynamical model predicts the shape of the function $\sigma(r,m)$, which 
can be used to explore the degree of energy equipartition in different radial shells. Equipartition is more efficient in the core of clusters, and the maximum degree is expected to be measured there.
Unfortunately, to date kinematic measurements in the core are difficult due to crowding effects and are limiting observers to look around it, as W22 do.
As a result, the measured degree of equipartition is less than the core one.
Furthermore, working with projected velocities also gives a smaller equipartition level than using 3D quantities. Indeed, at the projected radial distance $R$, the observer intercepts stars with a 3D radial distance $r\in[R,r_\mathrm{t}]$. Being the velocity dispersion $\sigma(r,m)$ a decreasing function of $r$, its value is lowered for each mass, in average, with respect to the 3D one. Being massive stars more centrally concentrated than less massive ones, the projection dumps more $\sigma(m)$ for low-mass stars and brings to a flatter profile, resulting in a lower equipartition degree with respect to the 3D case.
Both these effects bias the estimation of $m_\mathrm{eq}$, leading to higher values (i.e., lower equipartition). 
To visualize this, we plot in Fig.
\ref{fig:meq_r-R} the value of $m_\mathrm{eq}$ obtained by fitting with the Bianchini function the 3D and 2D theoretical profiles $\sigma(r,m)$ and $\sigma(R,m)$ respectively, for a particular choice of $\alpha$ and $\Phi_0$. 
Such analysis is possible due to the very good similarity between our prediction for $\sigma(m)$ and Eq. \eqref{eq:BianchiniFittingFunc}, which we check quantifying the uncertainty of the fitted $m_\mathrm{eq}$ and verifying its small value.
\begin{figure}[htbp]
    \centering
    \includegraphics[width=\linewidth]{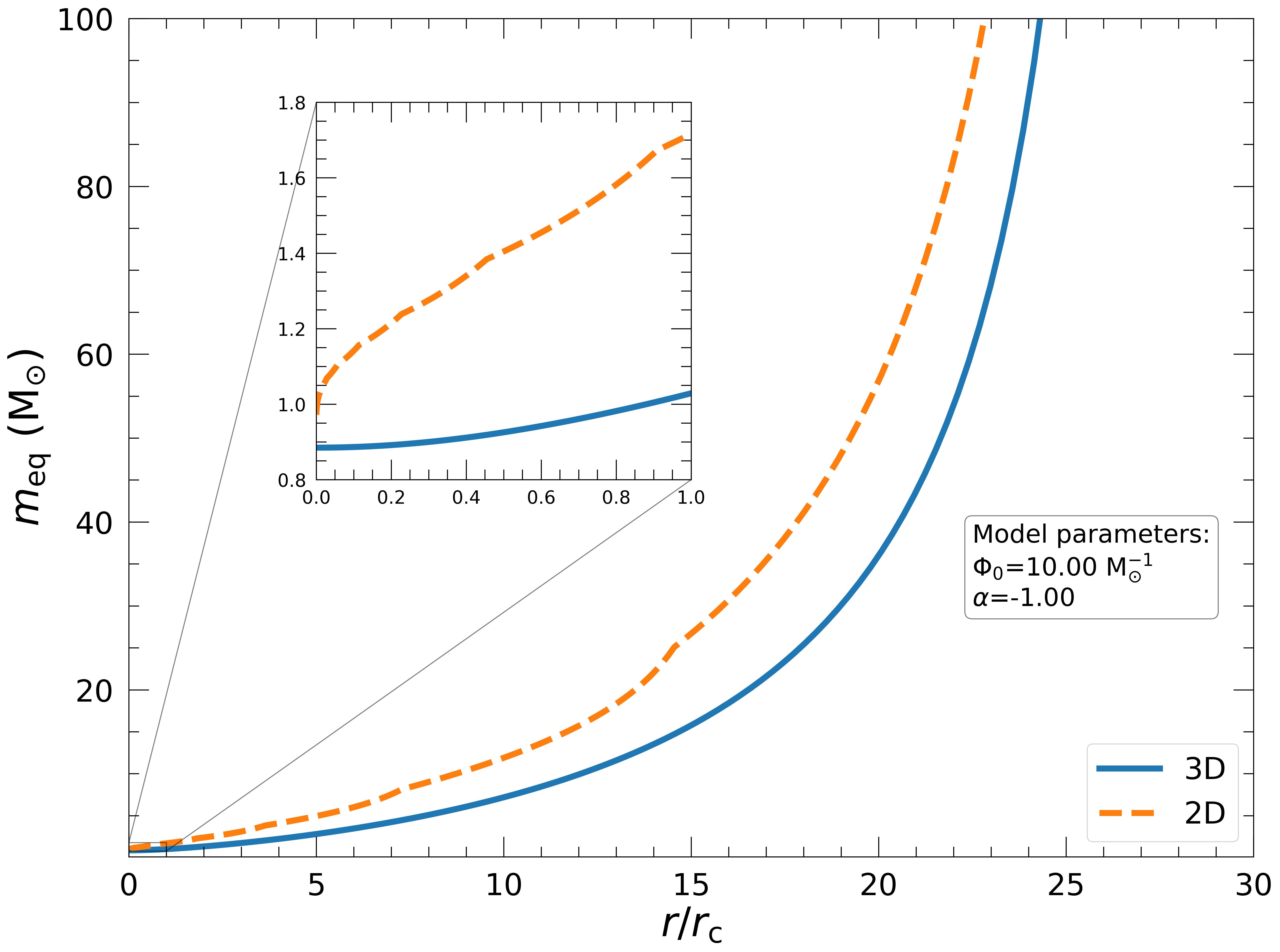}
    \caption{Radial and projected profile of the equipartition mass $m_\mathrm{eq}$, obtained by fitting the model prediction on the velocity dispersion with the Bianchini function at each radial coordinate, given in units of the core radius. The panel zooms into the region $r\leq r_\mathrm{c}$.}
    \label{fig:meq_r-R}
\end{figure}

As expected, $m_\mathrm{eq}$ increases with the distance from the center of the cluster, due to the decreasing efficiency of the relaxation process. 
A measure of the maximum degree of equipartition should be done with the 3D equipartition mass at least in the core $r<r_\mathrm{c}$, where an almost flat trend is seen, as shown in the zoom panel of Fig. \ref{fig:meq_r-R}. The projected profile of $m_\mathrm{eq}$ always gives bigger values, even in the core.
Every analysis of the projected velocity dispersion as function of stellar mass should consider that the estimation of equipartition obtained fitting data with the Bianchini function will give an overestimated $m_\mathrm{eq}$ (underestimating the degree of equipartition), with respect to the 3D value. This is not the case when the fit is done with our model, since the degree of equipartition depends on $\Phi_0$ that uniquely determines the velocity dispersion value $\sigma(r,m)$ in any position and for each mass, from which the projected one can be computed without any loss of information.\\
Furthermore, we give in Fig. \ref{fig:meq_r_alpha-PHI0} few 3D radial profiles of $m_\mathrm{eq}$ obtained changing $\Phi_0$ and $\alpha$, to show their effect.
\begin{figure}[htbp] 
    \centering
    \includegraphics[width=\linewidth]{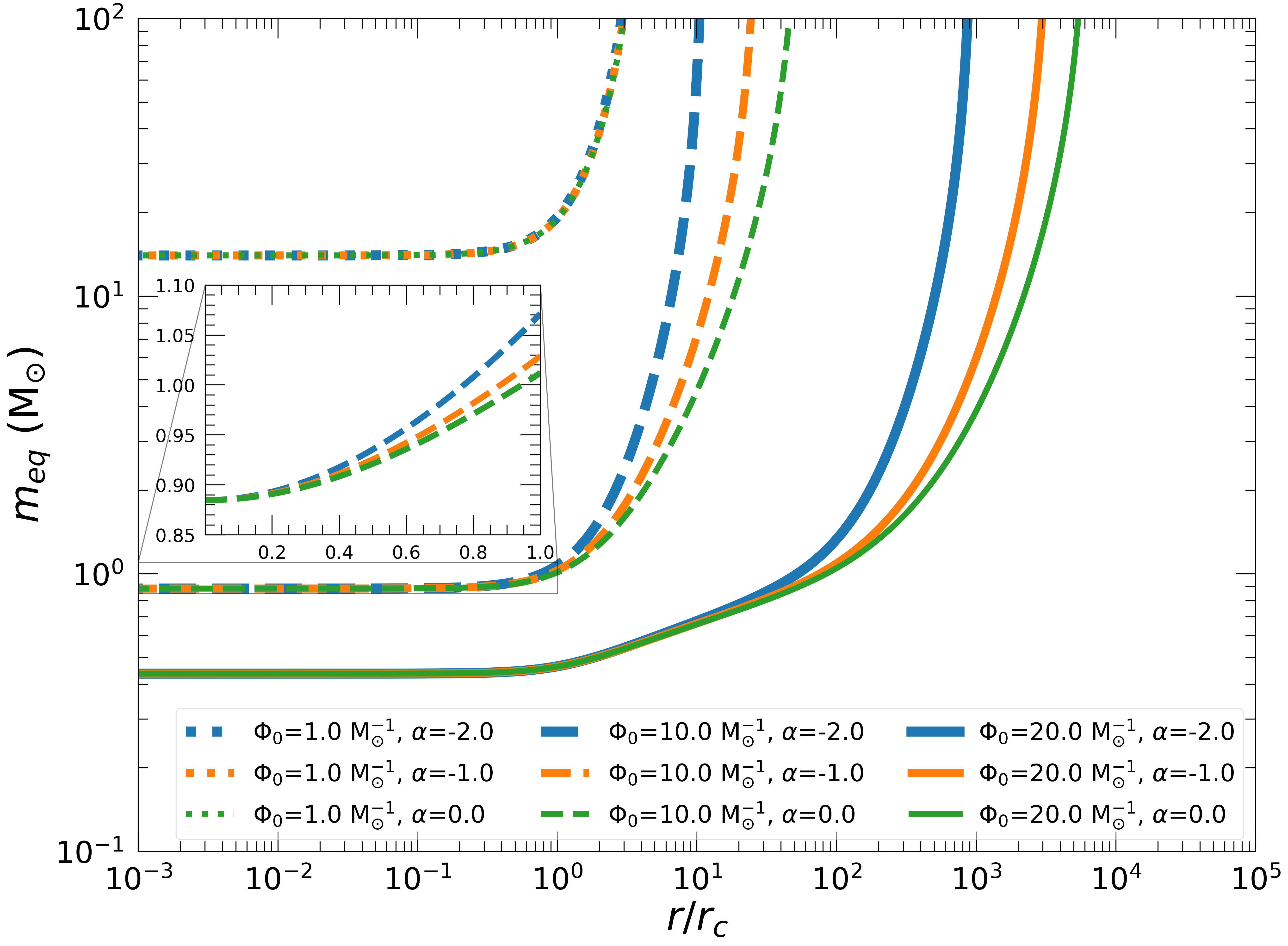}
    \caption{Theoretical radial profile of the equipartition mass, obtained for different values of the model parameters $\Phi_0$ (same line-style) and $\alpha$ (same colors and line-width). The panel zooms in the core region $r\leq r_\mathrm{c}$.}
    \label{fig:meq_r_alpha-PHI0}
\end{figure}
Here, for very low values of $\Phi_0$, namely a dynamically young cluster, the profile $m_\mathrm{eq}(r)$ starts from $\sim 10\, M_{\sun}$ in the central regions and grows with radius, almost insensible to $\alpha$ variations. With a greater $\Phi_0$, the effect of a different slope is clearly visible in the outer regions. A flatter mass function will produce a more extended profile, as a greater $\Phi_0$ do (due to the existent degeneration). Additionally, it also gives smaller values of $m_\mathrm{eq}$ at each radial coordinate than a steeper mass function, as already mentioned.\\
In the core, the value of $m_\mathrm{eq}$ is approximately constant and decreases with increasing $\Phi_0$, as expected from more advanced dynamical states.
The zoom panel of Fig. \ref{fig:meq_r_alpha-PHI0} gives the trend for $r\leq r_\mathrm{c}$, showing the small differences between the considered slopes. 
Such approximately flatness of $m_\mathrm{eq}(r)$ in the central region, a common behavior of other observable profiles like the surface density, can be used to quantify the maximum degree of equipartition reached by stars in the cluster.
Then, we build a relation between $m_\mathrm{eq}(r\leq r_\mathrm{c})$ and $\Phi_0$, which is highly useful in the perspective of quantifying the maximum degree of equipartition in terms of the equipartition mass. Indeed, the value of $\Phi_0$ can be constrained from other observables more easily measurable than the velocity dispersion as function of mass in the center, which suffers crowding.
Since the slope of the mass function has a negligible effect in the relation between $m_\mathrm{eq}$ and $\Phi_0$ in the core, we choose $\alpha =-0.5$ and plot such relationship in Fig. \ref{fig:meq-vs-PHI0_rc}, comparing with the estimates of $m_\mathrm{eq}$ and $\Phi_0$ already shown in Fig. \ref{fig:meq-vs-PHI0} and given in Table \ref{tab:params_alphaBaumgardt2023}. 
The plot shows how the parameters constrained from the projected $\sigma(m)$ observations lie in the region above the curve.
\begin{figure}[htbp]
    \centering
    \includegraphics[width=\linewidth]{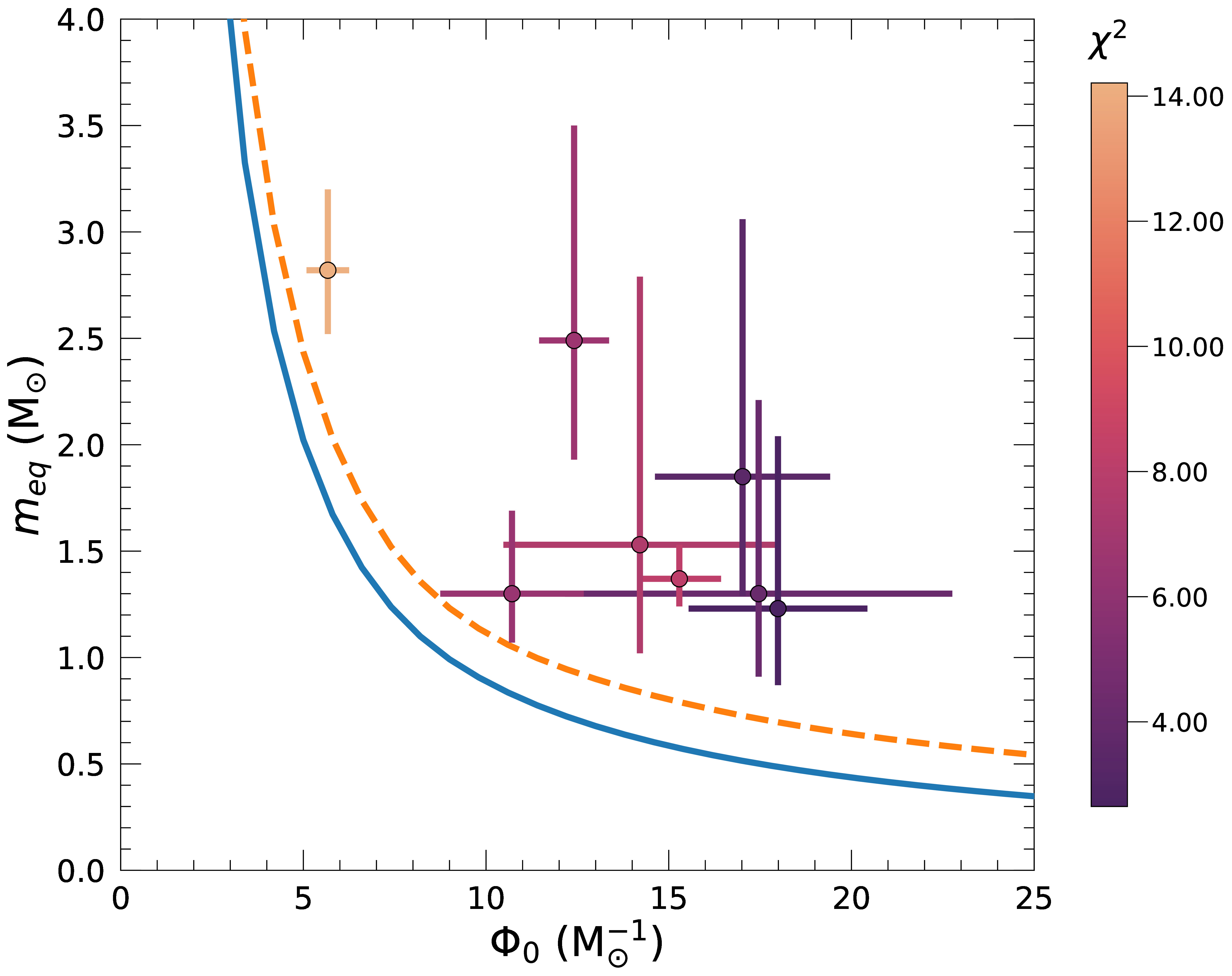}
    \caption{Equipartition mass $m_\mathrm{eq}$ in the core ($r\leq r_\mathrm{c}$) as function of $\Phi_0$. The blue continuous line and the orange dashed line are the 3D and 2D theoretical predictions, respectively. The circles with error bars are the same as in Fig. \ref{fig:meq-vs-PHI0}.}
    \label{fig:meq-vs-PHI0_rc}
\end{figure}
 This underlines that the estimated $m_\mathrm{eq}$ by W22 do not measure the maximum degree of equipartition. They suffer the projection effect and the radial shell selection, as already said.
 Indeed, the radial coverage of the data sample has a median radius beyond the core radius for most clusters. Only for two clusters, namely \object{NGC 5139} and \object{NGC 6656}, the selected shell lies around the core radius ($ 0.4\la r/r_\mathrm{c} \la 1.2$). Indeed, these clusters are also the closest points to the theoretical curve in Fig. \ref{fig:meq-vs-PHI0_rc}. \\
 Concerning the estimates of $\Phi_0$, we stress that this is not affected by projection or radial coverage, since it is a global parameter that identifies the equilibrium configuration with its profiles.
 As a result, once $\Phi_0$ is determined or constrained from observations, even with a limited radial coverage, the equipartition degree is known in every region of the cluster. It can also be presented in terms of the Bianchini equipartition mass, as we have shown here.

\subsection{Relation with structural parameters}
\label{subsec:RelationStructParams}
The dynamical state of GCs is related to the amount of relaxation they experienced during their life. A dynamically old cluster has seen several relaxation timescales $t_\mathrm{relax}$ during its evolution, a time that is shorter in the central regions, being gravitational interactions more frequent. 
In literature, the relaxation process is often quantified with the relaxation timescale in the core $t_\mathrm{rc}$ \citep{Djorgovski1993} and at half-mass radius \citep{BinneyTremaine2008}, which have an analytical approximated formula for their evaluation. This leads W22 to compare their prediction of $m_\mathrm{eq}$ with the number of core and median relaxation timescales $N_\mathrm{core}$ and $N_\mathrm{half}$, as shown in their Fig. 16.\\
From the \citet{King66} model, it is also expected that dynamically evolved clusters have a greater concentration $c=\log(r_\mathrm{t}/r_\mathrm{k})$ value, with the King radius $r_\mathrm{k}$ often replaced with the core radius $r_\mathrm{c}$. Indeed, a monotonically increasing relation exists between the concentration and the corresponding $W_0$ equilibrium parameter of the King model.\\
In the framework of our multi-mass dynamical model, the equilibrium configuration is identified by $\Phi_0$. However, the scaling radius of the model $r_\mathrm{k,u}$ loses, in principle, the relation with the core radius that the King model had. 
This leads us to define the equivalent King concentration $c_\mathrm{k} = \log(r_\mathrm{t}/r_\mathrm{k,u})$ and the theoretical concentration $c_\mathrm{th}=\log(r_\mathrm{t}/r_\mathrm{c,th})$, with $r_\mathrm{c,th}$ estimated from the surface density profile, assuming a proportionality relationship between the mass and luminosity density profiles (as outlined before). 
These concentrations have a small relative difference and both monotonically increase with $\Phi_0$, following a trend that depends on the slope $\alpha$. In the perspective of comparing with observations, we consider only $c_\mathrm{th}$.\\
About the relaxation timescale, the model derivation highlights a dependence on mass, consistent with what is observed. However, obtaining an analytical formula requires further investigation and development.
We are then limited to take the values of $N_\mathrm{core}$ and $N_\mathrm{half}$ from literature.\\
We expect that $\Phi_0$ increases with $N_\mathrm{core}$, $N_\mathrm{half}$ and $c_\mathrm{obs}=\log(r_\mathrm{t}/r_\mathrm{c})$. For the core and median relaxation timescales numbers we take the same values considered by W22, while for the concentrations we consider $c$ in the Harris catalog and its estimation obtained with two sources for the core radius, the one in the Harris catalog and the other by \citet{Baumgardt2018-2023} online catalog (which used the \citet{Spitzer1987} definition), while for the tidal radius we refer to \citet{Webb2013}, reported in the same catalog.\\
\begin{figure}[htbp]
    \centering
\includegraphics[width=0.95\linewidth]{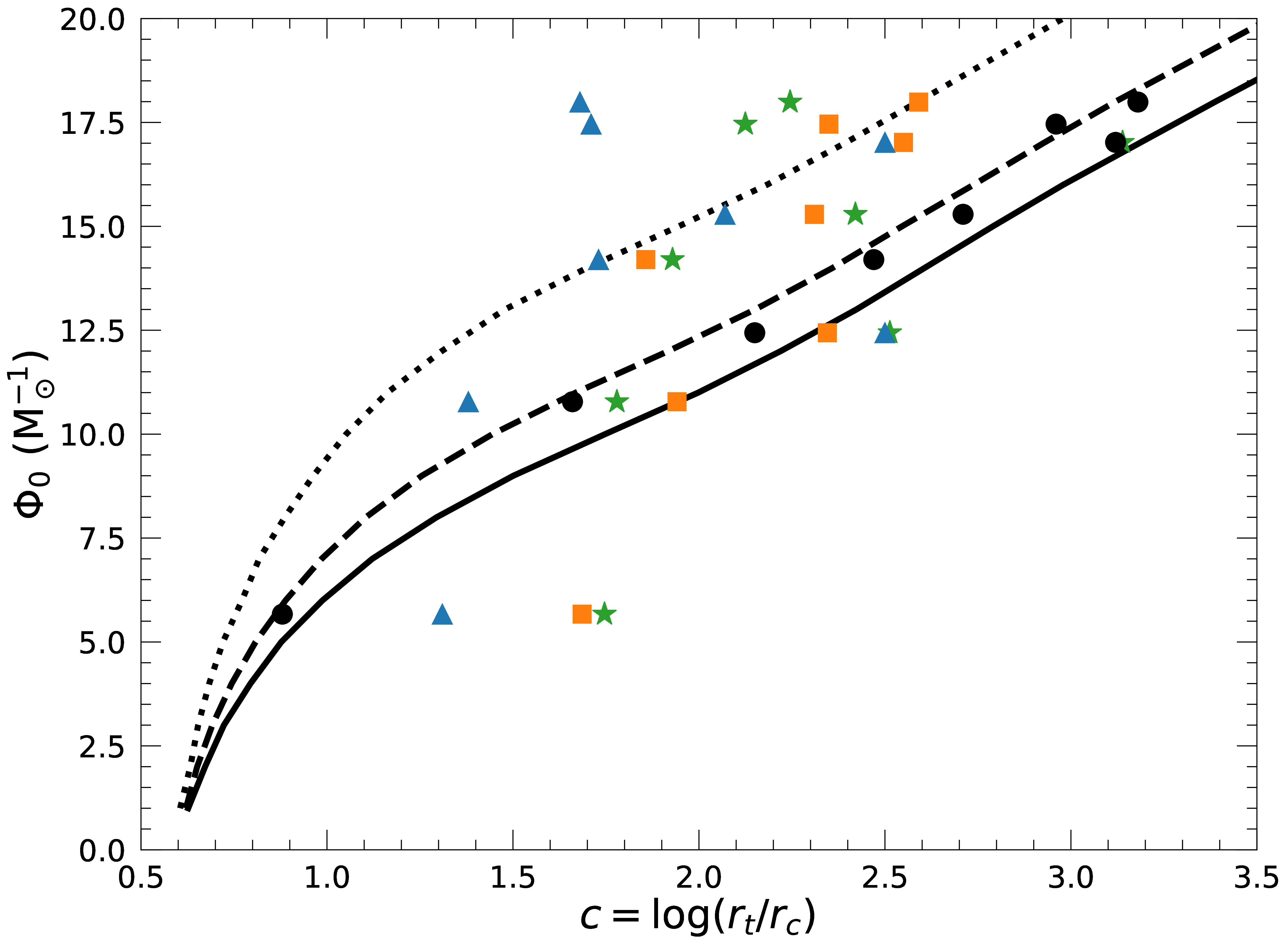}
    \caption{Relation between $\Phi_0$ and the concentration $c=\log{(r_\mathrm{t}/r_\mathrm{c})}$. The continuous, dashed and dotted lines represent, respectively, the theoretical prediction with a mass function slope $\alpha = 0.0$, $\alpha =-1.0$ and $\alpha=-2.0$, while the black circles show the constrained values of $\Phi_0$ and $c$ for the analyzed clusters.
    The blue triangles show the King concentration from the \citet{Harris1996} catalog (2010 edition), while the orange squares and green stars are computed using the core radius from \citet{Harris1996} and \citet{Baumgardt2018-2023} catalogs respectively, and the tidal radius from \citet{Webb2013}.}
\label{fig:PHI01par_c_Model_obs_multiplot}
\end{figure}

In Fig. \ref{fig:PHI01par_c_Model_obs_multiplot} the previously estimated values of $\Phi_0$ are given with the concentration. We show the theoretical relation for some choices of the slope $\alpha$, against the observational concentrations from different sources.
The predicted concentration grows continuously with $\Phi_0$, as already stated, but there are differences with the observational ones. From the model point of view, such discrepancy can be related to the fit quality and the value of the theoretical core radius. Low fit quality results can give an incorrect estimate of $\Phi_0$, which affects the vertical positions of each marker in the pictures. Concerning the core radius, the adopted choice is reasonable, since it is based on the assumption that the luminosity density profile is proportional to the density profile, with a multiplying factor only dependent on mass.
From the observational point of view, we first recall that the Harris catalog concentrations come by fitting the surface brightness profile of clusters with a King single-mass model \citep{King66}, that has been found in disagreement with observations in the outer regions, being clusters more extended, like already pointed out so far by \citet{DaCostaFreeman1976}. Indeed, the authors found that the outer regions of M3 can be described by a King profile with $c=1.98$, while the inner ones using $c=1.29$. This suggests that clusters are more extended and, consequently, have a higher concentration with respect to the single-mass model prediction. Multi-mass models can overcome such discrepancy, being more extended.
In principle, a concentration value completely based on observations requires the knowledge of both the core radius and the tidal radius. 
While the first is easier to measure, the latter is quite tricky, since a clear distinction between stars that belong to the cluster or are leaving it is difficult to infer.
This leads us to compute the concentration using the core and tidal radius from different sources \citep{Baumgardt2018-2023, Harris1996,Webb2013}. Indeed, \citet{Webb2013} found an empirical formula for the limiting radius of GCs through N-body simulations, which takes the orbital properties of the system in the galactic gravitational potential into account. 
This results in a greater tidal radius and consequently a greater concentration with respect to the King ones from the Harris catalog, as can be inferred from Fig. \ref{fig:PHI01par_c_Model_obs_multiplot}, comparing the orange squares and the green stars with the blue triangles.
Such tendency slightly erases the differences with the theoretical prediction, with the only exception of \object{NGC 5139} ($\omega$-Cen), which however is a complex object, as already mentioned, whose deviation is not so surprising.\\
Our analysis of few clusters reveals that there are important inhomogeneities among observational-based concentrations. Indeed, we can state that the estimate of the concentration suffers several sources of errors that can differ between clusters. 
The difficulty of measuring the tidal radius and, consequently, the concentration suggests to recover the role of theoretical models in predicting its value, as it was for the King model. The concentration value could be estimated from the relationship with the gravitational potential well through $\Phi_0$, to be constrained fitting observational profiles with model predictions.
Anyway, an extensive comparison between the theoretically predicted concentrations and the observational ones must be addressed in future works, with a greater GCs sample and heterogeneous observational sources. \\
\begin{figure}[htbp]
\centering
\begin{subfigure}{.45\textwidth}
  \centering
  \includegraphics[width=\linewidth]{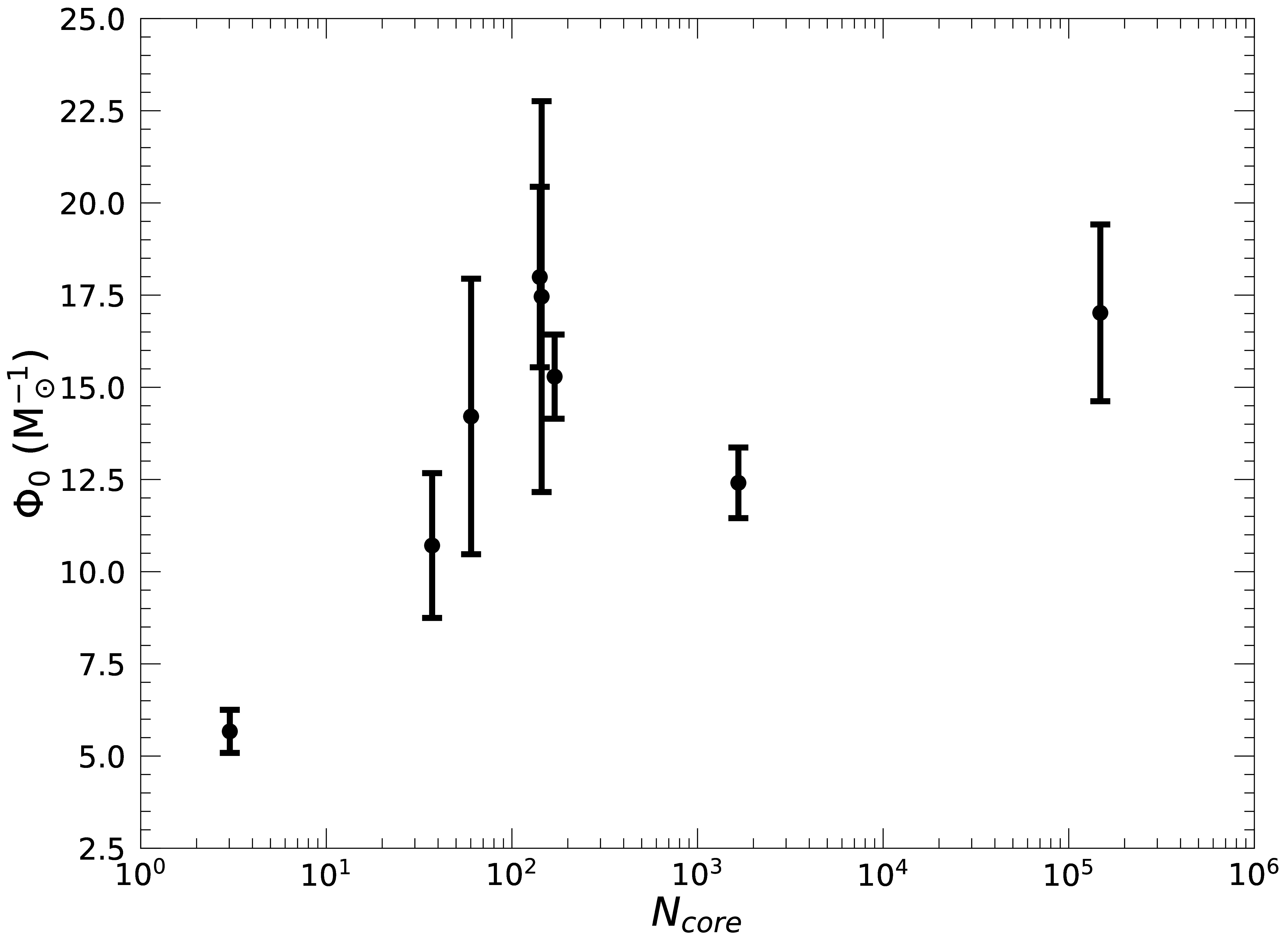}
  \label{fig:PHI0_Ncore}
\end{subfigure}%
\\
\begin{subfigure}{.45\textwidth}
  \centering
  \includegraphics[width=\linewidth]{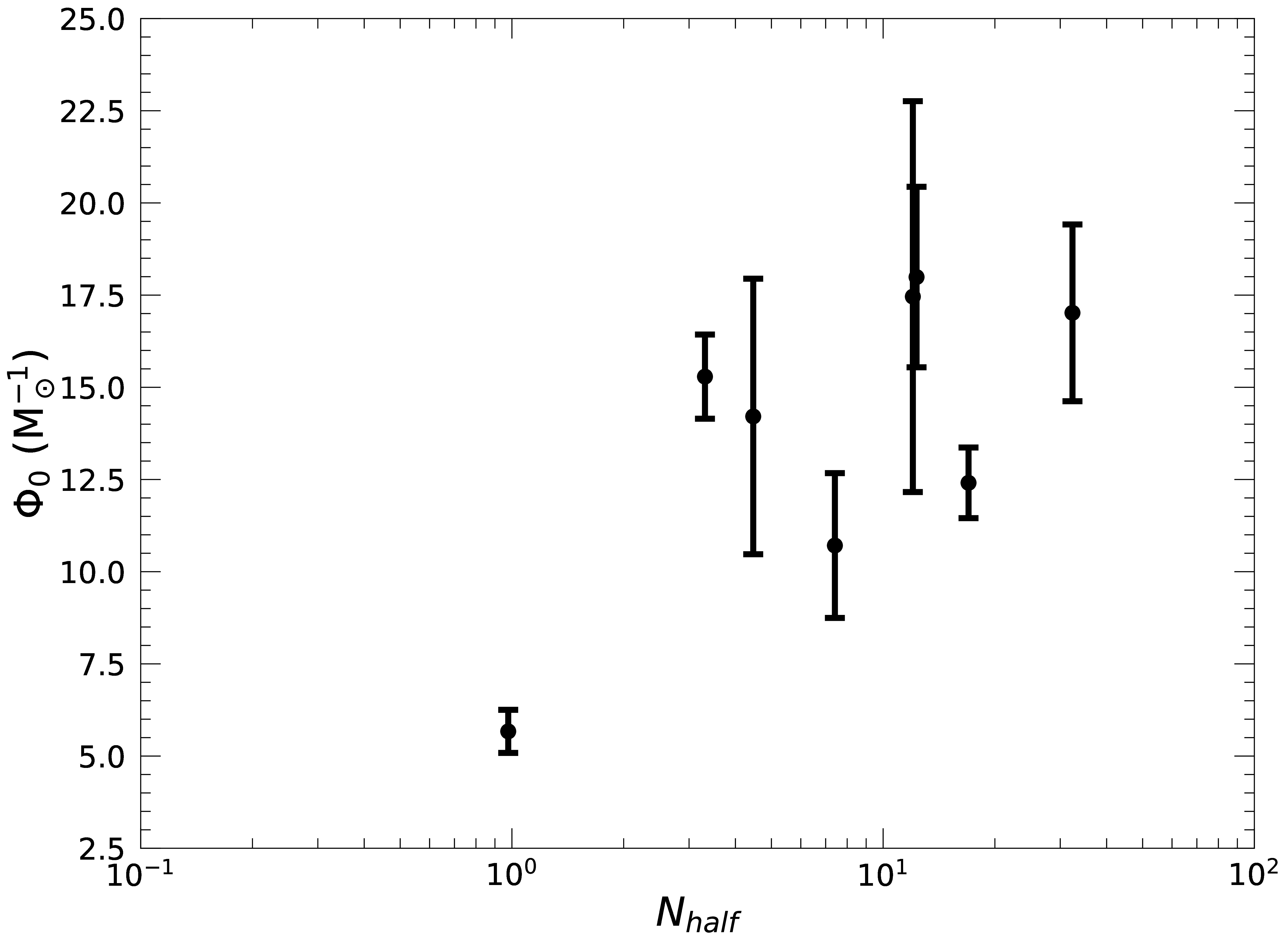}
  \label{fig:PHI0_Nhalf}
\end{subfigure}
\caption{Obtained values of $\Phi_0$ against the number of core relaxation timescales $N_\mathrm{core}$ (\textit{upper panel}) and the number of median relaxation timescales $N_\mathrm{half}$ (\textit{lower panel}) from \citet{Watkins2022}.}
\label{fig:PHI0_Ncore_Nhalf}
\end{figure}
In Fig. \ref{fig:PHI0_Ncore_Nhalf} we show the relation between the gravitational potential and the number of core and median relaxation timescales. The plot shows a tendency of having greater $\Phi_0$ values with a larger $N_\mathrm{core}$ and $N_\mathrm{half}$, as expected for more dynamically evolved states. A similar trend was found by W22 with the $m_\mathrm{eq}$ parameter. Unfortunately, the great errors and the absence of a theoretical prediction on $N_\mathrm{core}$ and $N_\mathrm{half}$ do not allow us to further discuss the obtained pattern. \\ 
Another quantifier of the dynamical state of GCs is the so-called dynamical-clock, mainly measured through the $A^+$ parameter that is the area between the cumulative distribution of Blue Straggler Stars (BSSs) and the reference stars \citep{Ferraro2012, Alessandrini2016, Lanzoni2016, Ferraro2020}. 
Only four of the analyzed GCs have a measure of $A^+$, which also well correlates with the number of core relaxation timescales and the core radius, as shown by \citet{Lanzoni2016} and \citet{Ferraro2018}.
We give in Fig. \ref{fig:PHI0_A+_logrc} the estimated $\Phi_0$ and the corresponding $A^+$ and $\log(r_\mathrm{c})$, where we use again the core radius from the Harris catalog and \citet{Baumgardt2018-2023} online catalog.\\
\begin{figure}[htbp]
\centering
\begin{subfigure}{.45\textwidth}
  \centering
  \includegraphics[width=\linewidth]{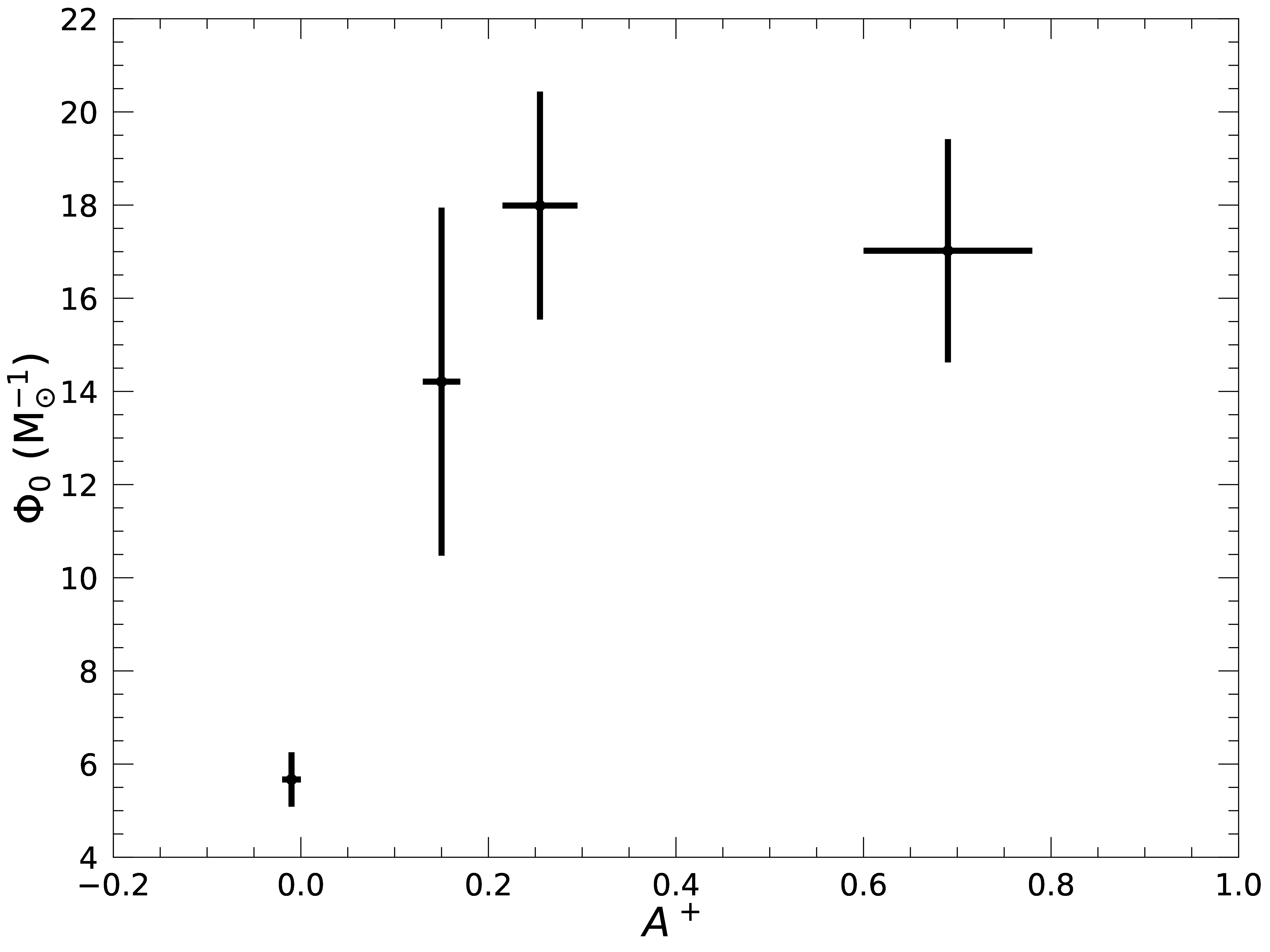}
  \label{fig:PHI0_A+}
\end{subfigure}%
\\
\begin{subfigure}{.45\textwidth}
  \centering
  \includegraphics[width=\linewidth]{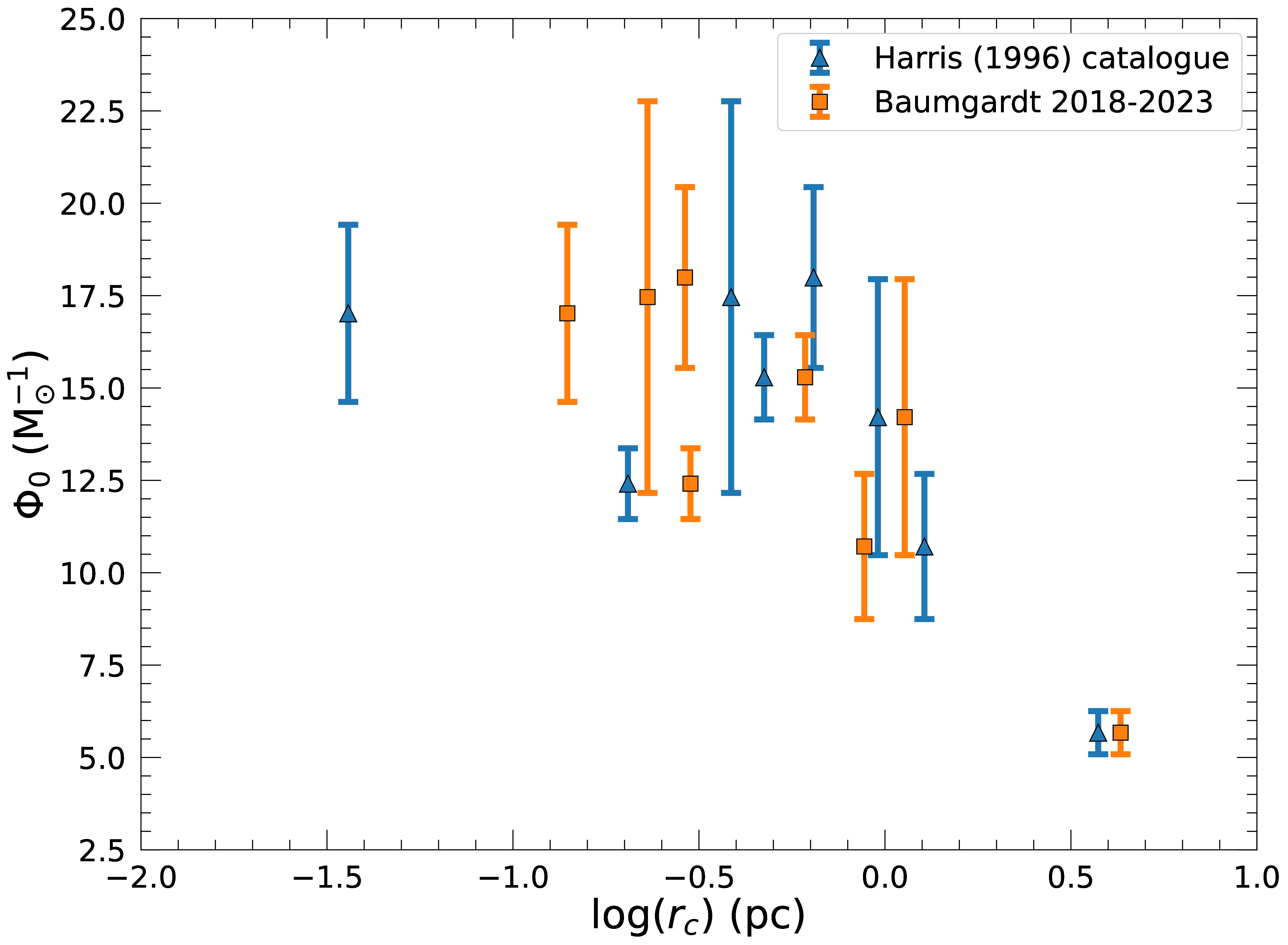}
  \label{fig:PHI0_logrc}
\end{subfigure}
\caption{\textit{Upper panel}: relation between $\Phi_0$ and the area $A^+$ between the cumulative distribution of Blue Straggler Stars and the reference stars.
  \textit{Lower panel}: $\Phi_0$ relation with the core radius, using values from \citet{Harris1996} catalog (blue triangles) and \citet{Baumgardt2018-2023} online catalog (orange squares).}
\label{fig:PHI0_A+_logrc}
\end{figure}
As for the number of core relaxation timescales, the $\Phi_0$ parameter grows at increasing values of $A^+$, both measuring the dynamical state of GCs, although the number of points is small and only a general increasing trend can be deduced. 
While $\Phi_0$ measures the gravitational potential well, the $A^+$ parameter tracks the segregation of BSSs with respect to the other stars, measuring the mass segregation process. 
They have a lower velocity dispersion value, corresponding to their expected higher mass with respect to reference stars, namely the main sequence turn-off mass \citep[see][and refs therein]{Baldwin+2016}.
In principle, the mass of BSSs depends on the formation mechanisms of such objects, which is still unknown. The most spread idea is that they are binary stars or a result of stellar coalescence. To precisely measure the dynamical state of the cluster, the formation of such objects should be mostly coeval with the cluster itself. Any late BSS, that is recently formed, would naturally be less segregated, then dumping the overall BSSs population segregation, resulting in a lower dynamical state for the cluster.
Indeed, the dynamical processes that bring the collisional system toward mass segregation and energy equipartition are phenomena that strongly depend on the mass of the object and the relation between its age and the relaxation timescale.
Our work, as well as several energy equipartition studies in literature, underlines that massive stars are faster in reaching equipartition than less massive ones. Consequently, the population of BSSs can have internal differences in their kinematic properties if a spread in mass exists. As we observe and predict a different degree of energy equipartition for reference stars, a similar phenomenology can state also for BSSs. 
One of the main strength of our modeling approach is the ability of predicting such kinematic differences for objects with whatever mass, once known their mass function and the structural parameter $\Phi_0$.
That would further broaden the role of BSSs in characterizing GCs dynamical state and understanding the complex internal dynamics, giving important insight in the energy equipartition and mass segregation topics. \\
Concerning the relation with the core radius, advanced dynamical states have a greater value of $\Phi_0$ and a lower $r_\mathrm{c}$, as shown in the lower panel of Fig. \ref{fig:PHI0_A+_logrc}, which also affects the concentration that will be higher. Indeed, as the relaxation proceed, the cluster is driven toward the core collapse phase and the gravothermal catastrophe.
Here, the structure of the system is altered and King-based models do not reproduce well the structural properties. 
However, the energy equipartition keeps going on in the survived core, where a higher degree of equipartition can be expected. 
As mentioned in Sect. \ref{subsec:VaryingPHI0}, the estimates of $\Phi_0$ from the observable $\sigma(m)$ may suffer from a systematic effect related to the adopted value for the core radius, which is neglected in our analysis. 
With simple order-of-magnitude calculations and a linear interpolation of the $\Phi_0$ -- $\log{(r_\mathrm{c})}$ relation, 
in the case of \object{NGC 5904} and \object{NGC 6397} a $10\%$ variation in the core radius is requested to get a systematic shift on $\Phi_0$ comparable to the estimated statistical error. For all the other clusters in our sample, the variation of $r_\mathrm{c}$ should be much larger.
Although the relative error for the core radius is typically smaller, as can be seen from Table \ref{tab:params_fitSBPs}, larger discrepancies are found when the Harris catalog values are compared to those of \citet{Baumgardt2018-2023}.

\subsection{Fitting surface brightness profiles}
\label{subsec:FittingSBPs}
To provide an independent determination of the structural parameter $\Phi_0$, we fit the SBPs data by \citet{Trager1995} with the prediction of the dynamical model, by using theoretical isochrones from \href{http://basti-iac.oa-abruzzo.inaf.it/index.html}{BaSTI}  \citep{Hidalgo+2018,Pietrinferni+2021,Salaris+2022,Pietrinferni+2024} for an age of 13 Gyrs, with $[\alpha/\element{Fe}]=+0.4$, $Y = 0.247$ and a different metallicity [\element{Fe}/\element{H}] for each cluster, taken from the Harris catalog, as described in Sect. \ref{subsubsec:SBPs}.\\
In Fig. \ref{fig:NGC6341_SBP} we plot the obtained best-fit profile for \object{NGC 6341}.
In Table \ref{tab:params_fitSBPs} we give the estimated $\Phi_0$ as well as the central surface brightness, the core and tidal radius and the normalized $\chi^2$ value for the analyzed clusters.\\
\begin{figure}[htbp]
    \centering
    \includegraphics[width=\linewidth]{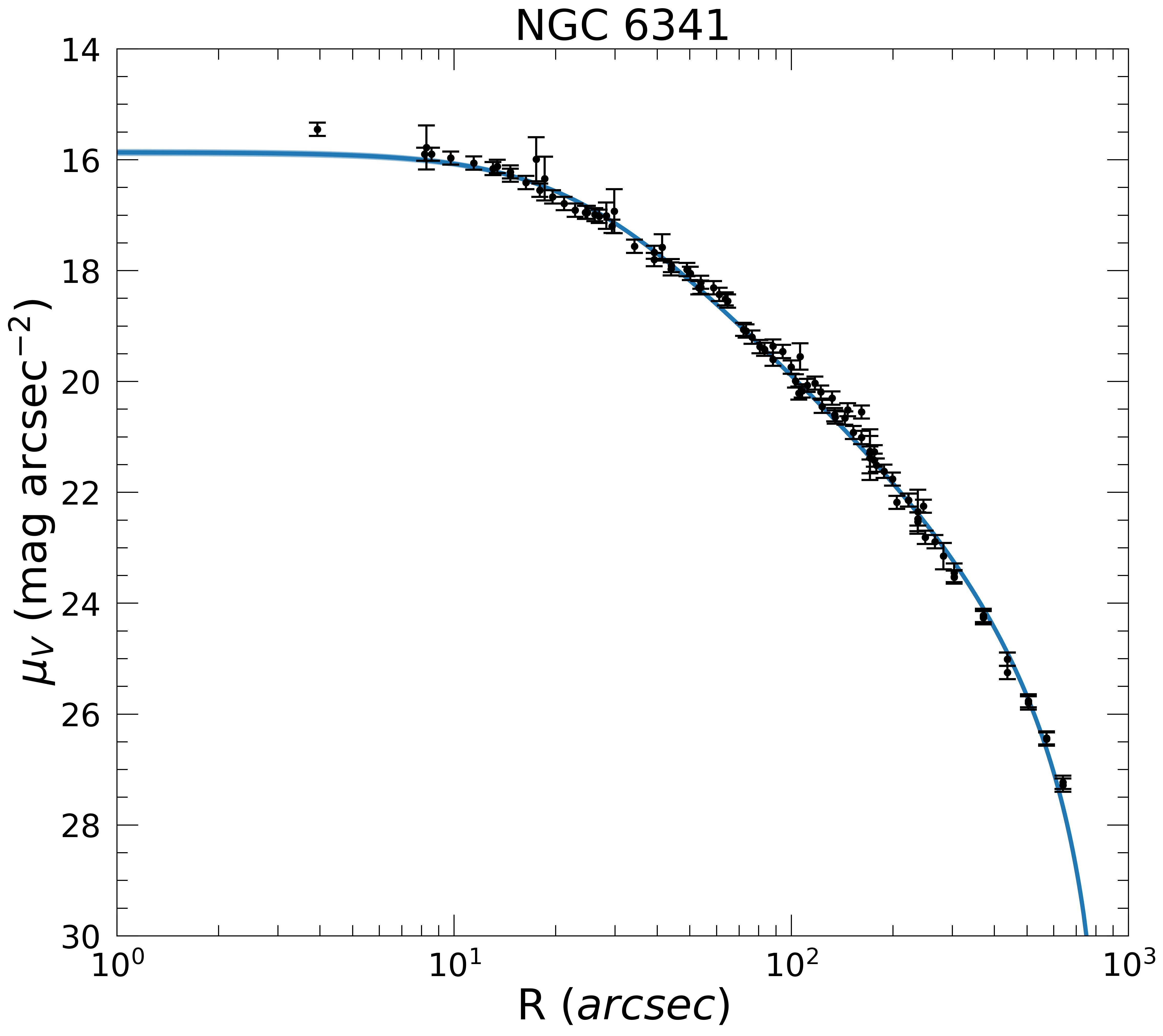}
    \caption{Surface brightness profile for \object{NGC 6341}. The black circles with error bars are the data from \citet{Trager1995} analyzed following the work by \citet{McLaughlin&vanderMarel2005} and \citet{Zocchi+2012}. The continuous blue line is our model best fit with its confidence band, obtained by assuming \citet{Baumgardt2023} mass function slope and adopting the BaSTI isochrones \citep{Hidalgo+2018,Pietrinferni+2021, Salaris+2022, Pietrinferni+2024} with 13 Gyr, $[\alpha/\element{Fe}]=+0.4$, $Y = 0.247$ and metallicity [\element{Fe}/\element{H}]=-2.31, taken from the \citet{Harris1996} catalog (2010 edition).}
    \label{fig:NGC6341_SBP}
\end{figure}
\begin{table*}[htbp]
    \centering
    \caption{Structural parameters obtained by fitting SBPs.}
    \label{tab:params_fitSBPs}
    \begin{tabular}[c]{lcr@{$\pm$}lr@{$\pm$}lr@{$\pm$}lr@{$\pm$}l r | r@{$\pm$}l}
    \hline \hline
    \multicolumn{1}{c}{{ID}} 
    & \multicolumn{1}{c}{\textbf{$\alpha^{(1)}$}} 
    &\multicolumn{2}{c}{$\Phi_0^\mathrm{13 Gyr} \,(M_{\sun}^{-1}$)} 
    & \multicolumn{2}{c}{$\mu_\mathrm{V,0}$ (mag$_\mathrm{V}$ arcsec$^{-2}$)} 
    & \multicolumn{2}{c}{$r_\mathrm{c}$ (arcsec)} 
    &\multicolumn{2}{c}{$r_\mathrm{t}$ (arcsec)} 
    & \multicolumn{1}{c}{ $\chi_\mathrm{norm}^2$} 
    &\multicolumn{2}{c}{$\Phi_0^\mathrm{11 Gyr}\,(M_{\sun}^{-1}$)} 
    \\
        \hline
\object{NGC 104}  & -0.65 &  13.55&0.03 & 14.50&0.02 & 26.27&0.20 & 3119&24   & 3.06 & 12.95&0.03\\
\object{NGC 5139} & -0.80 &  10.89&0.10 & 16.85&0.04 & 146.6&1.6  & 3206&35   & 1.63 & 10.37&0.09\\
\object{NGC 5904} & -0.76 &  12.75&0.12 & 16.29&0.04 & 33.4&0.34 & 1601&16   & 3.18 & 12.14&0.07\\
\object{NGC 6266} & -1.14 &  14.23&0.09 & 15.45&0.03 & 15.80&0.25 & 1133&18   & 2.93 & 13.63&0.08\\
\object{NGC 6341} & -0.82 &  13.04&0.10 & 15.87&0.03 & 20.79&0.22 & 885.4&9.3 & 1.95 & 12.39&0.07\\
\object{NGC 6397} & -0.32 &  14.38&0.20 & 16.86&0.04 & 30.97&0.76 & 3151&77   & 3.69 & 13.70&0.26\\
\object{NGC 6656} & -0.90 &  12.51&0.25 & 17.47&0.02 & 86.7&1.5   & 3027&51   & 1.06 & 11.97&0.26\\
\object{NGC 6752} & -0.67 &  15.09&0.07 & 15.47&0.02 & 17.95&0.19 & 2220&24   & 2.22 & 14.40&0.05\\ \hline
    \end{tabular}
    \tablefoot{Table columns: ID, mass function slope (1), estimated $\Phi_0$, central surface brightness $\mu_\mathrm{0,V}$, core radius $r_\mathrm{c}$, tidal radius $r_\mathrm{t}$ and normalized $\chi^2$ test value. The last column gives the estimated $\Phi_0$ for an 11 Gyr isochrone.}
    \tablebib{
(1)~\citet{Baumgardt2023}.
}
\end{table*}
The obtained best-fit curves reproduce well the observed data, with a very narrow error band and small uncertainties for the parameters. 
However, the obtained estimates neglect possible error sources, both statistical and systematic ones.
Regarding the conversion of the observed magnitudes into luminosities, several systematic effects can occur, as due to the extinction values or the distance modulus value as well as in the assumed solar V-magnitude. 
However, all these parameters produce an offset in the magnitude, but they do not affect our minimization procedure and the obtained $\Phi_0$.
This is not the case for the parameters related to the isochrones, such as the age and the adopted values of [$\alpha$/\element{Fe}], \element{He} abundance and [\element{Fe}/\element{H}]. 
As an example, we report in the last column of Table \ref{tab:params_fitSBPs} the effect of a different age in the estimated value of $\Phi_0$, using isochrones at 11 Gyr, keeping the previous adopted values for [$\alpha$/\element{Fe}], $Y$ and [\element{Fe}/\element{H}]. 
The overall impact of a reduced age is a smaller $\Phi_0$, resulting in a systematic effect. 
As already explained in Sect. \ref{subsubsec:SBPs}, a smaller age implies a larger maximum mass. Since more massive stars are more segregated and affect the luminosity distribution in the innermost region of the cluster, a steeper brightness profile is expected. Therefore, in order to reproduce the observed SBP, the effect of an increase in the maximum mass should be compensated by a reduction of the $\Phi_0$ value.
Note that we cannot distinguish from the 11 Gyr and 13 Gyr cases, because the implied variations of the observable parameters (such as the central surface brightness, core radius and tidal radius) are smaller than their respective statistical errors.
In addition, the isochrones vary from cluster to cluster also due to the different metallicity that we considered. 
The role of the metallicity, [$\alpha$/\element{Fe}], \element{He} mass fraction $Y$, and cluster age in shaping GCs SBPs, especially in the framework of multi-mass King-like models, requires a broader exploration and a dedicated work. The same holds for the mass function shape, that can alter the predictions due to its structural role, being a fundamental physical ingredient in such dynamical models. In fact, the massive stars content is extremely important in shaping the SBPs and thus the slope $\alpha$ is expected to play a relevant effect.\\
Finally, we compare the independent estimates of $\Phi_0$ obtained from internal kinematics, namely the velocity dispersion as function of stellar mass, with those from the SBPs. Concerning the latter, we first account for age variations by averaging the corresponding $\Phi_0$ values and computing the associated uncertainty. 
In Fig. \ref{fig:PHI0SBPs-sigmam} we plot the relative variation $\Delta\Phi_0/\langle \Phi_0 \rangle$, showing the $2\sigma$ confidence interval.
\begin{figure}
    \centering
    \includegraphics[width=\linewidth]{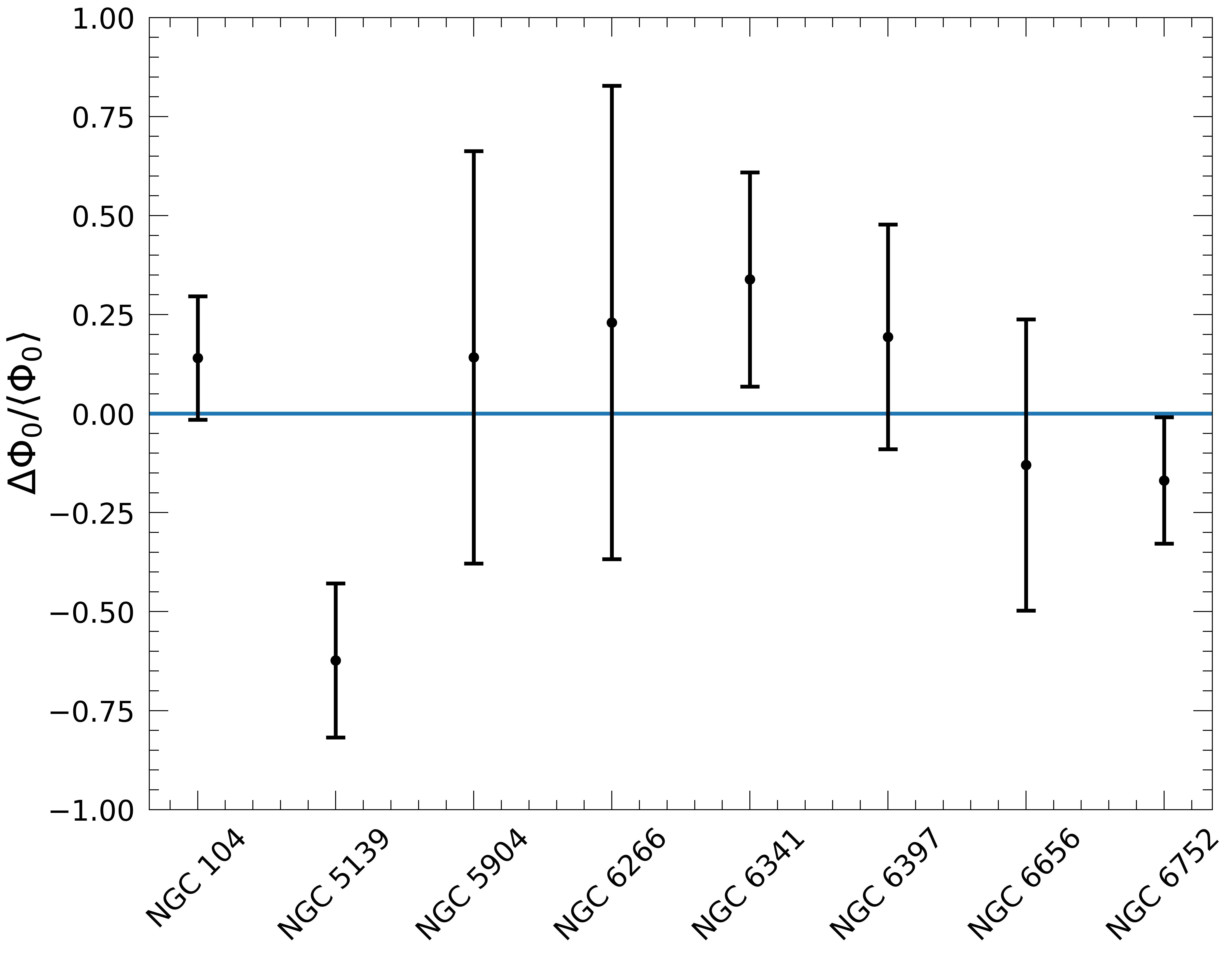}
    \caption{Relative variation in the estimates of the $\Phi_0$ parameter for each of the analyzed clusters, obtained by fitting the velocity dispersion dependence on stellar mass $\sigma(m)$ and the surface brightness profiles $I(R)$. The error bars are the $2\sigma$ level uncertainties.}
    \label{fig:PHI0SBPs-sigmam}
\end{figure}
The two estimates are compatible at $2\sigma$ level for most clusters, although some are compatible at $1\sigma$ confidence interval (namely \object{NGC 5904}, \object{NGC 6266} and \object{NGC 6656}). The larger deviation is obtained for \object{NGC 5139} ($\omega$-Cen). As already mentioned, this cluster is a much more complex object. It is probably the remnant of a dwarf galaxy. It hosts multiple stellar generations, also showing metallicity variations, for which multiple isochrones should be used and potentially multiple dynamical states for each of the subsystems that make up such an object.\\
The plot also highlights how multi-mass King-like dynamical models can predict both internal kinematics and the surface brightness profiles. 
However, as already outlined, there are several possible sources of error that could further increase our uncertainties in the $\Phi_0$ estimates. 
A full exploration of the assumptions made in the SBPs fitting procedure is needed to better constrain the $\Phi_0$ value.
Only afterward, its relation with the other structural quantities such as concentration, number of relaxation times, core radius and $A^+$ can be addressed similarly to what done in Sect. \ref{subsec:RelationStructParams}.


\section{Conclusions}
In this work, we explore the energy equipartition degree in GCs by means of a multi-mass King-like dynamical model for collisional and phase-space limited systems. Such a theoretical description offer a fundamental way to predict several dynamical processes occurring in GCs, such as energy equipartition, mass segregation and evaporation. These phenomena alter the properties of clusters, modifying their spatial and kinematic structure.
Here we focus on the equipartition degree, determining the dynamical state of few GCs by means of model parameters.\\
We fit the velocity dispersion dependence on stellar mass measured through HST proper motion by \citet{Watkins2022} with our prediction, discussing the relation with the \citet{Bianchini2016} fitting function and the equipartition mass $m_\mathrm{eq}$. We obtain a similar confidence level and a relation between the equipartition mass and the model parameter $\Phi_0$, that measures the gravitational potential difference between the center and the edge of the cluster.
An increasing value for such quantity describes a more dynamically old cluster, producing a larger degree of energy equipartition, associated with a lower value of $m_\mathrm{eq}$.\\
We confirm that energy equipartition in the analyzed GCs is only partial, even in the central regions. The equipartition degree looks similar in the core, while strongly dumps in the outer regions. More massive stars are closer to equipartition than less massive ones, as already found and outlined in other works.\\
We extend our fitting procedure by adding to the parameter space the slope of the mass function $\alpha$, a required input for the model, initially taken from \citet{Baumgardt2023}. This procedure provides similar results regarding the estimate of $\Phi_0$, but prevents any constraint on the mass function slope from data. Although we obtain a degeneration between the two parameters, the effect of varying the gravitational potential well alters more the observable $\sigma(m)$ than the variations in the mass function, when reasonable (i.e., $\alpha \in [-2.0,0]$).\\
We carefully discuss the implication of quantifying the degree of energy equipartition through the Bianchini fitting function with respect to our dynamical model prediction, when working with a restricted radial shell and projected quantities. 
We find that the estimated $m_\mathrm{eq}$ from the projected dispersion is higher than the three-dimensional one. Moreover, for radial shells overcoming the core radius, it underestimates the maximum degree of equipartition, reached in the core. On the contrary, the equilibrium parameter $\Phi_0$ uniquely defines both projected and 3D radial theoretical profiles, without suffering shell selection and projection effects.
We also get that variations in the slope $\alpha$ affect the equipartition degree mainly in the outer regions, while its effect is relatively small in the core. Here, by fitting theoretical profiles with the Bianchini function, we obtain a mostly constant $m_\mathrm{eq}$, suggesting using this value as a measure of the maximum degree of energy equipartition in clusters. However, taking advantages of dynamical models like ours offers the opportunity to quantify the equipartition degree in the core of clusters by means of structural parameters. Here, observational data suffer several limitations and a theoretical tool predicting observables can give an important support to the astronomical community.\\
We compare the estimates of $\Phi_0$ with other structural properties of GCs, which depend on the dynamical state of the system.
The theoretical relation between $\Phi_0$ and the concentration is presented, as well as the estimated $\Phi_0$ against observational values for the concentration, obtained from different sources \citep{Harris1996,Baumgardt2018-2023,Webb2013}.
The spreads among observations as well as the differences with respect to the theoretical prediction underline the presence of several error sources. We advise recovering the predicting role of models concerning the determination of the concentration value for GCs.
An increasing trend is also seen between $\Phi_0$ and the number of core and median relaxation timescales.
A similar tendency occurs also with the area $A^+$ between Blue Straggler Stars and reference stars cumulative distributions \citep{Lanzoni2016, Ferraro2018}, although the number of points is small (only four clusters).
All such structural quantities increase for dynamically old systems, as $\Phi_0$.
Conversely, the core radius decreases with the dynamical state, a trend we clearly see between $\Phi_0$ and $r_\mathrm{c}$. 
These outcomes strongly suggest including $\Phi_0$ among the properties that track the dynamical age of GCs.
However, a statistically satisfying comparison between structural parameters of Galactic GCs and our model ones must be addressed in future works, with a wider sample of clusters.\\
Finally, we successfully fit the surface brightness profiles by \citet{Trager1995}, following a procedure similar to that described by \citet{TrentiVanDerMarel2013} and \citet{Zocchi+2012}. To make the theoretical prediction, we need the surface density profile and a mass-luminosity relation, that we take from theoretical isochrones \citep{Hidalgo+2018,Pietrinferni+2021,Salaris+2022,Pietrinferni+2024}. We obtain an estimate for $\Phi_0$, the central surface brightness, the core and tidal radius as well. We discuss the possible effects that can alter our predictions, mainly in the mass-luminosity relation. A different age for the isochrone introduces a systematic error in the determination of $\Phi_0$ that is larger than the statistical uncertainty inherent to the fitting procedure. 
We average these contributions to get $\Phi_0$ estimates from the fitting procedure on each cluster SBP. We compare these values with the ones computed by fitting the velocity dispersion -- mass relation, showing that they are compatible at $2\sigma$ level for almost all clusters. The larger deviation is obtained for $\omega$-Cen, whose complexity suggests that our fitting procedure needs to be extended to consider multiple stellar populations, with a different chemical content, metallicity and consequently different isochrones.\\
The obtained results strongly underline the central role of dynamical models in predicting GCs phenomenology, such as the highly debated energy equipartition process.
With the increasing amount of information coming from internal kinematics observations, effort is required in advanced physical modeling. It can provide strong tools to unveil the mechanisms behind several dynamical phenomena such as segregation and evaporation, as well as anisotropic velocity distributions and internal rotation.
However, the determination of the dynamical state and the prediction on different observables suffers from several sources of error, both for the internal kinematics and for the surface brightness profiles, where the role of the assumptions made regarding structural properties and the mass-luminosity relation must be further explored.\\

\begin{acknowledgements}
MT thanks Michele Bellanzini for important hints that expanded this work.
The authors would thank the anonymous referee for the useful indications and comments that improved this paper.
\end{acknowledgements}

\bibliographystyle{bibtex/aa} 
\bibliography{biblio.bib}


\begin{appendix}
\label{sec:Appendix}

\section{Model derivation}
\label{sec:App-modelderivation}
The distribution function (DF) that defines our model is given in Eq. \eqref{eq:mmDF}. It comes from an analytical derivation, which recall the steps from the dynamical friction \citep{Chandrasekhar1943ApJ} to the \citet{King66} model.
In \citet{King65}, the author obtained a DF as an approximated solution of the Fokker-Planck equation, valid for collisional systems. Then, he generalized and used it in \citet{King66}. Our DF keeps the memory of the King one, but takes the mass distribution into account, as was done by \citet{DaCostaFreeman1976}. They presented a discrete model, where each mass class has a DF with an energy cut-off and a weight factor, to be constrained from observations.
In principle, the assumption that each mass has that specific DF analytical shape must be proved similarly, solving the Fokker-Planck equation in a multi-mass system.\\
The procedure that lead to the King DF starts from the Boltzmann equation for collisional systems and follows the \citet{Chandrasekhar1943RevModPhys} development: assuming local approximation and low energy exchanges, the equation is written in terms of the diffusion coefficients, which quantify the dynamical friction that a test star with mass $m$ suffers due to field stars with mass $m_\mathrm{a}$. Following \citet{BinneyTremaine2008}, the \citet{SpitzerHarm1958} expression of the Fokker-Planck equation is obtained. King was able to solve such equation in 1965 with an approximation. He originally obtained a DF where both masses $m$ and $m_\mathrm{a}$ appear, but then he takes them equal.\\
In principle, the same approach can be followed again, but taking care of considering the field stars' mass function when evaluating the diffusion coefficients, as well as their velocity distribution, assumed Maxwell-Boltzmann like. 
A further needed step is the distinction between dynamical quantities and thermodynamic ones. In particular, when using a Maxwell-Boltzmann DF for field stars, which is $\propto \exp{[-v^2/(2\sigma_\mathrm{a}^2)]}$, one gets the scaling factor $\sigma_\mathrm{a}$ which is related to the thermodynamic temperature $\theta$ (kept constant) and the field stars' mass, namely $\sigma_\mathrm{a}^2 = k_\mathrm{B}\theta/m_\mathrm{a}$. 
In this framework $\sigma$ is the 1D velocity dispersion of the Maxwell-Boltzmann DF, and it depends only on the mass, while there is another quantity, which comes at the end of the procedure, that is the kinetic one.
In the limit of a Maxwell-Boltzmann DF for the test star (that describes an isothermal sphere) the kinetic quantity converges to the thermodynamic one.
This distinction allows us to explicit mass dependence in the obtained generalized expression of the Fokker-Planck equation, which finally states
\begin{equation}
    \frac{\mathrm{d}f}{\mathrm{d}t} = \frac{1}{t_\mathrm{R}(m)}\frac{1}{x^2}\frac{\partial}{\partial x}\left[ 2x\mathcal{G}(x,m)\left(2xf + \frac{\partial f}{\partial x}  \right)\right],
    \label{eq:mmFP}
\end{equation}
where $f=f(x,m;t)$ is the DF of the test star, $x=v^2/(2\sigma^2)$ and 
\begin{equation}
    \frac{\mathcal{G}(x,m)}{t_\mathrm{R}(m)}=\int_{\Delta m_\mathrm{a}}\frac{1}{\tau_\mathrm{R, a}}\left(\frac{m}{m_\mathrm{a}} \right)^{3/2} \left[ \frac{2}{\sqrt{\pi}x_\mathrm{a}^2}\int_0^{x_\mathrm{a}}y^2 \mathrm{e}^{-y^2}\mathrm{d}y\right] \mathrm{d}m_\mathrm{a},
\end{equation}
with $x_\mathrm{a}=v^2/(2\sigma_\mathrm{a}^2)=m_\mathrm{a}v^2/(2k_\mathrm{B}\theta)$ and $\tau_\mathrm{R, a}=\tau_\mathrm{R}(m,m_\mathrm{a})$ the relaxation time for binary encounters between $m$ and $m_\mathrm{a}$, with the latter distributed in the interval $\Delta m_\mathrm{a}$. Here, $t_\mathrm{R}(m)$ is the relaxation time of the mass $m$, due to gravitational encounters with all the field stars.\\
Following \citet{King65}, Eq. \eqref{eq:mmFP} can be solved similarly assuming $f(x,m;t)=g(x,m)\exp{[-\lambda(m)\, t/t_\mathrm{R}(m)]}$, with $\lambda$ the evaporation rate of stars with mass $m$ and $g(x,m)=A(m)\Bar{g}(x,m)$, that brings the dependence on $x$ and the dimensions in the multiplying factor $A(m)$. 
With an expansion in power series of the evaporation rate, with boundary conditions $\Bar{g}(0)=1$, $\Bar{g}'(0)=0$ and $\Bar{g}(x_\mathrm{e})=0$ where $x_\mathrm{e}$ is the cut-off, we obtain the approximated solution for $g(x,m)$ in the central region, where the treatment is valid, that is
\begin{equation}
    {g}(x,m) = k(m)\left[ \exp\left({-x^2}\right) -\exp{\left(-x_\mathrm{e}^2\right)}\right],
    \label{eq:g(x,m)}
\end{equation}
with $k(m)$ that gathers all the mass-dependent multiplying factors. 
From the reasonable assumption that field stars have a Maxwell-Boltzmann DF, to be recovered in the limit of an infinite escape velocity (i.e., the isothermal sphere), and since the DFs we are dealing with give the number of stars in the infinitesimal volume $\mathrm{d}^3v\,\mathrm{d}^3r\,\mathrm{d}m$, it results that
\begin{equation}
    k(m) = \left(\frac{m}{2\pi k_\mathrm{B}\theta}\right)^{3/2}\frac{\mathrm{d}n_0(m)}{\mathrm{d}m}\left[1-\exp{(-x^2_\mathrm{e})}\right]^{-1},
\end{equation}
where $n_0(m)$ is the number density of stars with mass between $m$ and $m+\mathrm{d}m$ in the central region and $\mathrm{d}n_0/\mathrm{d}m = \xi_0(m)/V$ with $\xi_0(m)$ the mass function and $V$ the volume of the central region. \\
The final step is the generalization to different radial regions. Applying the Jeans theorem leads to the DF in Eq. \eqref{eq:mmDF} that is given in terms of the kinetic energy and cut-off energy.

\section{Additional plots}
\label{sec:App-additionalplots}
In this section, we report some additional plots.\\
In Fig. \ref{fig:additional_sigmam} we give the velocity dispersion as function of stellar mass $\sigma(m)$ for the analyzed GCs, without \object{NGC 6397} (given in Fig. \ref{fig:NGC6397sigmam_1par}), with the best-fit model as output of the first fitting procedure, described in Sect. \ref{subsec:VaryingPHI0}, that takes the mass function slopes of GCs from \citet{Baumgardt2023} and constrain $\Phi_0$ and $\sigma_0$ from W22 projected dispersion data. It is also shown the fit result with the Bianchini fitting function. The respective parameters are in Table \ref{tab:params_alphaBaumgardt2023}.\\
In Fig. \ref{fig:additional_contours} we show the contour plots obtained for the same GCs, following our second approach of introducing the mass function slope to the parameter space (Sect. \ref{subsec:VaryingPHI0alpha}).\\
In Fig. \ref{fig:additional_SBPs} we report the SBPs obtained for each GC without \object{NGC 6341} (given in Fig. \ref{fig:NGC6341_SBP}), by fitting \citet{Trager1995} data with the model prediction as described in Sect. \ref{subsubsec:SBPs}, by assuming the mass function slopes from \citet{Baumgardt2023}, BaSTI isochrones \citep{Hidalgo+2018,Pietrinferni+2021, Salaris+2022, Pietrinferni+2024} at 13 Gyr, with $[\alpha/\element{Fe}]=+0.4$, $Y=0.247$ and [\element{Fe}/\element{H}] from the Harris catalog. 
It is important to note that the data from \citet{Trager1995} are a collection of heterogeneous observations, and some clusters have multiple values for the same radial coordinate, also showing different trends.
This likely degrades the quality of the fit and may lead to inaccurate values for the parameters. For example, while the high value of the normalized $\chi^2$ for \object{NGC 104} and \object{NGC 5904} can be related to small uncertainties in the data points, \object{NGC 6397} clearly has multiple data trends (see Fig. \ref{fig:additional_SBPs} in Appendix \ref{sec:App-additionalplots}). 
Consequently, not only we get a large $\chi^2$ value, but also the estimated core radius is an order of magnitude larger than that from the Harris catalog. 
This suggests that the points with smaller errors in the inner region are more reliable, which is also confirmed by the SBP by \citet{Noyola2006}, which looks more regular in the region $R<10^2$ arcsec. Thus, for this cluster our fitting procedure is not reliable and needs to be repeated by increasing the quality of the data, such as changing the cut-off value for $w_i$ and including the data from \citet{Noyola2006}.
Furthermore, it must be considered that \object{NGC 6397} is a post-core-collapse object, for which King-like models are not appropriate in describing the structure of the system, especially in the inner regions.\\
Finally, Fig. \ref{fig:chisq_1par_RelErrsigmam} gives the relation between the $\sigma(m)$ relative error on data against our $\chi^2$ value, while Fig. \ref{fig:sigmam_3D_eqlimit} underlines how the predicted velocity dispersion as function of the mass tends to the equipartition limit $\sigma(m) \propto m^{-1/2}$ for increasing values of $\Phi_0$.

\begin{figure*}[htbp]
\centering
\begin{subfigure}{.39\textwidth}
  \includegraphics[width=\linewidth]{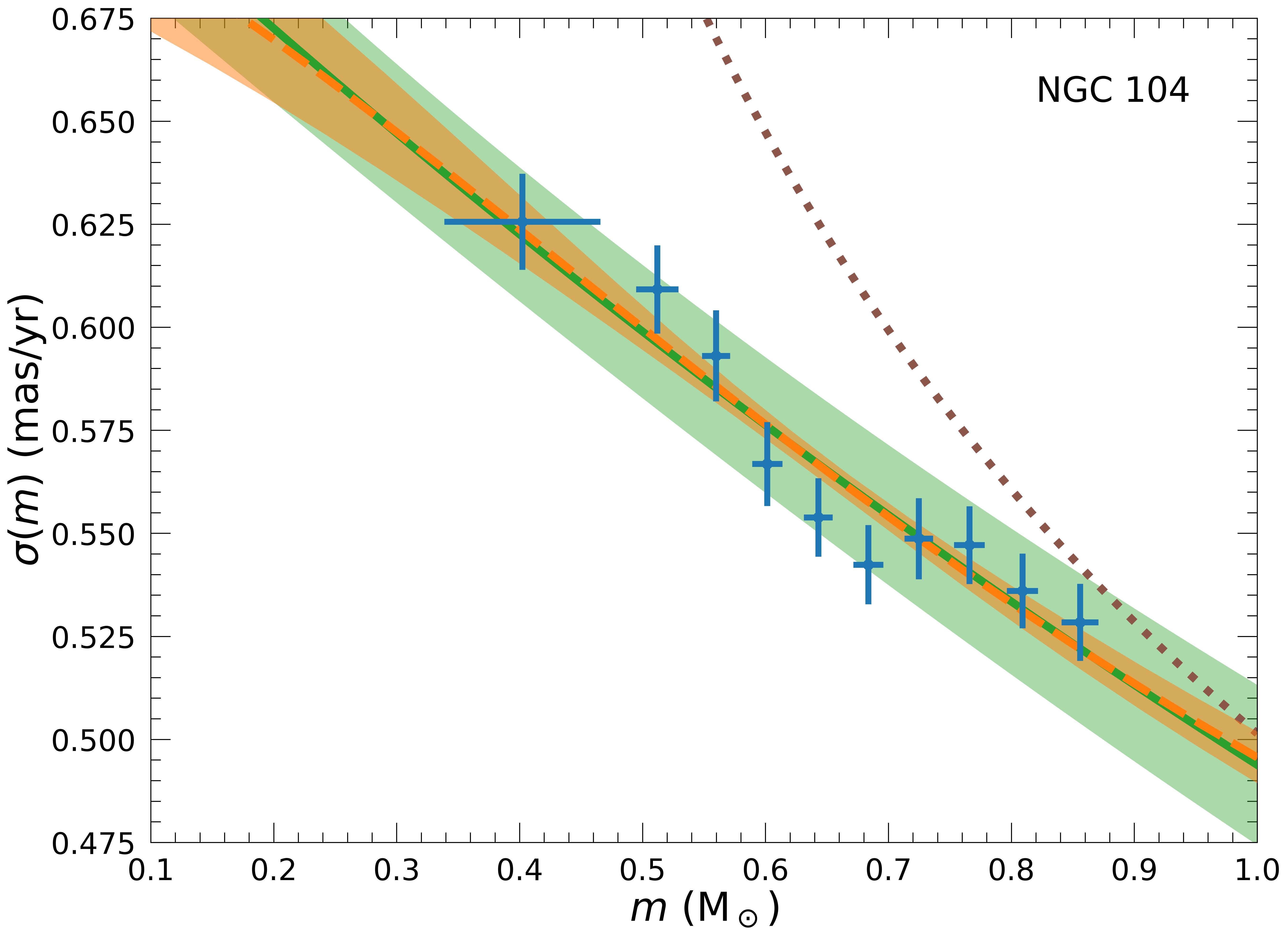}
    \label{fig:NGC104sigmam_1par}
\end{subfigure}%
\hspace{0.05\textwidth}
\begin{subfigure}{.39\textwidth}
  \centering
  \includegraphics[width=\linewidth]{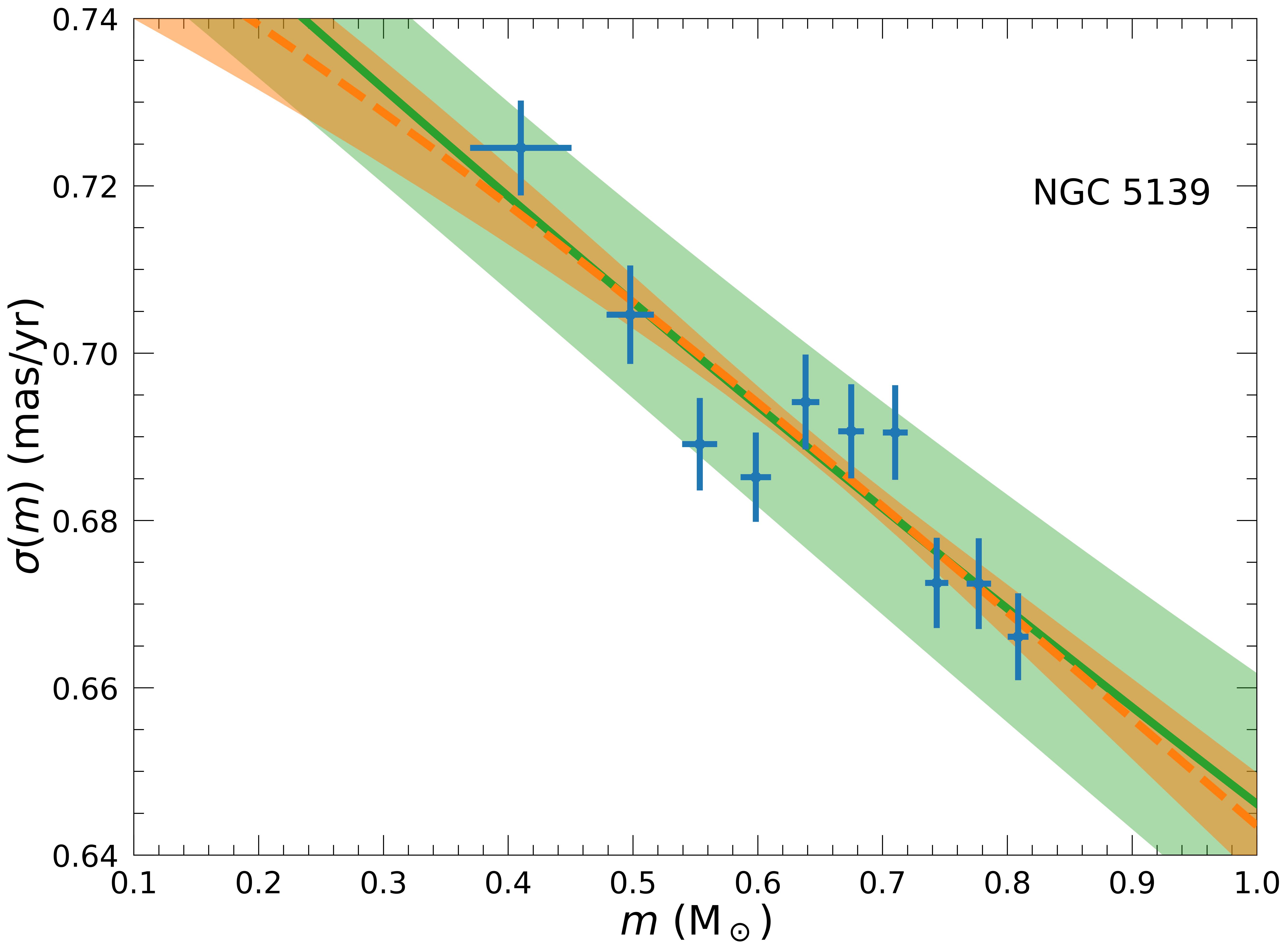}
    \label{fig:NGC5139sigmam_1par}
\end{subfigure}\\
\centering
\begin{subfigure}{.39\textwidth}
  \centering
  \includegraphics[width=\linewidth]{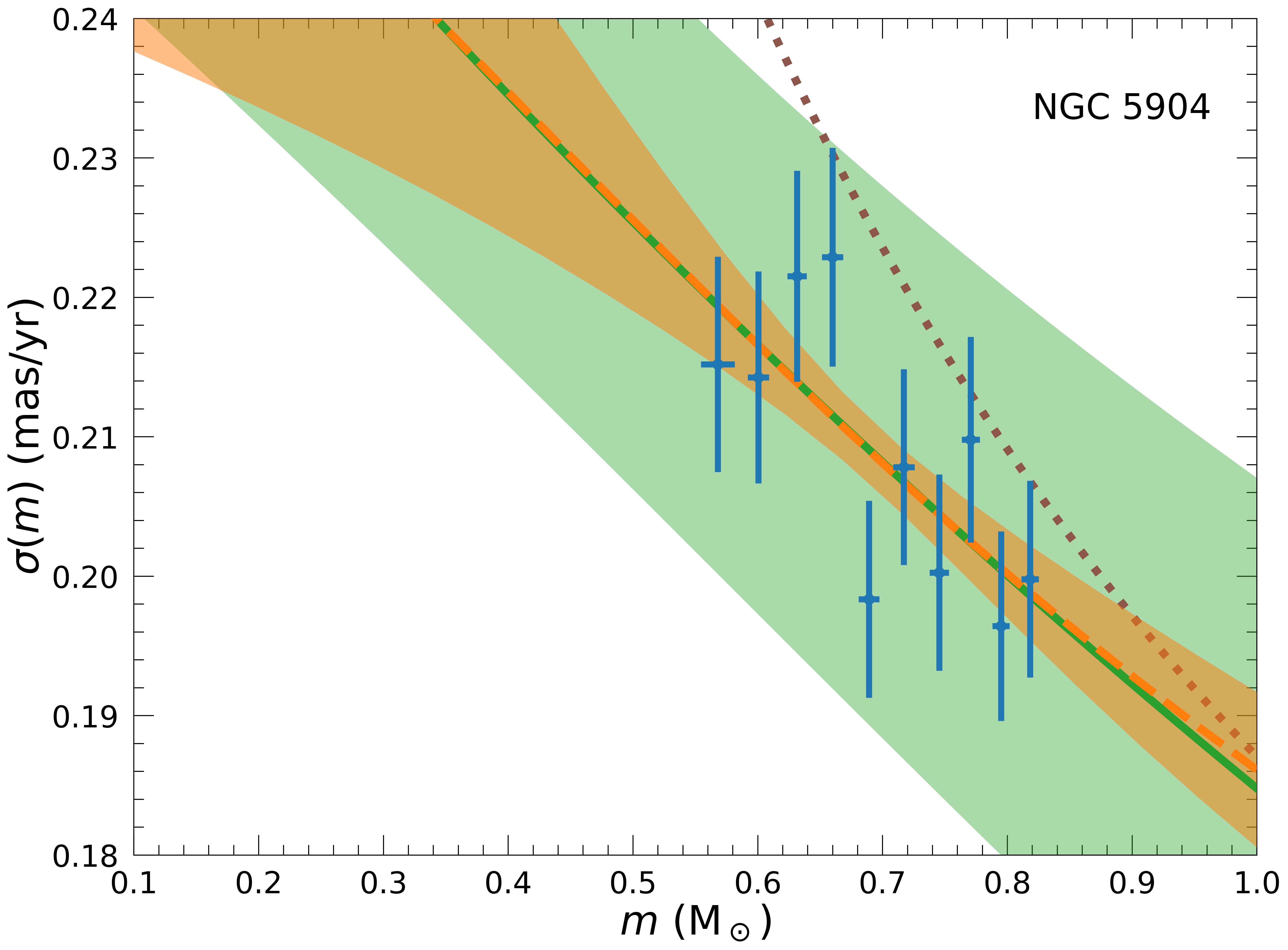}
    \label{fig:NGC5904sigmam_1par}
\end{subfigure}%
\hspace{0.05\textwidth}
\begin{subfigure}{.39\textwidth}
  \centering
  \includegraphics[width=\linewidth]{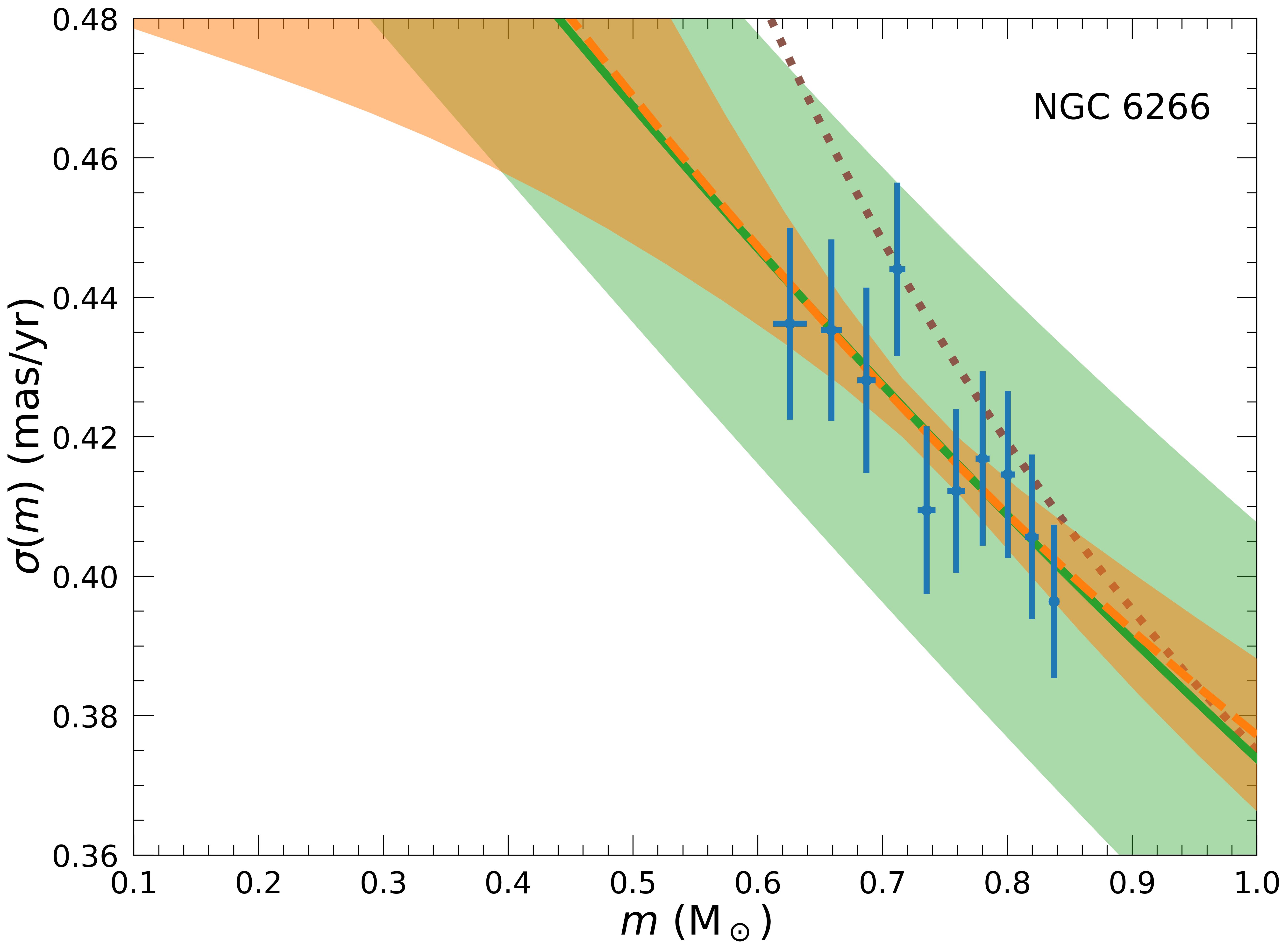}
    \label{fig:NGC6266sigmam_1par}
\end{subfigure}\\
\centering
\begin{subfigure}{.39\textwidth}
  \centering
  \includegraphics[width=\linewidth]{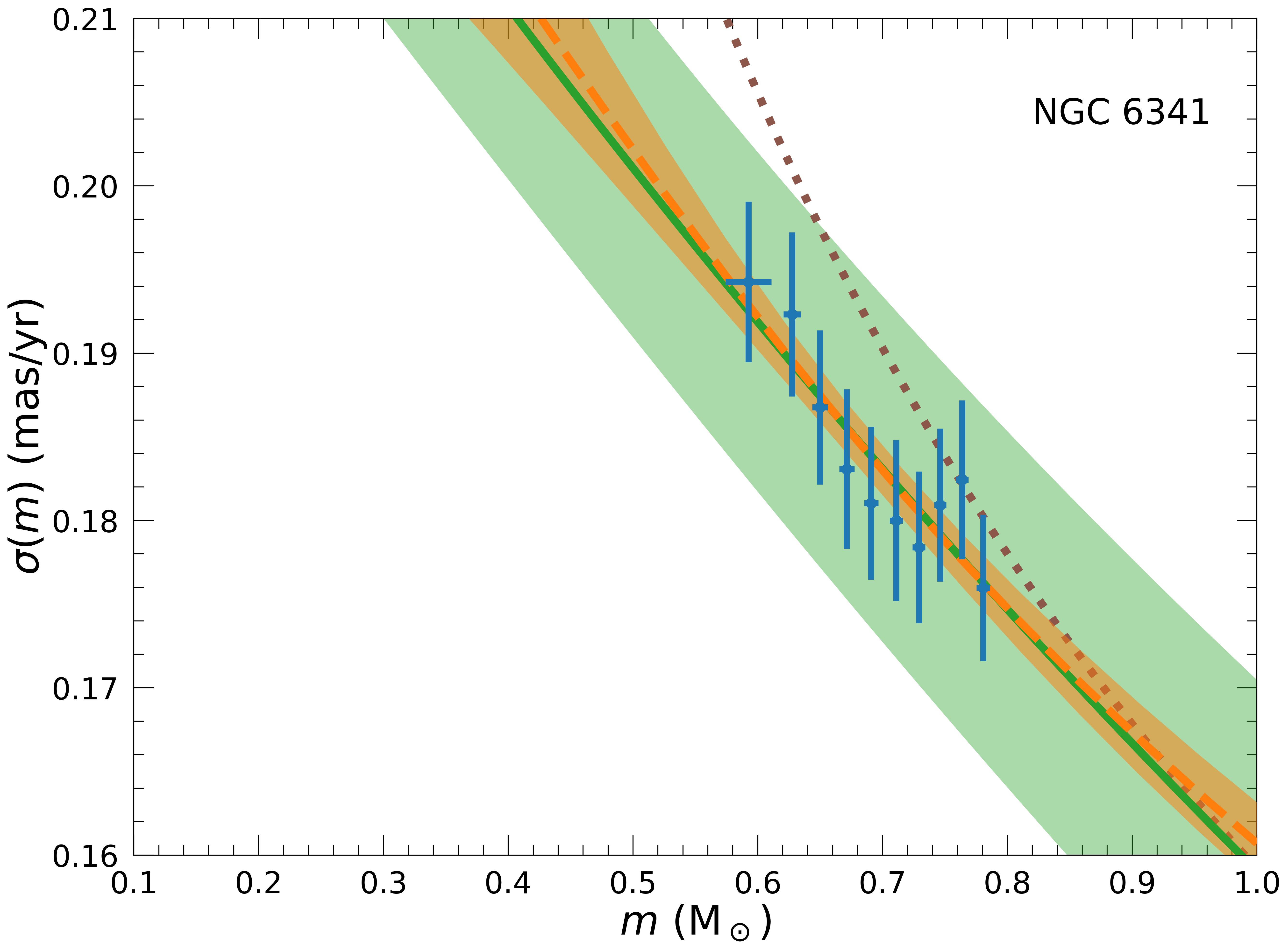}
    \label{fig:NGC6341sigmam_1par}
\end{subfigure}%
\hspace{0.05\textwidth}
\begin{subfigure}{.39\textwidth}
  \centering
  \includegraphics[width=\linewidth]{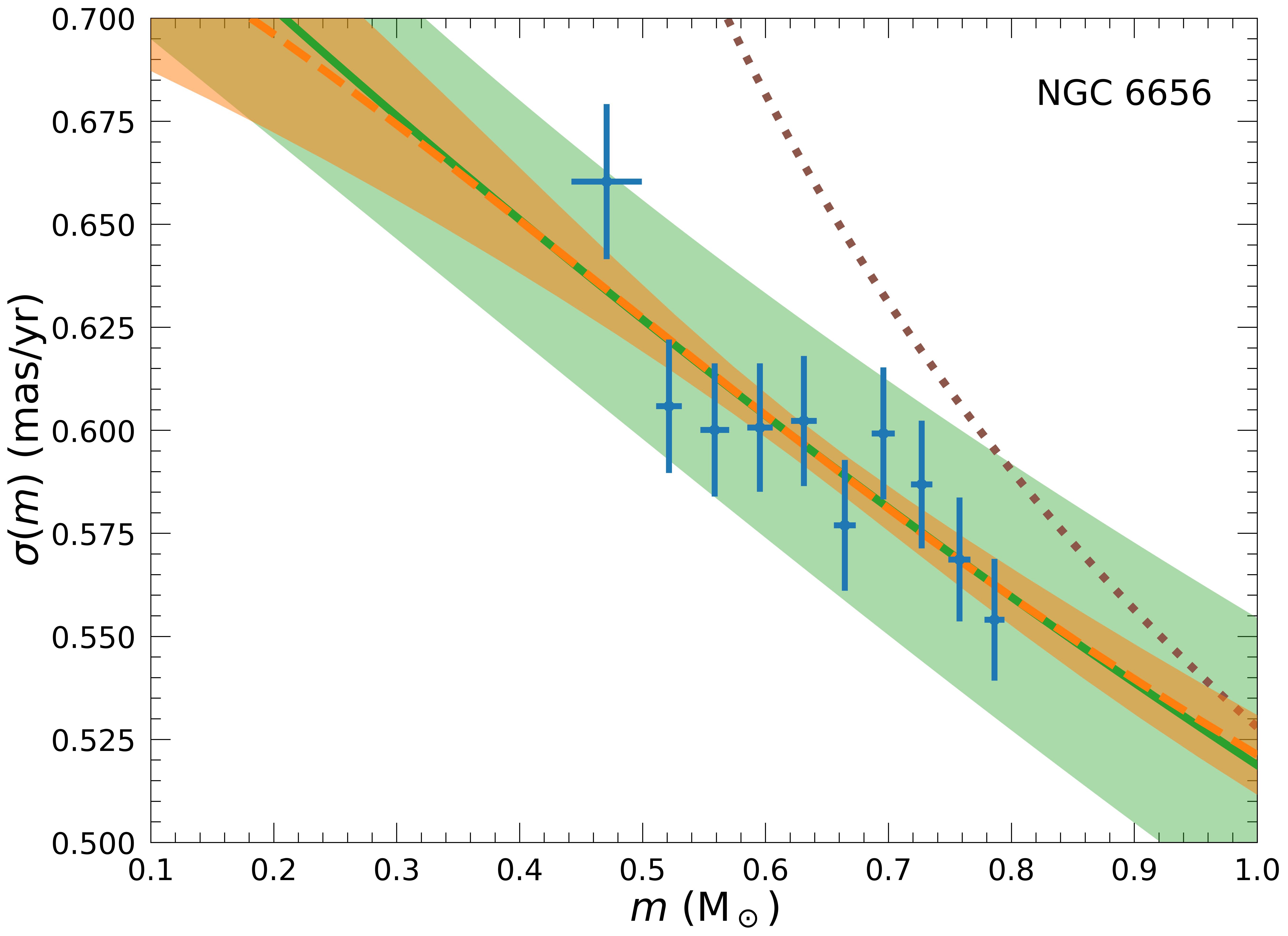}
    \label{fig:NGC6656sigmam_1par}
\end{subfigure}\\
\centering
\begin{subfigure}{.39\textwidth}
  \centering
  \includegraphics[width=\linewidth]{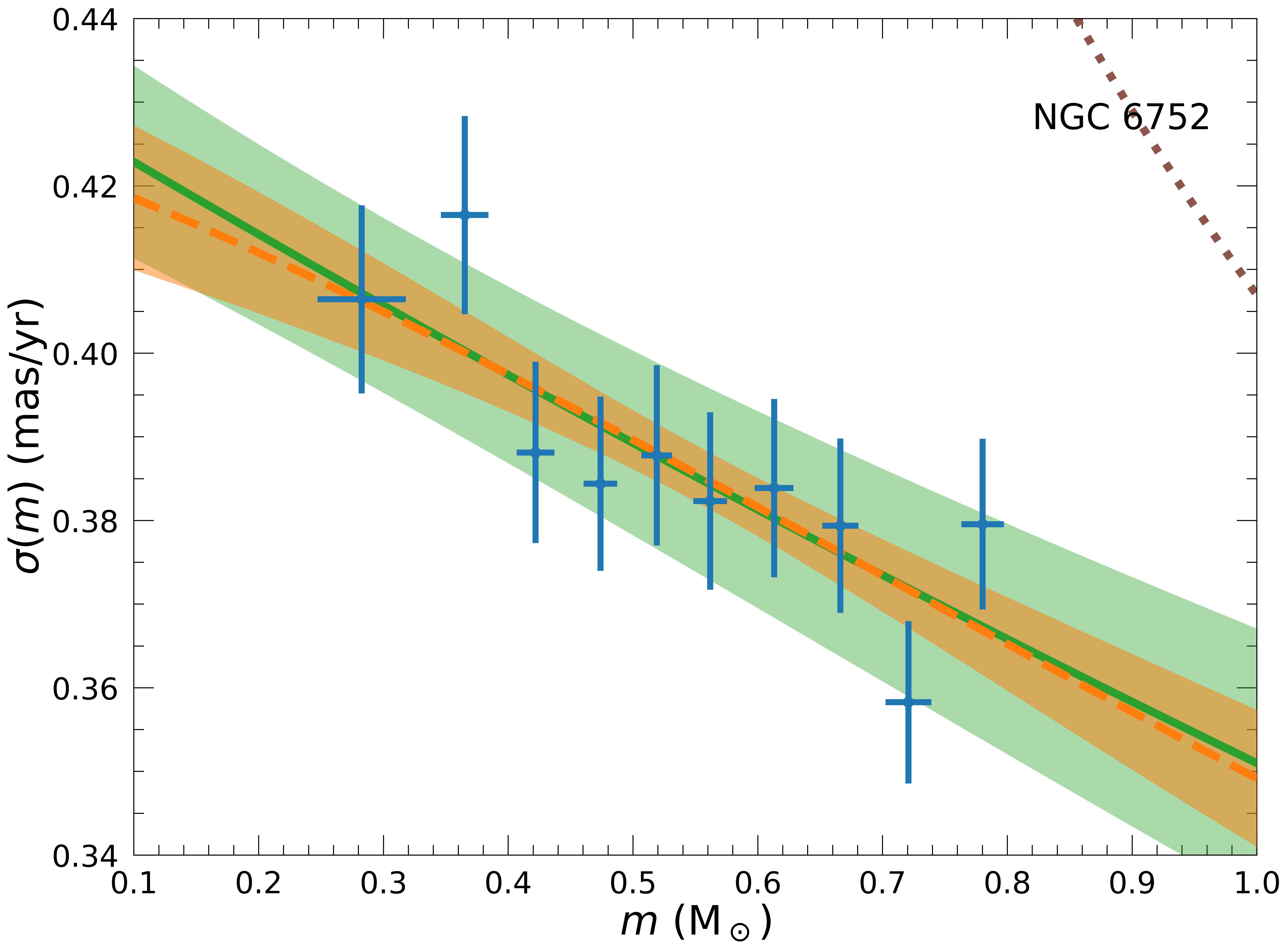}
    \label{fig:NGC6752sigmam_1par}
\end{subfigure}%
\caption{Velocity dispersion as function of stellar mass for the analyzed clusters (without \object{NGC 6397} given in Sect. \ref{subsec:VaryingPHI0}). The error bars are the binned data from \citet{Watkins2022}, the continuous green line is our best-fit with the \citet{Bianchini2016} fitting function with its error band, the dotted brown line is the equipartition limit and the dashed orange line is our model best-fit assuming \citet{Baumgardt2023} mass function slope, with its confidence band. Estimated parameters are given in Table \ref{tab:params_alphaBaumgardt2023}.}
    \label{fig:additional_sigmam}
\end{figure*}

\begin{figure*}[htbp]
\centering
\begin{subfigure}{.39\textwidth}
  \centering
  \includegraphics[width=0.95\linewidth]{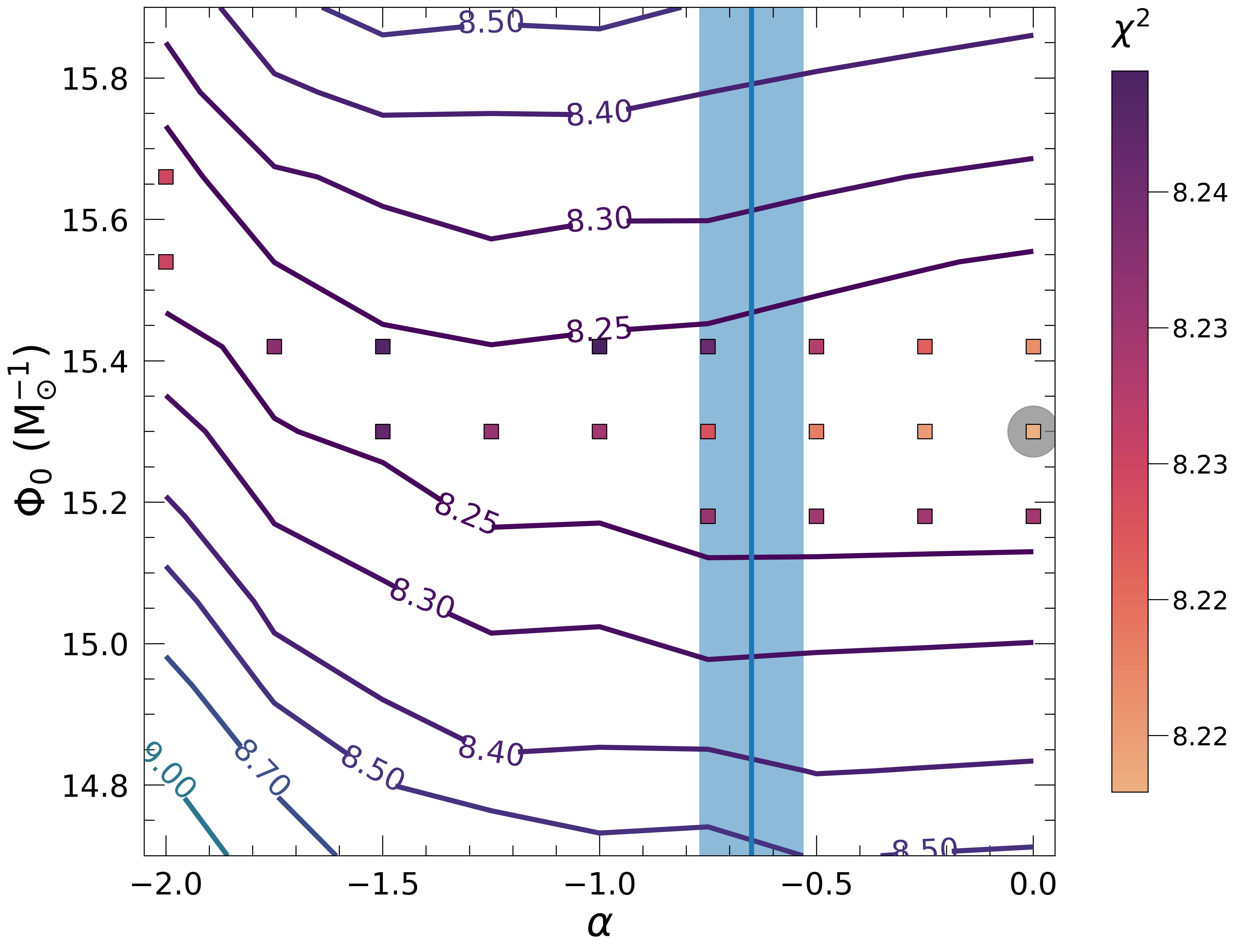}
    \caption{NGC 104}
    \label{fig:NGC104contour}
\end{subfigure}%
\hspace{0.05\textwidth}
\begin{subfigure}{.39\textwidth}
  \centering
  \includegraphics[width=0.95\linewidth]{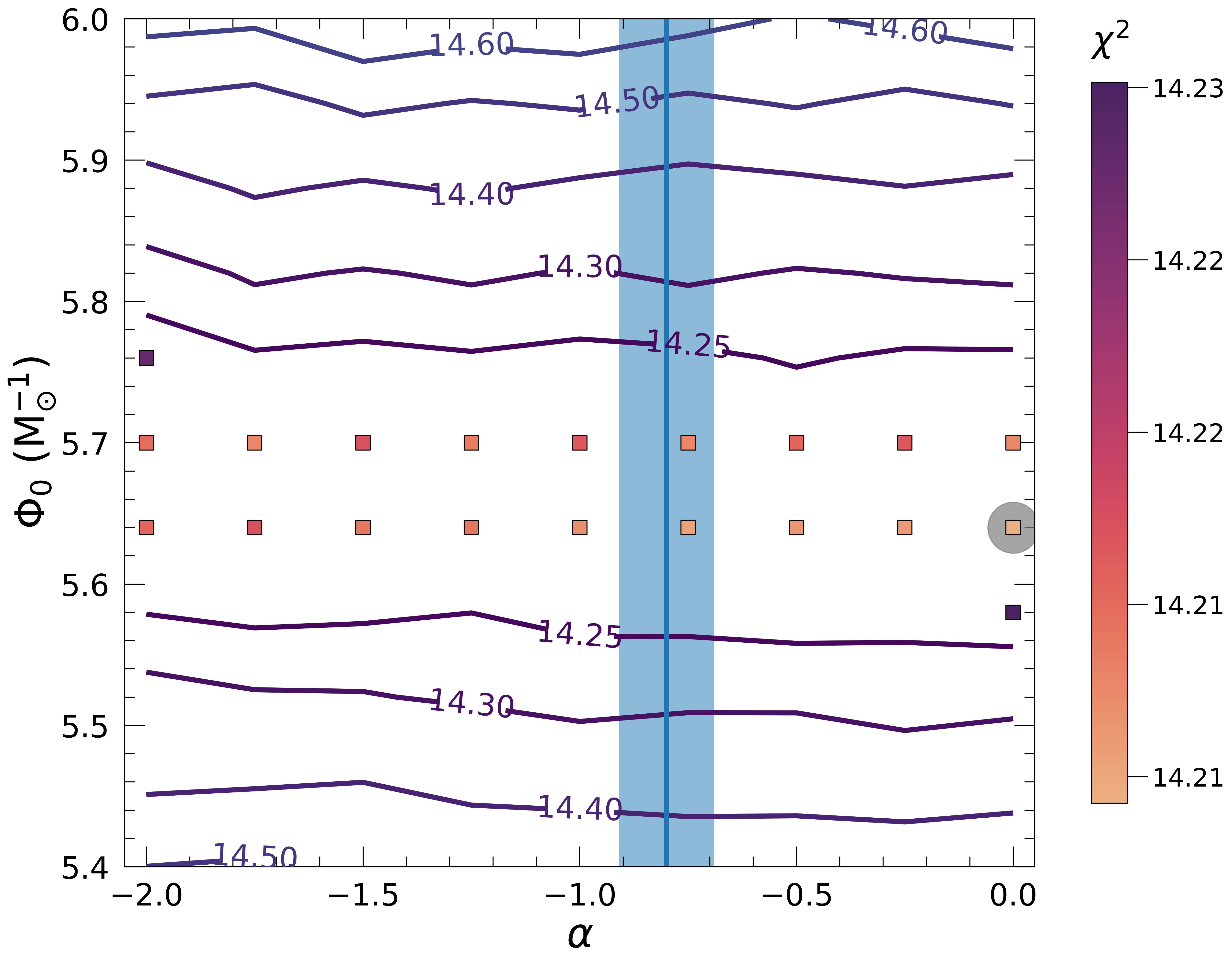}
    \caption{NGC 5139}
    \label{fig:NGC5139contour}
\end{subfigure}\\
\centering
\begin{subfigure}{.39\textwidth}
  \centering
  \includegraphics[width=0.95\linewidth]{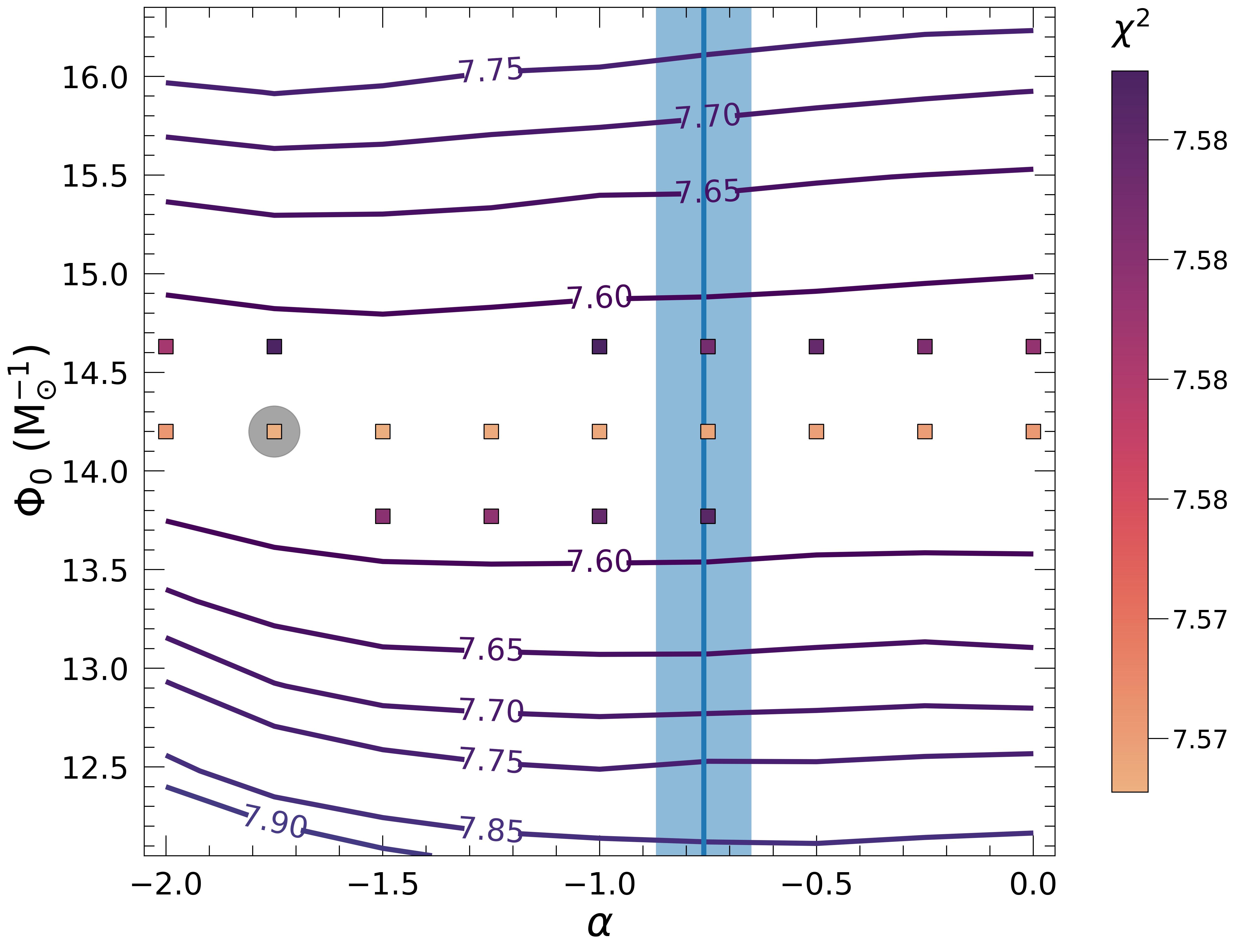}
    \caption{NGC 5904}
    \label{fig:NGC5904contour}
\end{subfigure}%
\hspace{0.05\linewidth}
\begin{subfigure}{.39\textwidth}
  \centering
  \includegraphics[width=0.95\linewidth]{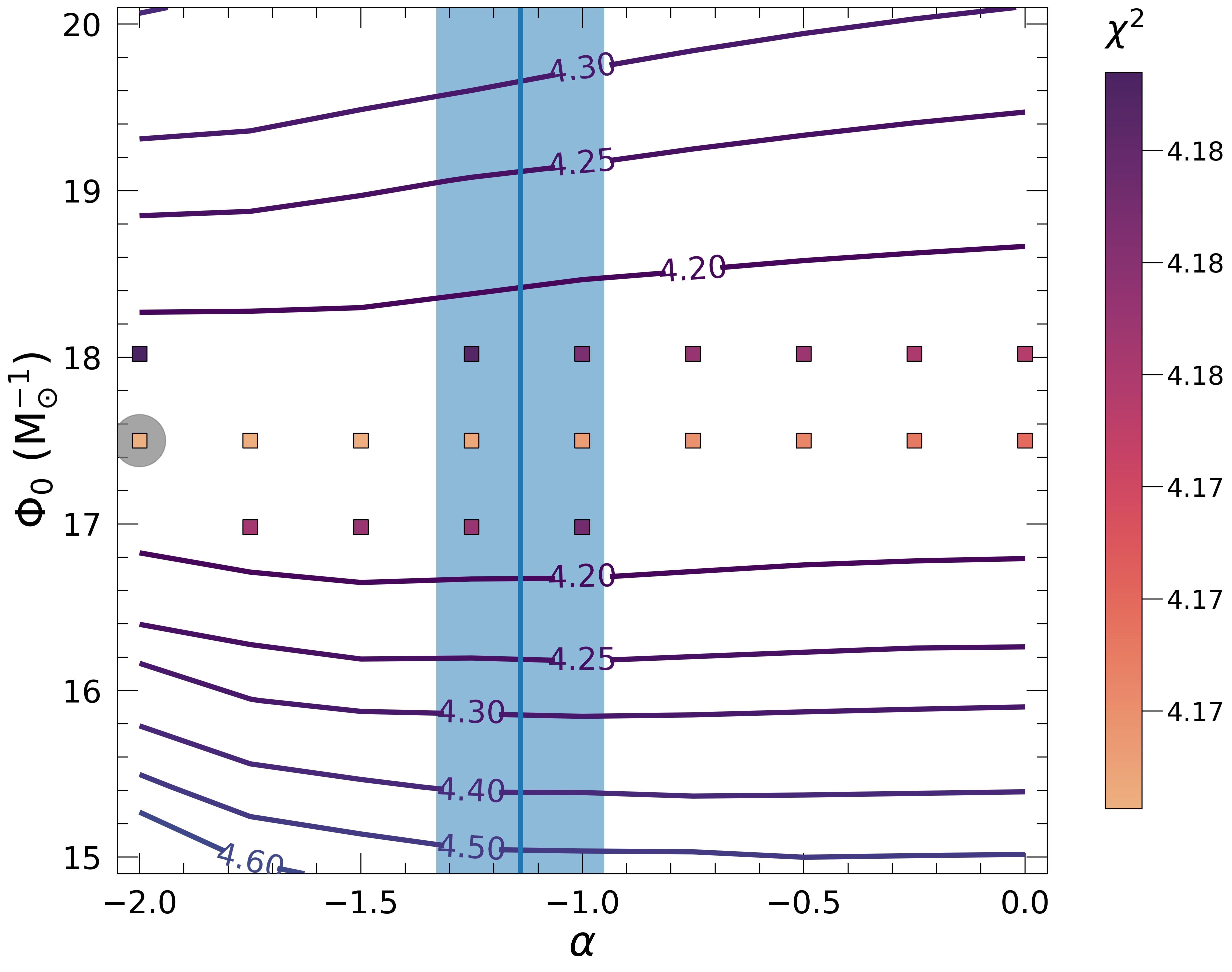}
    \caption{NGC 6266}
    \label{fig:NGC6266contour}
\end{subfigure}\\
\centering
\begin{subfigure}{.39\textwidth}
  \centering
  \includegraphics[width=0.95\linewidth]{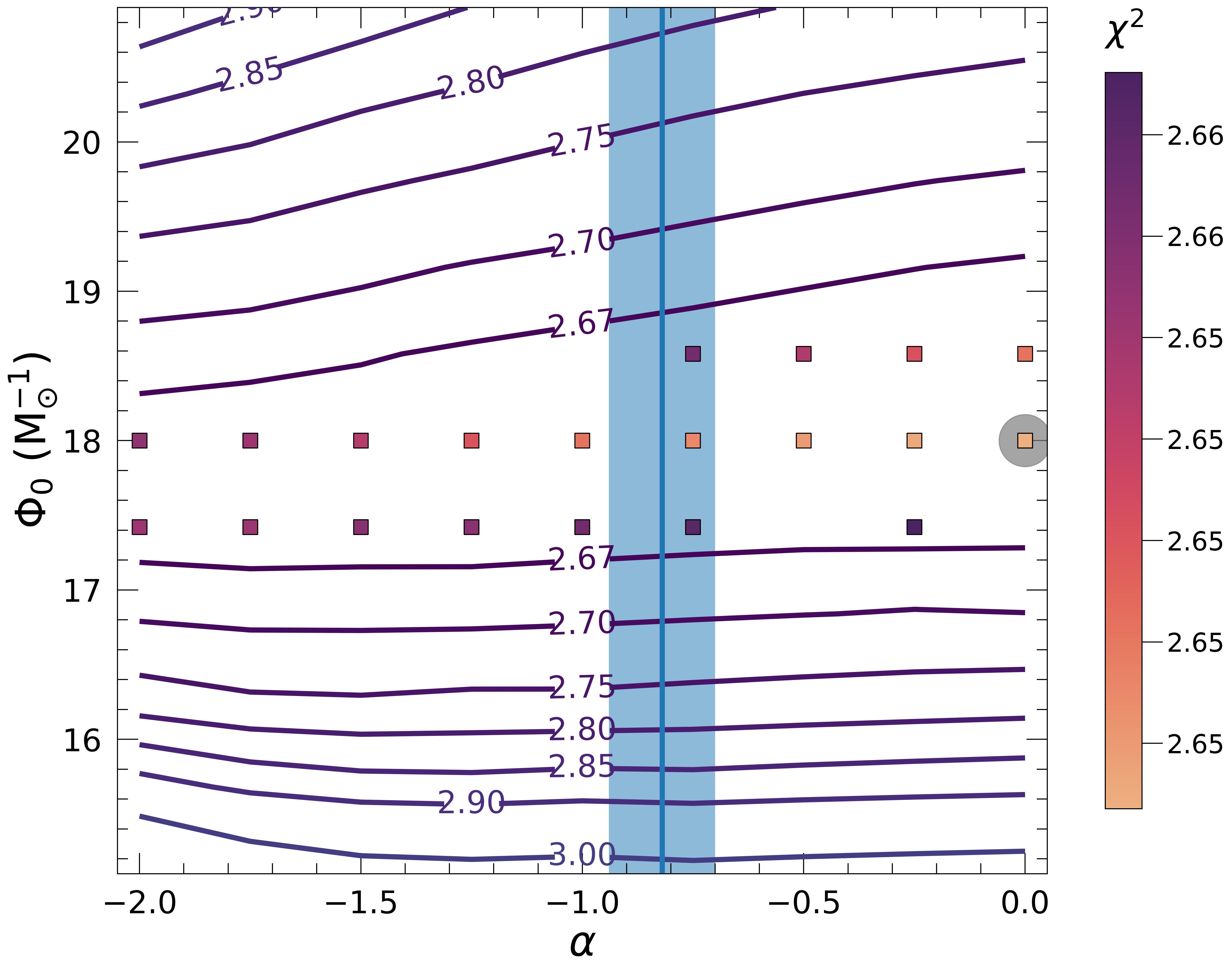}
    \caption{NGC 6341}
    \label{fig:NGC6341contour}
\end{subfigure}%
\hspace{0.05\linewidth}
\begin{subfigure}{.39\textwidth}
  \centering
  \includegraphics[width=0.95\linewidth]{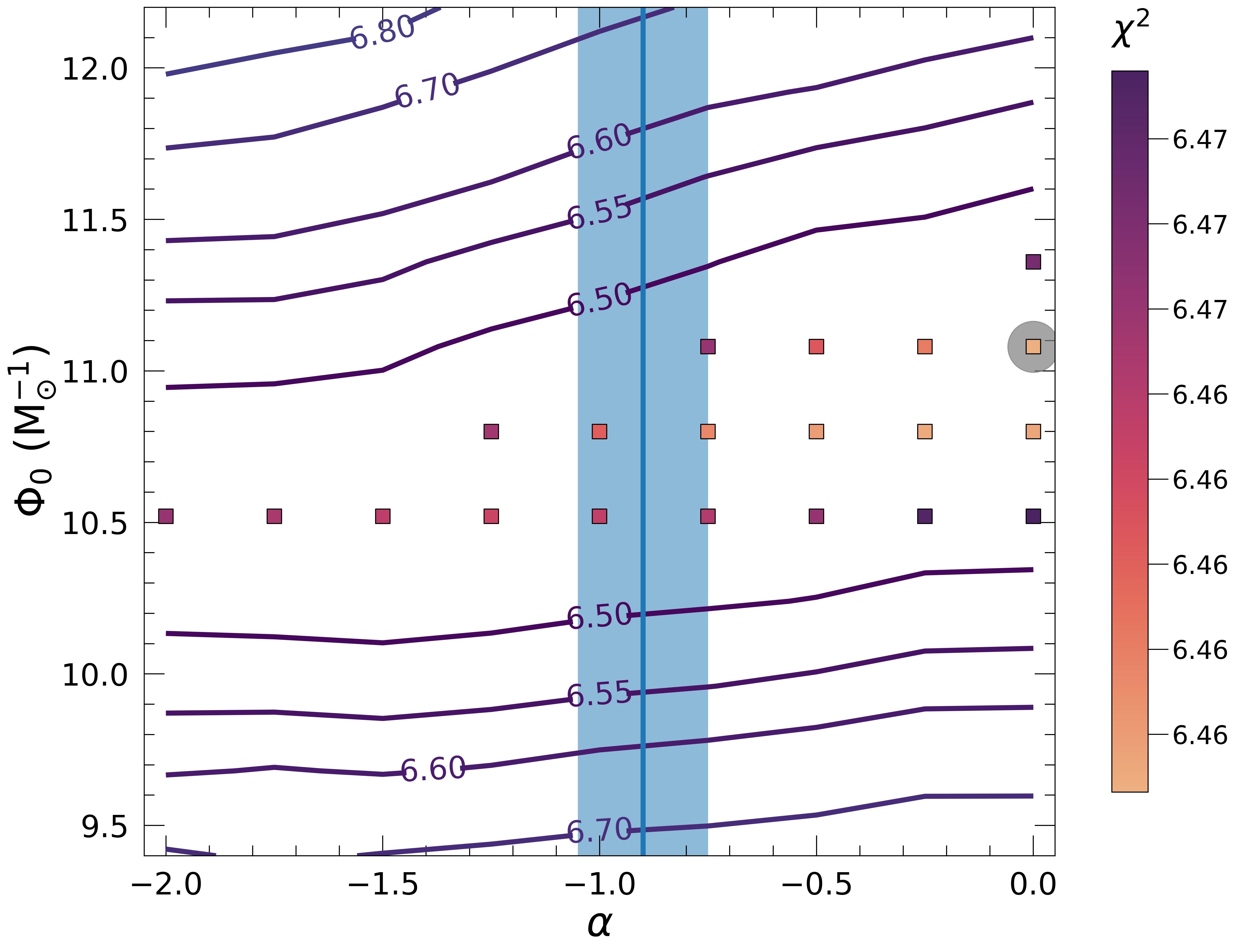}
    \caption{NGC 6656}
    \label{fig:NGC6656contour}
\end{subfigure}\\
\centering
\begin{subfigure}{.39\textwidth}
  \centering
  \includegraphics[width=0.95\linewidth]{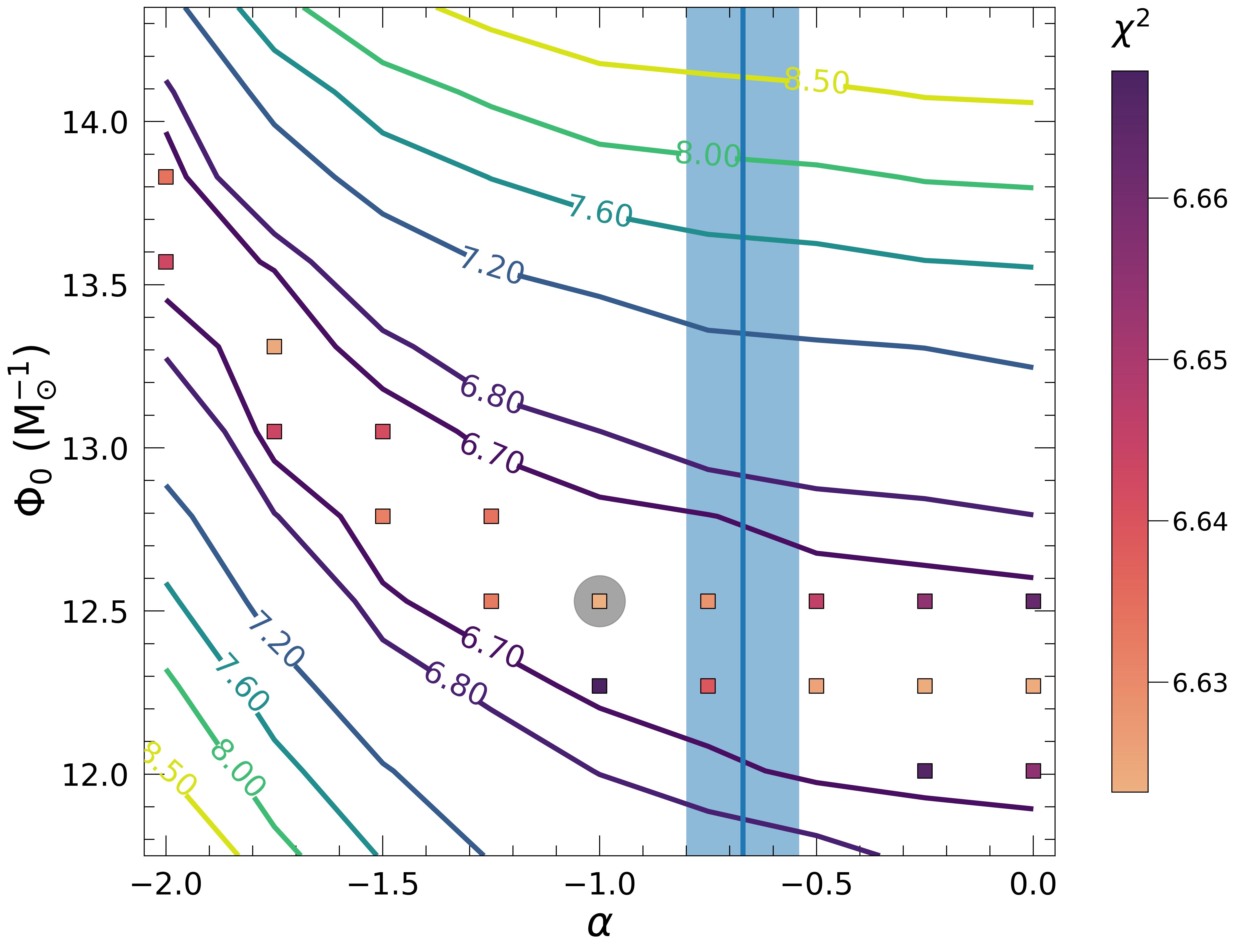}
    \caption{NGC 6752}
    \label{fig:NGC6752contour}
\end{subfigure}%
\caption{Contour plots for analyzed GCs (without \object{NGC 6397} given in the main text) in the $\Phi_0$-$\alpha$ plane showing the $\chi^2$ levels. The blue band gives the slope from \citet{Baumgardt2023}. Squares indicate locations around the $\chi^2$ minimum, identified by a big gray circle.}
\label{fig:additional_contours}
\end{figure*}

\begin{figure*}[htbp]
\centering
\begin{subfigure}{.32\textwidth}
  \includegraphics[width=\linewidth]{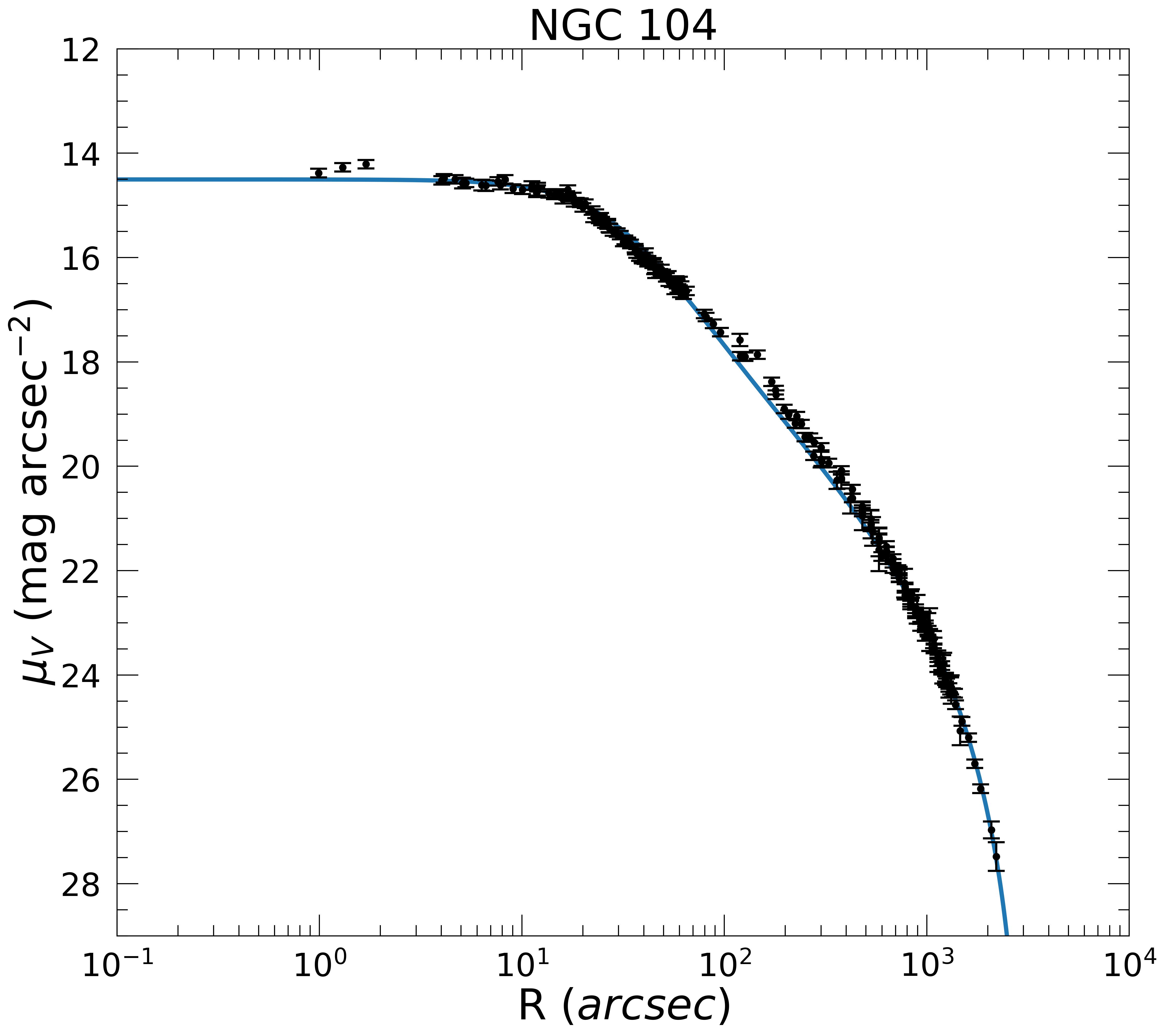}
    \label{fig:NGC104_SBP}
\end{subfigure}%
\hspace{0.05\textwidth}
\begin{subfigure}{.32\textwidth}
  \centering
  \includegraphics[width=\linewidth]{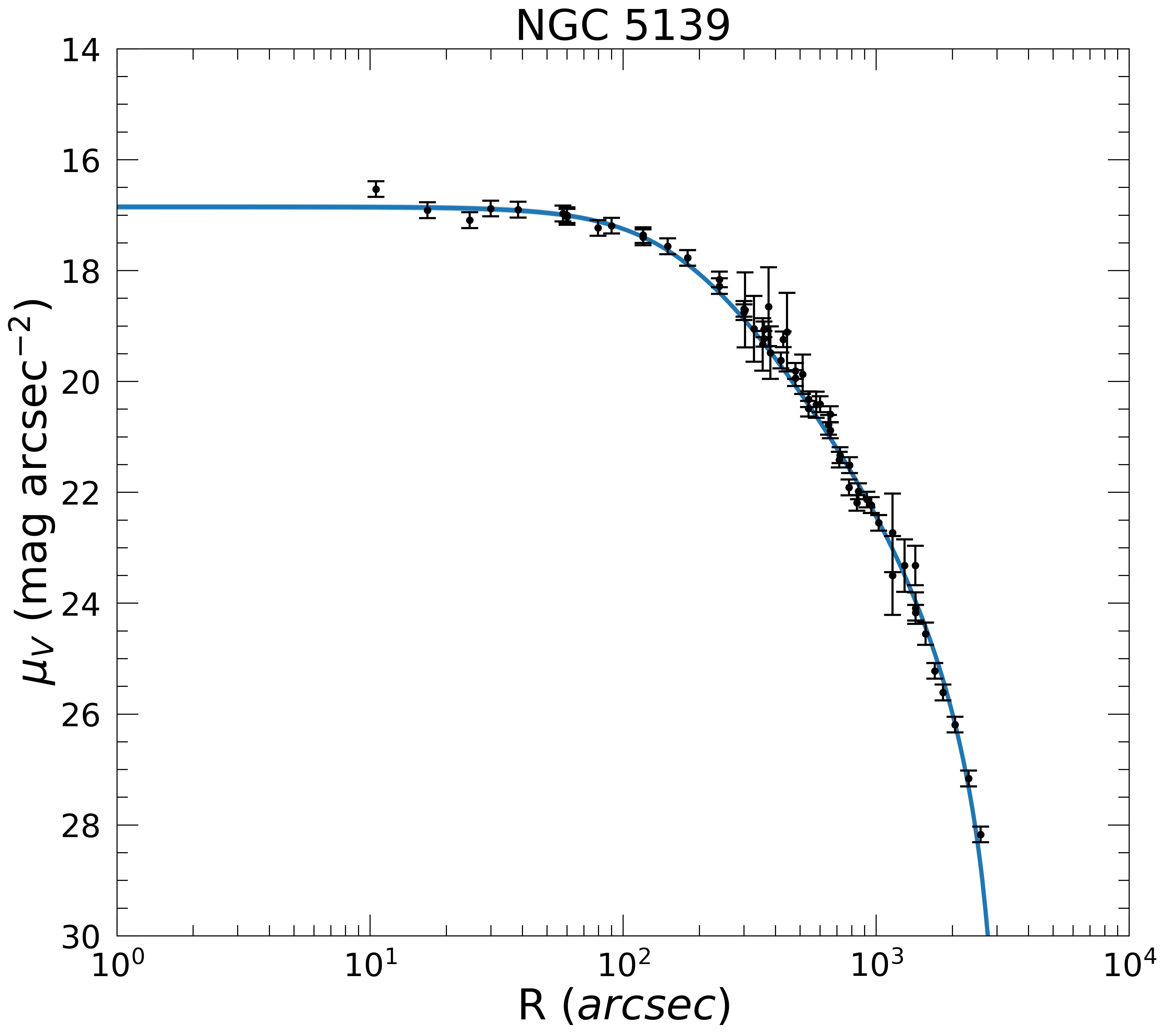}
    \label{fig:NGC5139_SBP}
\end{subfigure}\\
\centering
\begin{subfigure}{.32\textwidth}
  \centering
  \includegraphics[width=\linewidth]{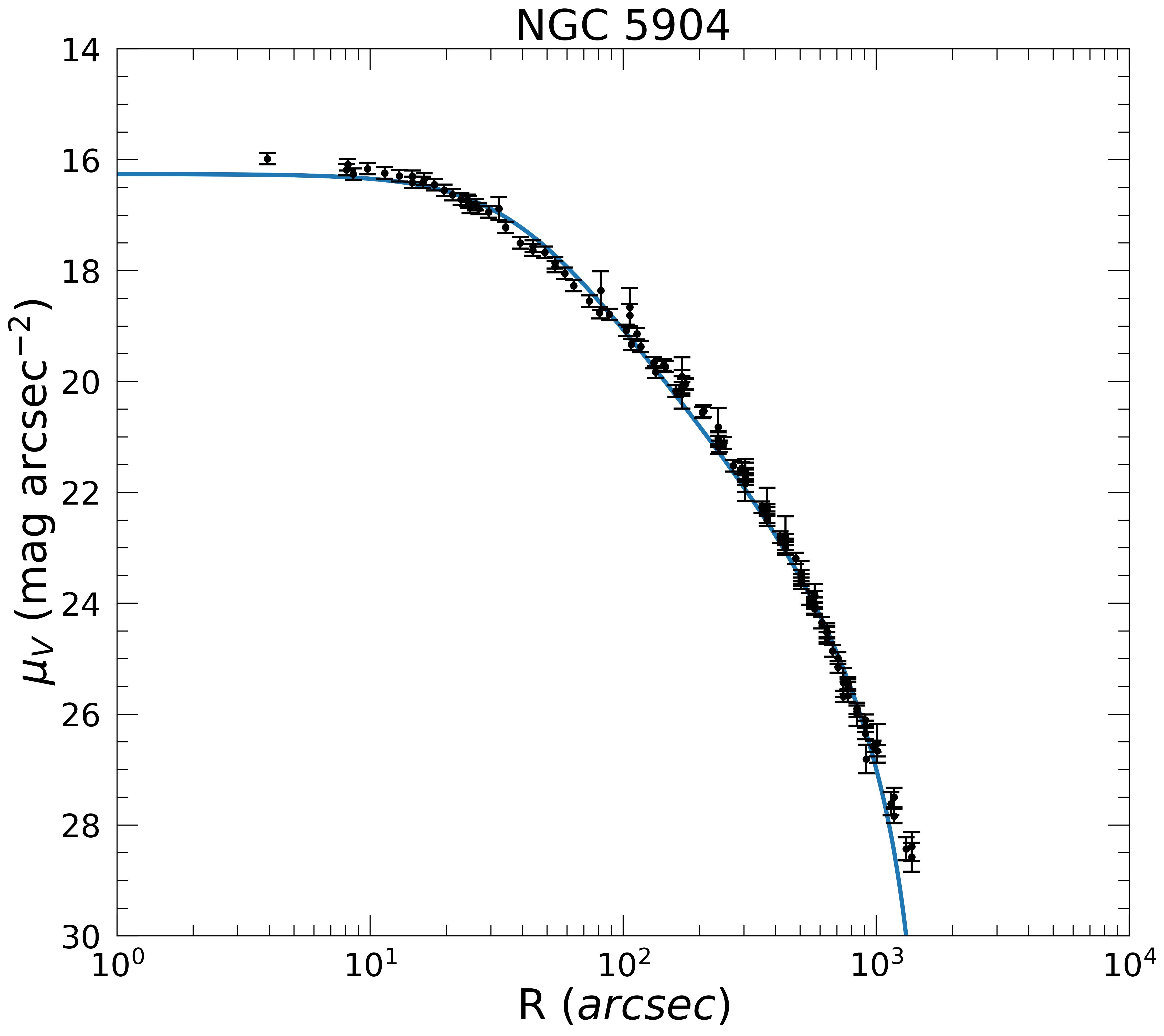}
    \label{fig:NGC5904_SBP}
\end{subfigure}%
\hspace{0.05\textwidth}
\begin{subfigure}{.32\textwidth}
  \centering
  \includegraphics[width=\linewidth]{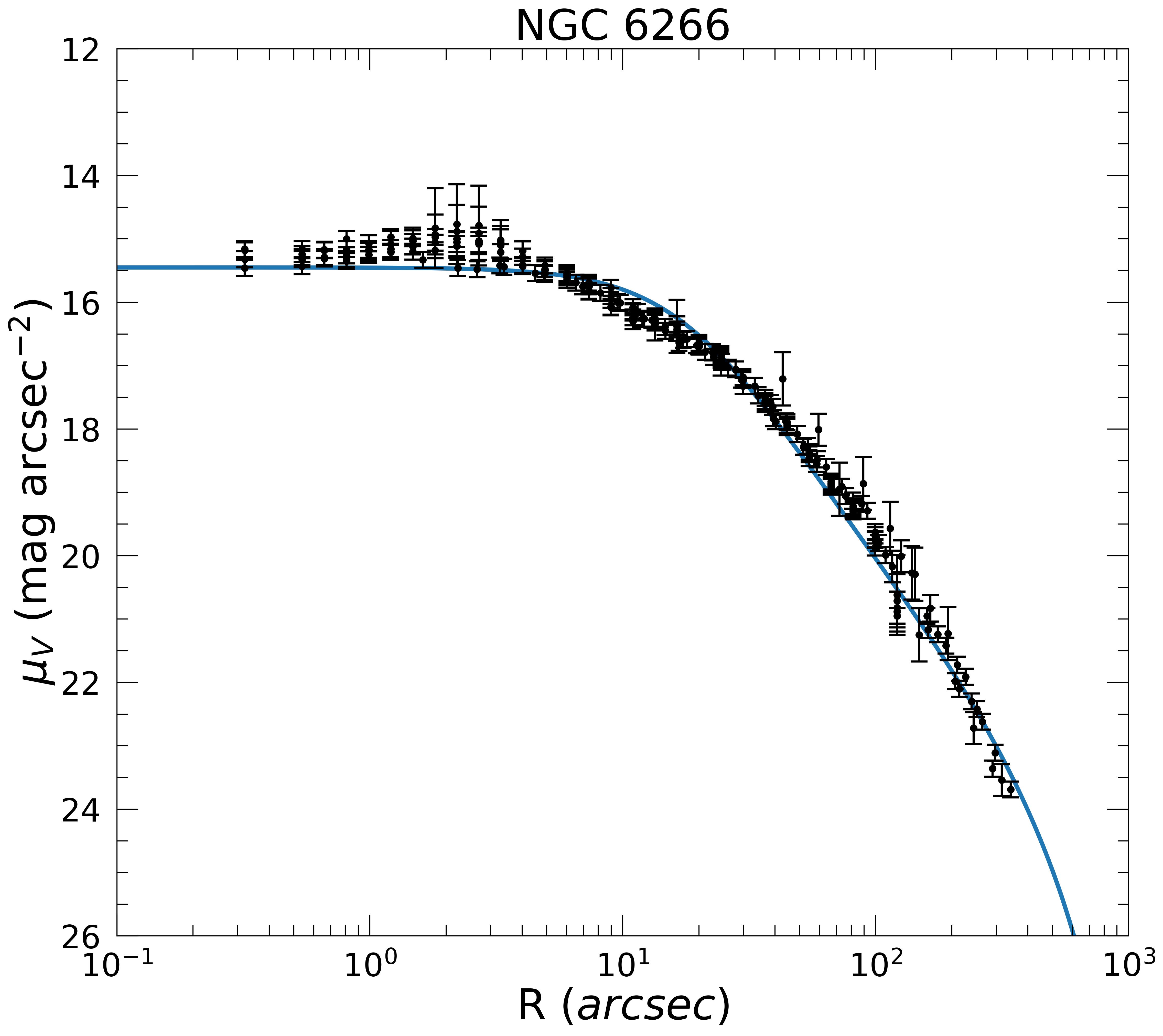}
    \label{fig:NGC6266_SBP}
\end{subfigure}\\
\centering
\begin{subfigure}{.32\textwidth}
  \centering
  \includegraphics[width=\linewidth]{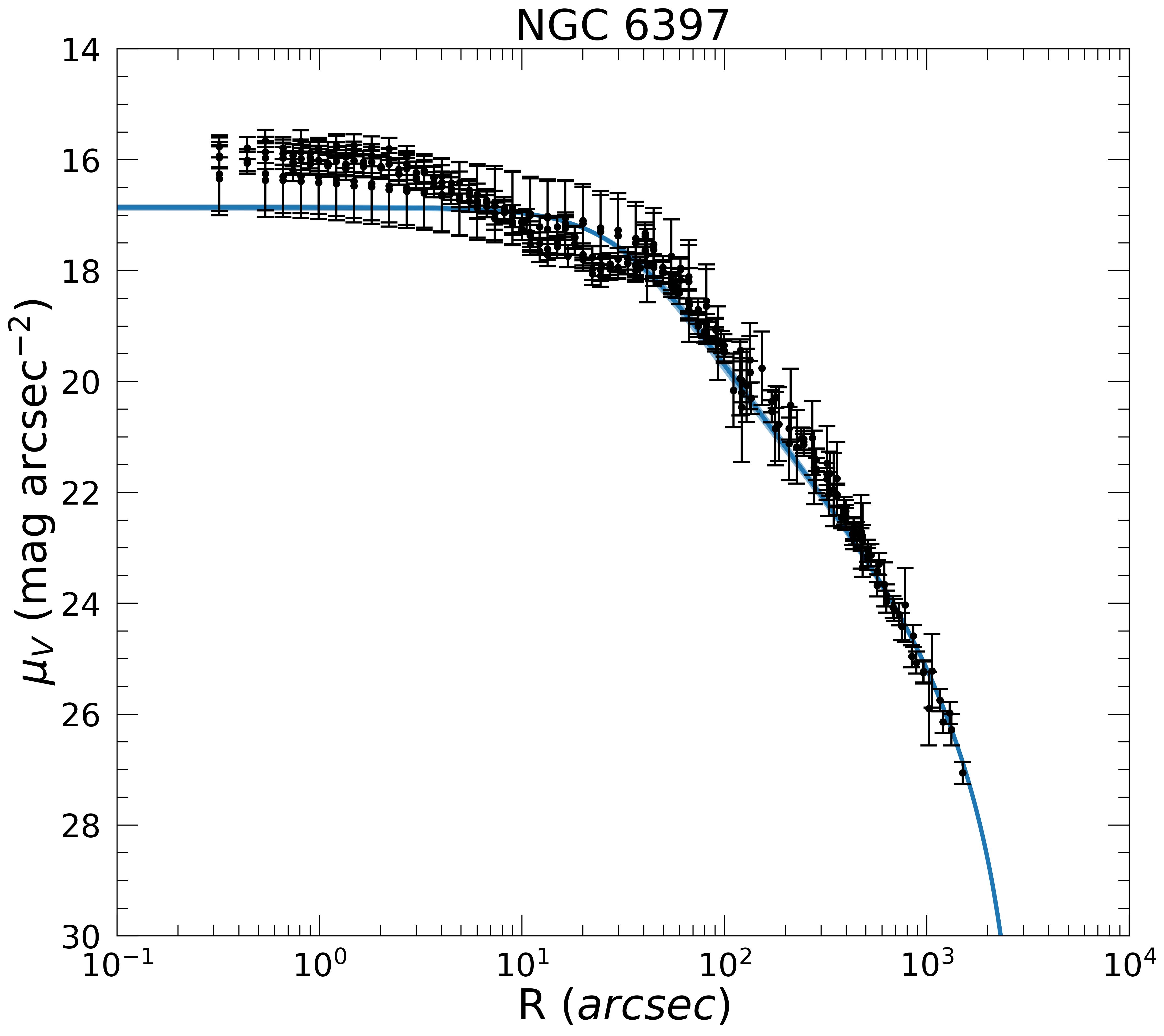}
    \label{fig:NGC6397_SBP}
\end{subfigure}%
\hspace{0.05\textwidth}
\begin{subfigure}{.32\textwidth}
  \centering
  \includegraphics[width=\linewidth]{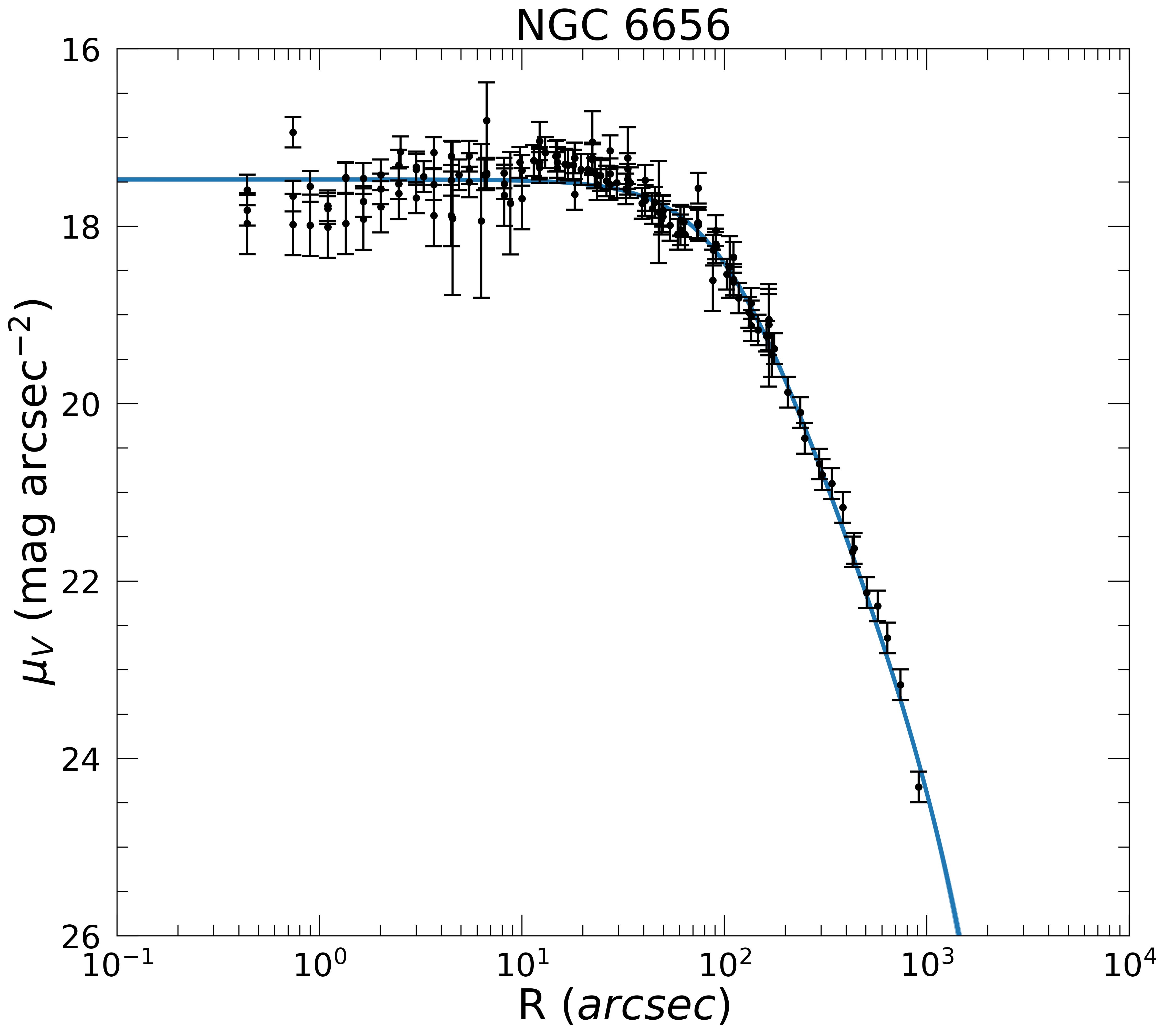}
    \label{fig:NGC6656_SBP}
\end{subfigure}\\
\centering
\begin{subfigure}{.32\textwidth}
  \centering
  \includegraphics[width=\linewidth]{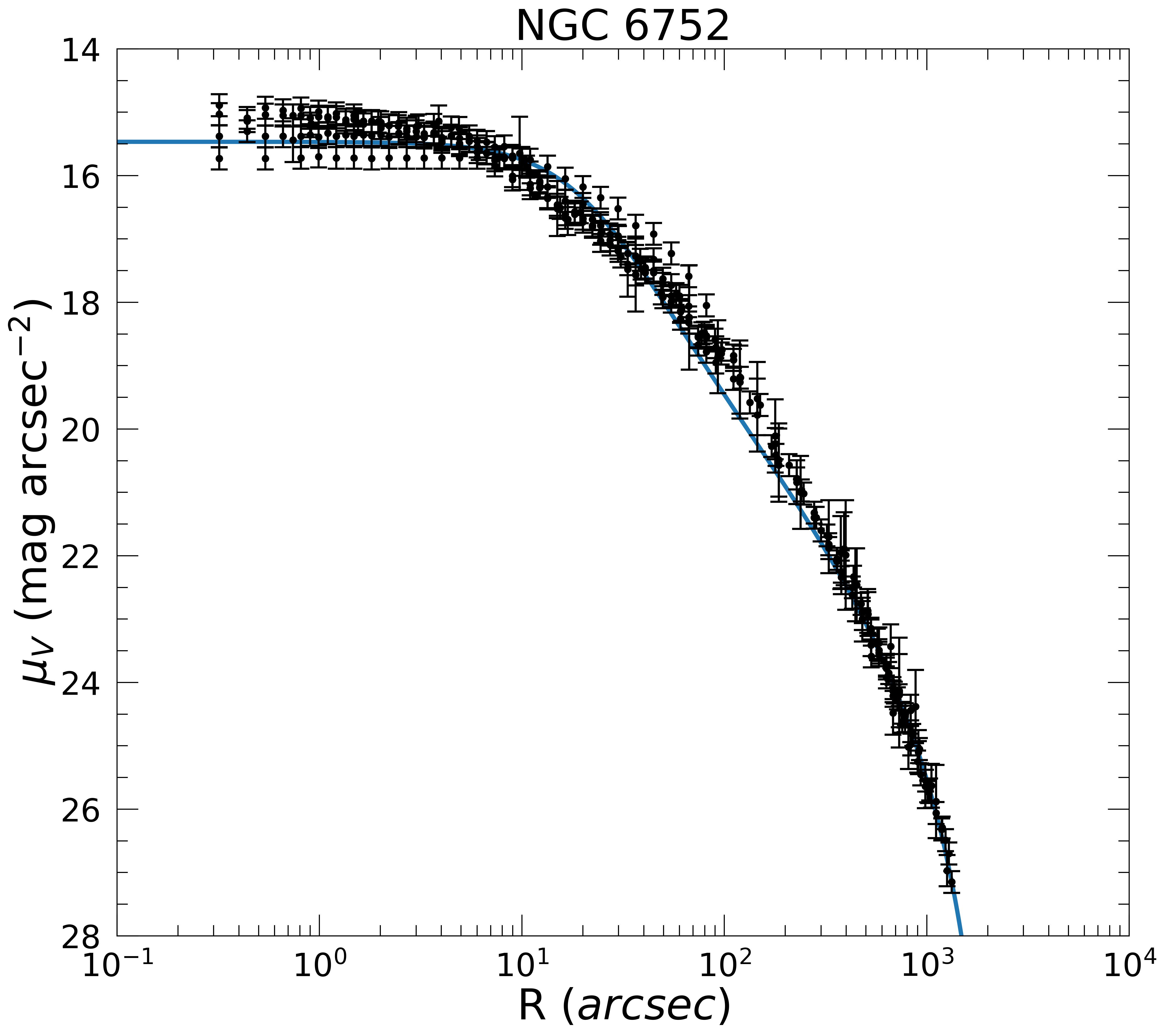}
    \label{fig:NGC6752_SBP}
\end{subfigure}%
\caption{Surface brightness profiles for the analyzed clusters (without \object{NGC 6341} given in Sect. \ref{subsec:FittingSBPs}). The dots with error bars are the data from \citet{Trager1995} analyzed following the work by \citet{McLaughlin&vanderMarel2005} and \citet{Zocchi+2012}. The continuous blue line is our model best-fit with its confidence band, obtained assuming \citet{Baumgardt2023} mass function slope and theoretical BaSTI isochrones \citet{Hidalgo+2018,Pietrinferni+2021, Salaris+2022, Pietrinferni+2024} with 13 Gyr, $[\alpha/\element{Fe}]=+0.4$, $Y = 0.247$ and metallicity [\element{Fe}/\element{H}] taken from the \citet{Harris1996} catalog (2010 edition). The estimated parameters are given in Table \ref{tab:params_fitSBPs}.}
    \label{fig:additional_SBPs}
\end{figure*}

\begin{figure}[htbp]
    \centering
    \includegraphics[width=\linewidth]{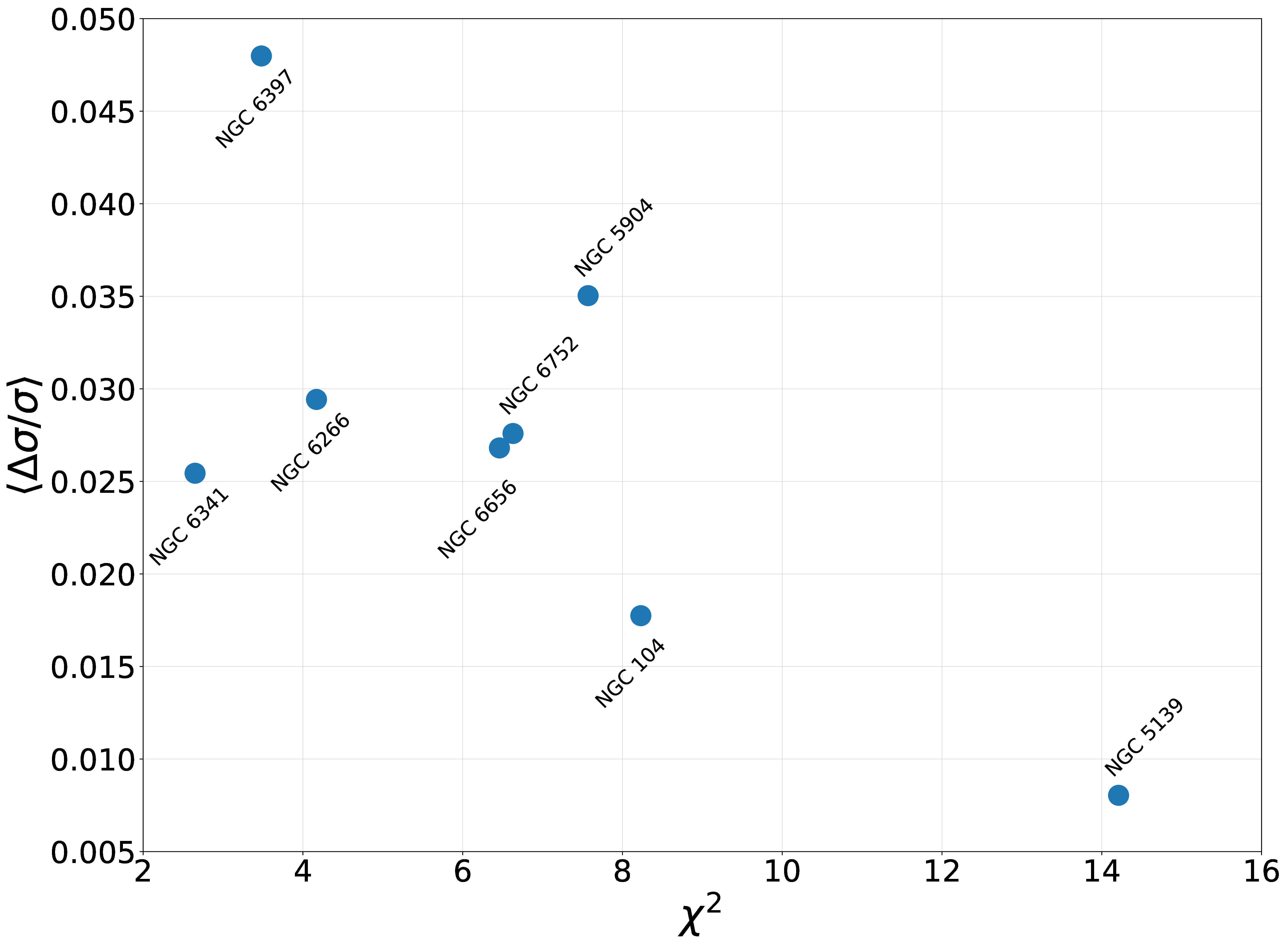}
    \caption{Relation between the $\chi^2$ test statistic of our fitting procedure in Sect. \ref{subsec:VaryingPHI0} with the average relative error on $\sigma(m)$ from \citet{Watkins2022}.}
    \label{fig:chisq_1par_RelErrsigmam}
\end{figure}

\begin{figure}[htbp]
    \centering
    \includegraphics[width=\linewidth]{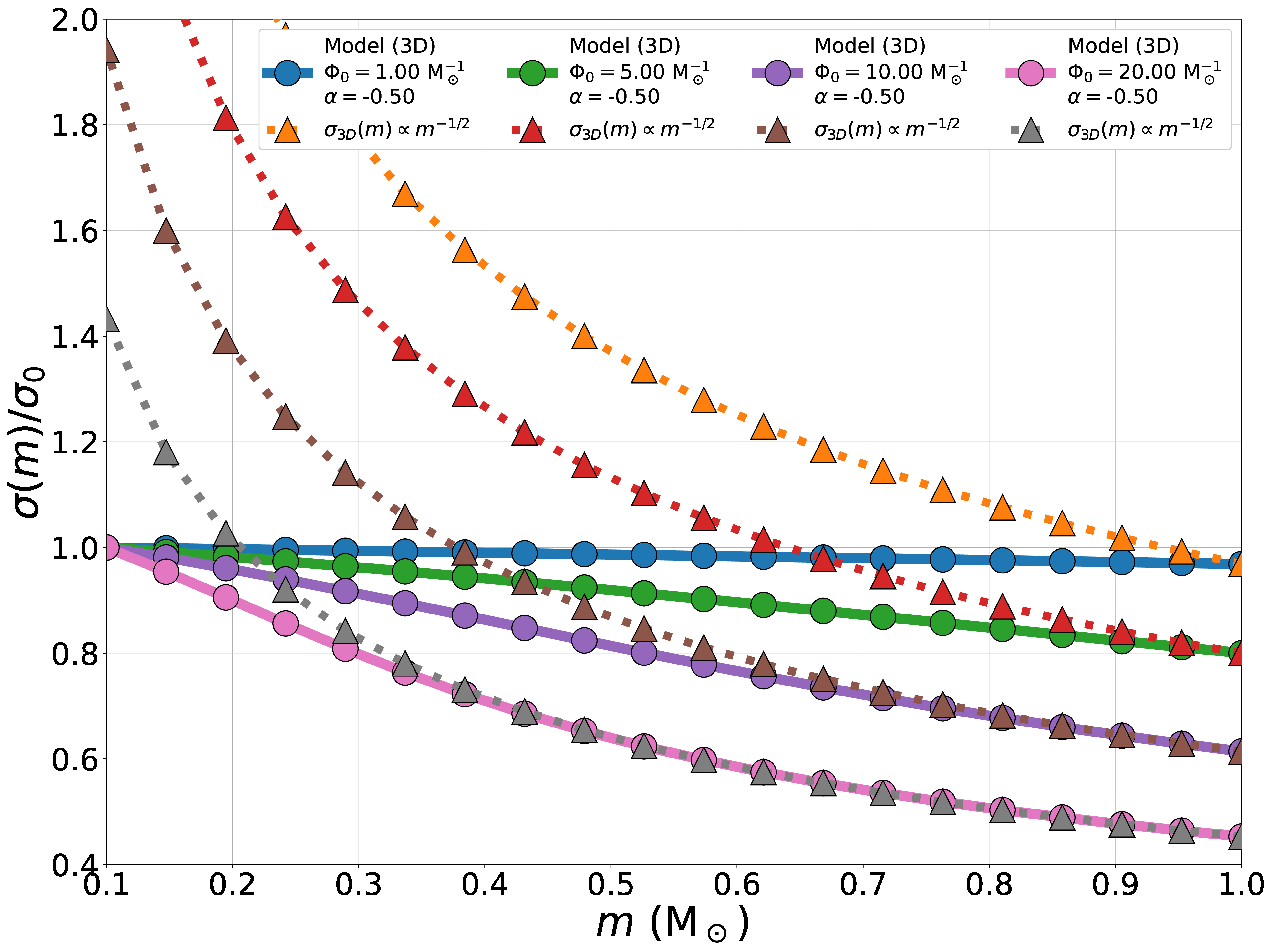}
    \caption{Model prediction of the velocity dispersion as function of stellar mass for different values of $\Phi_0$ (continuous lines with circles), compared to the complete equipartition limit (dashed lines with triangles). The plots are obtained averaging the dispersion radial profile in the core.}
    \label{fig:sigmam_3D_eqlimit}
\end{figure}

\end{appendix}


\end{document}